\documentclass[12pt]{article}
\usepackage[utf8]{inputenc}
\usepackage{cite}
\usepackage{amsmath,amssymb,amsthm} 
\usepackage{hyperref}
\usepackage{graphicx}
\usepackage{caption}
\usepackage{subcaption}
\usepackage[usenames, dvipsnames]{color} 

\newcommand{\be}{\begin{equation}}
\newcommand{\ee}{\end{equation}}
\newcommand{\bea}{\begin{eqnarray}}
\newcommand{\eea}{\end{eqnarray}}
\newcommand{\ol}{\overline}
\newcommand{\diff}{\mathrm{d}}

\begin{document}

\thispagestyle{empty}
\vspace*{-.2cm}
\noindent

\vspace*{1.2cm}
\begin{center}

{\Large\bf Euclidean wormholes, baby universes, \\[.4cm] and their impact on  particle physics and cosmology}
\\[2cm]

{\large Arthur Hebecker, Thomas Mikhail, Pablo Soler \\[6mm]}

{\it
Institute for Theoretical Physics, University of Heidelberg, 
Philosophenweg 19,\\ D-69120 Heidelberg, Germany\\[3mm]

{\small\tt (\,a.hebecker,\,t.mikhail,\,p.soler~@thphys.uni-heidelberg.de\,)} }\\[.5cm]
July 2, 2018
\\[1.6cm]
\end{center}

\begin{abstract}

The euclidean path integral remains, in spite of its familiar problems, an important approach to quantum gravity. One of its most striking and obscure features is the appearance of gravitational instantons or wormholes. These renormalize all terms in the Lagrangian and cause a number of puzzles or even deep inconsistencies, related to the possibility of nucleation of ``baby universes''. In this review, we revisit the early controversies surrounding these issues as well as some of the more recent discussions of the phenomenological relevance of gravitational instantons. In particular,  wormholes are expected to break the shift symmetries of axions or Goldstone bosons non-perturbatively. This can be relevant to large-field inflation and connects to arguments made on the basis of the Weak Gravity or Swampland conjectures. It can also affect Goldstone bosons which are of physical interest in the context of the strong CP problem or as dark matter.

\end{abstract}

\newpage

\tableofcontents

\section{Introduction}
It is reasonable to think that a consistent theory of quantum gravity has to allow for topology change. Indeed, if the euclidean path integral has any relevance at all, then it appears unnatural to forbid 4-manifolds with non-trivial topology. After all, they are locally indistinguishable from $\mathbb{R}^4$. Further evidence in favor of topology change comes, for example, from string theory: String interactions and loops rely entirely on toplogy change in the worldsheet theory, the latter being a relatively well-understood examples of 2d quantum gravity. In addition, 10d supergravity theories with their stringy UV completion involve controlled examples of topology change. These occur if one dynamically moves through special loci in Calabi-Yau moduli space, e.g. through a conifold point. 

\begin{figure}[ht]
	\centering
	\includegraphics[width=0.4\linewidth]{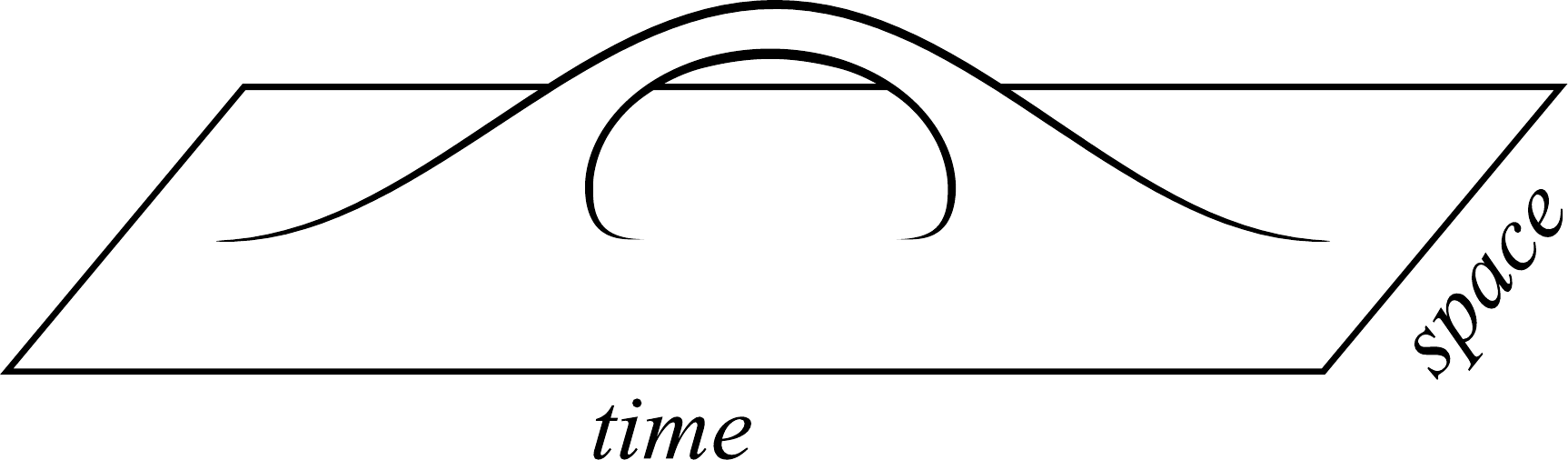}
	\caption{Wormhole corresponding to the creation and absorption of a baby universe.}
	\label{fig:sw}
\end{figure}

However, our point of departure will be more simple minded, focusing on topology change in 4d effective quantum gravity. Consider the evolution of 3d spatial manifolds in time. It is natural to think that in the course of this evolution an $\mathbb{R}^3$ can transit to an $\mathbb{R}^3$ plus an $S^3$ `baby universe', which subsequently reunite becoming again an $\mathbb{R}^3$ (cf.~Fig.~\ref{fig:sw}). This can be viewed as a tunneling transition, which gains quantitative support from the existence of a corresponding euclidean solution -- the Giddings-Strominger wormhole~\cite{Giddings:1987cg}. While topology change has been discussed before~\cite{Wheeler:1955zz, Regge:1961px, Hawking:1979zw, Ellis:1983jz, Lavrelashvili:1987jg, Hawking:1987mz, Hawking:1988ae}, the Giddings-Strominger solution~\cite{Giddings:1987cg} and especially the application to the cosmological constant problem suggested by Coleman~\cite{Coleman:1988tj} led to an enormous spike of activity~\cite{Coleman:1988cy, Hawking:1988wm, Giddings:1988cx, Lee:1988ge, Rubakov:1988jf, Klebanov:1988eh, Fischler:1988ia, Kaplunovsky, Preskill:1988na, Grinstein:1988wr, Nielsen:1988kf, Choi:1989ck, Preskill:1989zu, Gilbert:1989nq,  Rey:1989mg, Grinstein:1988eb, Grinstein:1988ja, Coleman:1989ky, Burgess:1989da, Brown:1989df, Polchinski:1989ae, Abbott:1989jw, Duff:1989ah, Coleman:1989zu, Tamvakis:1989mj, Grinstein:1989pc, Giddings:1989bq, Hawking:1989vs, Tamvakis:1989aq, Carlip:1989qq, Coleman:1990tz, Grinstein:1990zb, Hawking:1990in, Hawking:1990jb, Hawking:1990ue, Hawking:1991vs, Lyons:1991im,Linde:1991sk,Twamley:1992hu} (see~\cite{Coleman:1991rs} for an early overview).

As part of these investigations, severe problems in the resulting picture of a macroscopic spacetime surrounded by baby universes were uncovered~\cite{Fischler:1988ia, Kaplunovsky, Polchinski:1989ae, Hawking:1989vs}. While the interest has then subsided, important results have continued to appear over the years~\cite{Kallosh:1995hi, Gibbons:1995vg,Barcelo:1995gz, Nirov:1995cc, Rubakov:1996cn, Rubakov:1996br, Green:1997tv, Rey:1998yx, Gutperle:2002km,Maldacena:2004rf, Bergshoeff:2004fq, Bergshoeff:2004pg, Collinucci:2005opa, Bergshoeff:2005zf, Dijkgraaf:2005bp, ArkaniHamed:2007js,Bergman:2007ss,Chiodaroli:2008rj,Chiodaroli:2009cz, Cortes:2009cs, Mohaupt:2010du,Betzios:2017krj}. It has, however, neither been shown that wormholes and baby universe are unphysical nor has a satisfactory overall picture been developed. Thus, euclidean wormholes or gravitational instantons have remained a lurking fundamental issue in our understanding of quantum gravity. We emphasize that this issue is not easily dismissed as a problem of the UV completion. On the contrary, large wormholes tend to be as puzzling as small ones, such that the problems appear to be there even in the low-energy effective theory.\footnote{In this review we focus on large wormholes. An interesting and closely related topic, which lies beyond the scope of this work, are topological fluctuations of spacetime at small scales (the Planck scale~\cite{Wheeler:1955zz,Hawking:1979zw} or string scale~\cite{Iqbal:2003ds}) as constituents of a microscopic description of quantum gravity.}

More recently, the interest in wormholes has been renewed in the context of large-field inflation, axion-physics, and the widespread excitement (see e.g.~\cite{Cheung:2014vva, delaFuente:2014aca, Rudelius:2015xta, Montero:2015ofa, Brown:2015iha, Bachlechner:2015qja, Hebecker:2015rya,Brown:2015lia, Heidenreich:2015wga, Junghans:2015hba, Kooner:2015rza, Kaloper:2015jcz, Kappl:2015esy, Choi:2015aem, Klaewer:2016kiy}) about the Weak Gravity Conjecture and the Landscape/Swampland paradigm \cite{ArkaniHamed:2006dz, Vafa:2005ui, Ooguri:2006in,Ooguri:2016pdq,Brennan:2017rbf}. This is natural since wormholes have the potential to break global symmetries, such as the shift symmetry of the axion. In addition, they may be considered the macroscopic, gravitational version of instantons in pretty much the same way as charged black holes are the macroscopic version of charged particles. Thus, the interest in the Weak Gravity Conjecture and its implications for phenomenology naturally lead to an enhanced interest in (euclidean) wormholes~\cite{Montero:2015ofa, Hebecker:2015rya, Heidenreich:2015nta, Harlow:2015lma, Hebecker:2016dsw, Hertog:2017owm, Alonso:2017avz, Ruggeri:2017grz,Shiu:2018wzf}. 

Our review is motivated in several ways: First, as just explained, it is timely to reconsider the wormhole issue in view of the growing interest in generic quantum gravity constraints on effective field theories. Second, the unsolved problems from the 90's are, in our opinion, as important as ever. Additionally, one of the main phenomonelogical targets in the otherwise rather theory-driven wormhole debate have always been axions.\footnote{
We 
will use the name axion for any shift-symmetric periodic scalar, even if unrelated to QCD.
}
Since axions are becoming more and more central in Beyond-the-Standard-Model research, scrutinizing their generic features is of particular importance. 
Finally, we believe that the post-90's theoretical developments of AdS/CFT, holography and (gravitational) entanglement have not yet been fully exploited in the context of euclidean wormholes. Thus, significant technical progress may be expected concerning the fundamental issues raised by those objects. 

In the long run, we can think of two different outcomes: On the one hand, wormhole effects may turn out to be absent from certain theories, in particular from the 4d quantum gravity describing the real world. This would solve many puzzles. Advocates of this possibility have to address a number of questions. In particular, what is the specific mechanism behind this `wormhole censorship'? As we will argue, it appears difficult to imagine such a mechanism which would not also forbid topology change in general. This, of course, would be a radical step. Related to this: How can we forbid wormholes in 4d while maintaining their central role in the 2d quantum gravity known as string theory? Furthermore, if wormholes are forbidden, what is the generic gravitational effect responsible for the breaking of global shift symmetries of axions? On the other hand, if wormholes exist, they represent a radical departure from standard interpretations of effective field theories. As we will describe, the correct understanding of their effects requires solving numerous fundamental problems.  In the hope that these questions can be successfully addressed in the near future, we consider it worthwhile summarizing the state of the art and describing the main puzzles and open issues posed by wormholes.

We start in Sect.~\ref{sec:GravitationalInstantons} by recalling how instantons (of either gauge-theoretic or stringy nature) generate a potential for any scalar to which they are minimally coupled. We then describe the famous Giddings-Strominger solution~\cite{Giddings:1987cg}, which corresponds to a throat with cross section $S^3$, connecting two points in $\mathbb{R}^4$ (cf.~Fig.~\ref{fig:sw}). The throat is supported by $H_3$ flux, and the dual of $H_3$ is the field strength of an axion. This axion then naturally couples to the two wormhole ends, which can locally be interpreted as instanton and anti-instanton. The axionic shift symmetry is potentially broken by a `dilute gas' of such wormholes. We also briefly comment on dilatonic instantons as they generically arise in string theory, emphasizing that it has by now been established that wormhole solutions do really arise in string-derived models~\cite{Tamvakis:1989aq, Bergshoeff:2004pg, ArkaniHamed:2007js, Hertog:2017owm}.

Next, in Sect.~\ref{sec:WormholeEffects}, we discuss how the low-energy effective action is corrected by wormholes (of Giddings-Strominger type and, more generally, by any `spacetime handles' of the form displayed in Fig.~\ref{fig:sw}). We follow the pioneering work by Coleman \cite{Coleman:1988tj} and Preskill \cite{Preskill:1988na}. Crucially, in contrast to instantons, wormholes induce a bilocal action, which has the potential to break locality or even quantum coherence. However, the bilocal correction can be turned into a local one by introducing appropriate auxiliary integration variables ($\alpha$ parameters). Alternatively, this can be captured by thinking in terms of a `state of baby universes', the absorption and emission of which is described by operators $a^\dagger$ and $a$. In this language, the $\alpha$ parameters are simply the eigenvalues of $\hat{\alpha}=a+a^\dagger$. If the (infinitely many) $\alpha$ parameters take definite and not excessively large values, effective 4d locality and the dilute gas approximation are  maintained. However, exact predictivity for Lagrangian parameters on the basis of some underlying microscopic theory is lost.

Section~\ref{sec:PhenoApplications} is devoted to phenomenological applications. The early literature focuses on the indeterminacy of effective coupling constants. In particular, Coleman argued that the cosmological constant is statistically driven to zero value by the distribution of $\alpha$ parameters and their interplay with large-scale 4d gravity~\cite{Coleman:1988tj}. The violation of axionic shift symmetries and other global symmetries has also been studied from the beginning (see e.g.~\cite{Rey:1989mg}). More recently, the shift symmtry of a large-$f$ axion has been discussed in the context of wormholes and their interplay with the Weak Gravity Conjecture~\cite{Montero:2015ofa, Hebecker:2015rya, Heidenreich:2015nta}. We review some of this discussion, pointing out in particular difficulties in making strong, generic arguments against large-field axionic inflation~\cite{Hebecker:2016dsw}. Additionally, we discuss possible wormhole effects on axions with $f< M_P$ (including but not limited to the QCD axion) following \cite{Alonso:2017avz}. These may be relevant to ultralight dark matter, axion stars and black hole superradiance.

Open conceptual issues are the main subject of Sect.~\ref{sec:IssuesQuestions}. There are many of those, making the whole subject interesting but at the same time very difficult. We start with the FKS catastrophe~\cite{Fischler:1988ia, Kaplunovsky}, which turns Coleman's cosmological constant calculation into an argument for an overdensity of large wormholes. We go on to briefly discuss the generic problems of euclidean quantum gravity and, in particular, the negative-mode problems possibly affecting the Giddings-Strominger solution~\cite{Rubakov:1996cn,Hertog:2017owm, Alonso:2017avz}. Finally, we discuss the quantum cosmology involving macroscopic universes and a baby universe state. This can be relatively well undestood in a 1d toy model, but becomes already rather complicated in 2d quantum gravity. The latter case has of course received particular attention since its `large universe' may be the worldsheet of a fundamental string, while the baby universe state is represented by the dynamical target space of string theory. Finally, we analyse the Wheeler-DeWitt perspective as well as issues arising in the AdS/CFT paradigm. We conclude in Sect.~\ref{conc}.

\section{From instantons to wormholes}
\label{sec:GravitationalInstantons}
In this section we describe the simplest wormhole configurations, extrema of the euclidean action of Einstein gravity coupled to axionic fields (and possibly dilatons). We start with a brief description of the related but much better understood case of flat spacetime, where instantons arise as euclidean saddle points of gauge theories.
\subsection{Instantons}
Let us start by recalling the familiar case of a 4d gauge theory with
\be
{\cal L}=\frac{1}{2g^2}\,\mbox{tr}\,F_{\mu\nu}F^{\mu\nu}\,.
\ee
For simplicity the gauge group is taken to be $SU(2)$. The euclidean path integral necessarily involves certain finite action configurations (instantons) for which the field strength is non-zero in the vicinity of some point $x_0\in \mathbb{R}^4$ and falls off quickly as $|x-x_0|\to \infty$. Moreover, the value of
\be
n= \frac{1}{8\pi^2} \int\,\mbox{tr} (F\wedge F)
\ee
is integer, with $n=\pm 1$ characterizing a single instanton or anti-instanton (see e.g. \cite{Coleman:1978ae,Vainshtein:1981wh,Tong:2005un,Bianchi:2007ft,Vandoren:2008xg}). The minimal action for such $n=\pm 1$ configurations is
\be
S= \frac{8\pi^2}{g^2}\,.
\ee
The underlying solutions have 8 moduli: the components of $x_0$, a size modulus, and three zero modes associated to global $SU(2)$ transformations. 

\begin{figure}[ht]
	\centering
	\includegraphics[width=0.4\linewidth]{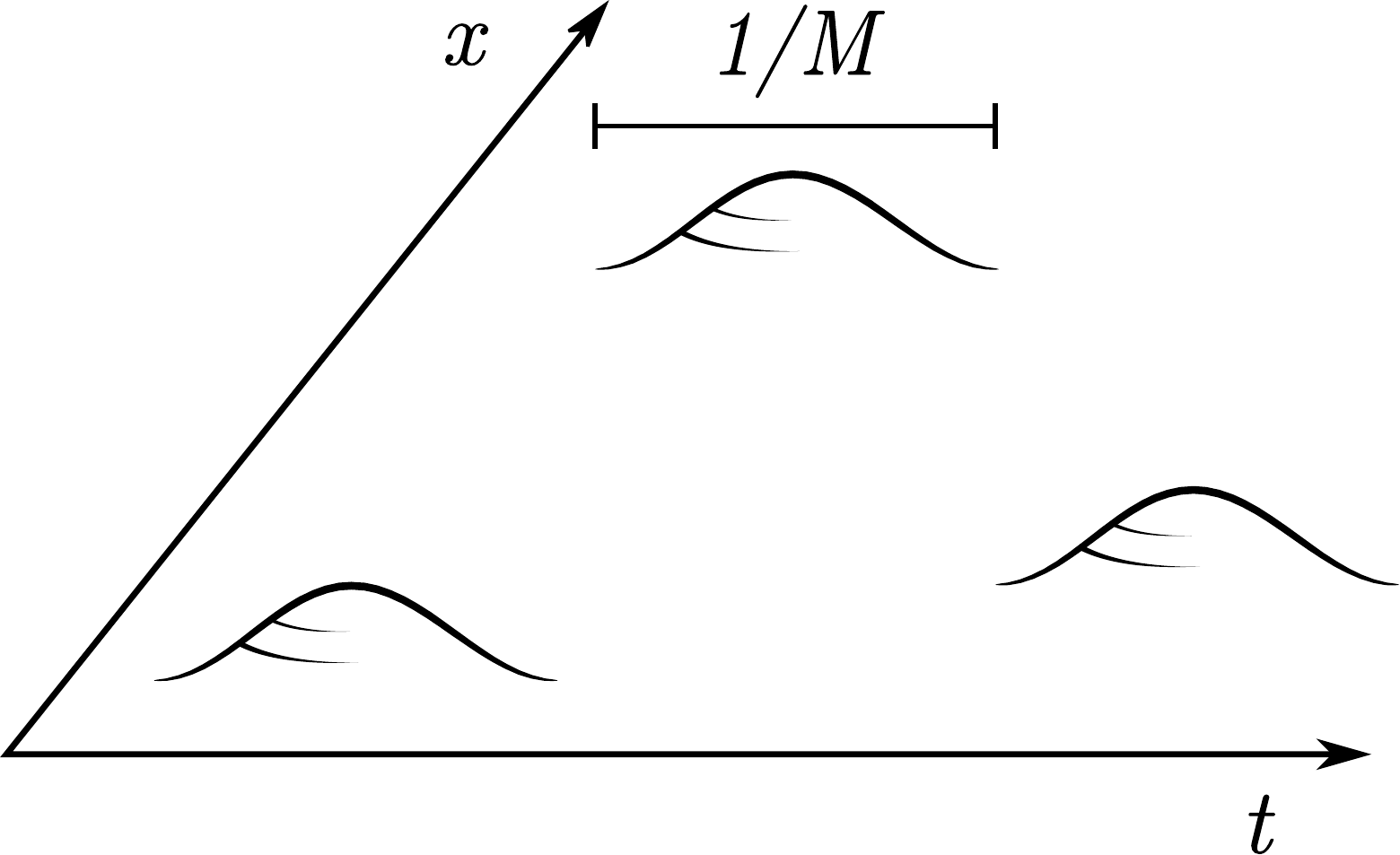}
	\caption{Gauge theory instantons as well-separated, localized lumps of field strength.}
	\label{fig:instantons}
\end{figure}

In calculating the partition function of the theory, one has to sum over any number of such instanton or anti-instanton configurations and integrate over all their moduli. This can be done very explicitly (see below) in the so-called dilute instanton gas approximation, i.e. assuming that the regions where $F$ is significantly non-zero are much smaller than their distance. Unfortunately, this clashes with the fact that a large contribution comes from very extended instantons, making the calculation e.g. in the practically interesting case of QCD non-trivial. A relevant toy model can however be obtained by Higgsing the gauge theory at $M\gg\Lambda$, with $\Lambda$ the confinement scale. 
The largest instantons now have size $\sim 1/M$ and the dilute gas approximation can be parametrically controlled (cf.~Fig. \ref{fig:instantons}).

\begin{figure}[ht]
	\centering
	\includegraphics[width=0.3\linewidth]{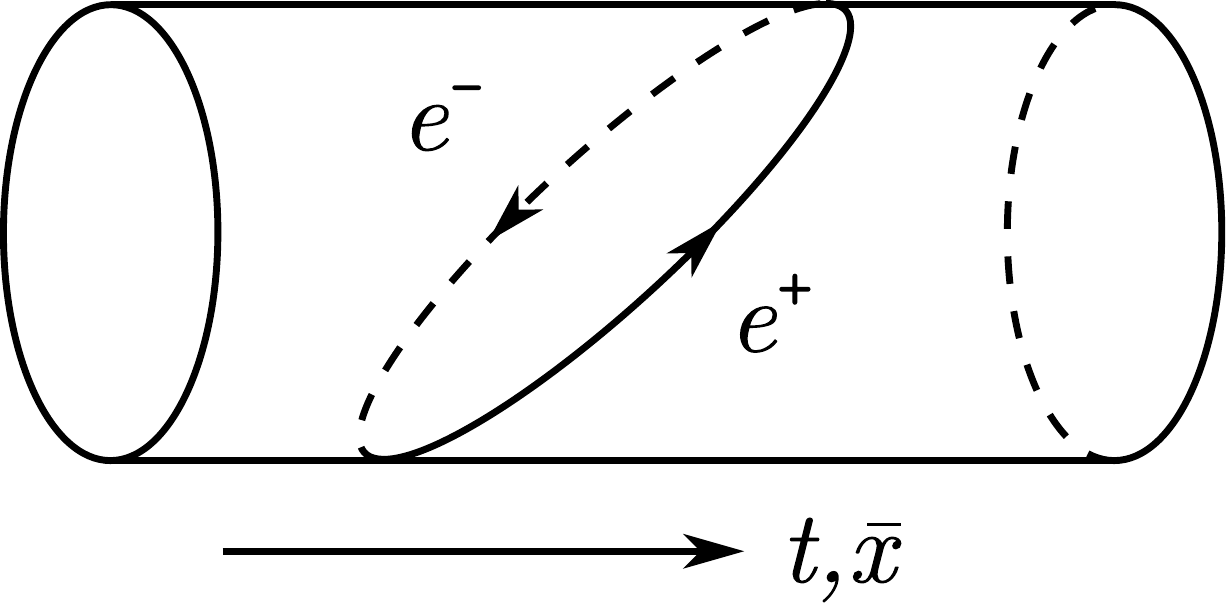}
	\caption{Euclidean brane instanton as particle-antiparticle fluctuation wrapping the compact space.}
	\label{fig:strins}
\end{figure}

Another equally familiar case is that of stringy or exotic instantons. To recall this case, start with the toy model of a 5d gauge theory on $\mathbb{R}^{1,3}\times S^1$. Clearly, if charged particles exist, this theory has tunneling processes in which a particle-anti-particle pair emerges from the vacuum and annihilates after passing around the $S^1$ in opposite directions (cf.~Fig.~\ref{fig:strins}). In the euclidean theory, this corresponds to a $0$-brane wrapped on the $S^1$ at some point $x_0\in \mathbb{R}^4$. The generalization to string compactifications with appropriate D$p$-branes (or E$p$-branes, with `E' for euclidean) wrapped on $(p+1)$-cycles of the compact space is obvious (for reviews see e.g.~\cite{Akerblom:2007uc,Blumenhagen:2009qh,Ibanez:2012zz}).

Crucially, in both of the above examples a shift symmetric, periodic scalar coupling to the instantons is naturally expected to be present. In the first case, it is the analogue of the QCD axion, coupling through
\be
{\cal L}\supset \theta\,\mbox{tr} (F\wedge F)/8\pi^2\,.
\ee
In the second case, it is the `Wilson-line' scalar descending from the 5d gauge field or, more generally, the 4d scalar descending from the Ramond-Ramond $C_{p+1}$-form field dimensionally reduced on the appropriate $(p+1)$ cycle.

For us, the above prelude serves only to motivate the following model theory of {\it generic} (or {\it fundamental}) instantons: It is defined by the partition function 
\be
Z\!\!=\!\!\!\int \!\! D\phi D\theta \,e^{-S[\phi,\theta]}\,\,
\!\sum_{n=1}^{\infty}\sum_{\ol{n}=1}^\infty 
\frac{1}{n!\ol{n}!}\prod_{i=1}^n \!\left( \int \!d^4x_i\,M^4\,e^{-S_I+i\theta(x_i)}\right)
\!\prod_{\ol{\imath}=1}^{\ol{n}}\!\left( \int \!d^4x_{\ol{\imath}}\,M^4\,e^{-S_I-i\theta(x_{\ol{\imath}})}\right)\!,
\label{zins}
\ee
which can of course be extended to a prescription for calculating Greens functions in the usual way. In this theory, the instantons are fundamental, zero-dimensional objects coupling to the axion-like field (just axion from now on) in the mathematically natural way: The axion is interpreted as a zero-form gauge potential which simply has to be evaluated at the position of the charged object (in the stringy language a D$(-1)$ brane).\footnote{
Note 
that this coupling remains imaginary even in the euclidean formulation. A pragmatic way to see this is to recall that $\theta$ is introduced as a periodic variable. A possibly deeper way is to think of instantons as tunneling events in the lorentzian theory and of $\exp(i\theta)$ as a relative phase between initial and final state. The latter is of course not affected by Wick rotation. 
} 
Furthermore, $\phi$ stands for all other fields in the model and $S_I$ is the instanton action. It arises (together with the typical instanton scale $M$) as the tunneling suppression factor $M^4\exp(-S_I)$, which can also be interpreted as the instanton density. 

Famously, the instanton and anti-instanton sum exponentiate and the two exponents involving $\theta$ combine to produce a cosine. This gives
\be\label{eq:instantonpotential}
Z=\int D\phi D\theta \,\exp\left(-S[\phi,\theta]+\int d^4x\, 2M^4e^{-S_I}\cos(\theta(x))\right)\,.
\ee
We emphasize that, apart from possible corrections to the dilute gas approximation, this is exact. Furthermore, it can be easily extended to situations in which the instantons couple, in addition to the necessary topological coupling to the zero-form $\theta$, to other fields. For example, $S_I$ may depend on the background values of some of the degrees of freedom denoted by $\phi$. 

\subsection{Giddings-Strominger solution}

At the end of the previous section, we advertised the point of view that instantons coupled to axions are a limiting case of the general concept of a $p$-form gauge theory: In this case $p=0$ and the charged object is zero-dimensional. By analogy to the gauge theory, one then expects the existence of objects akin to black branes. In other words, there might exist purely gravitational solutions charged under the axion which represent the continuation of instantons into the high-mass (or high-tension) regime. 

An object which fulfills such an expectation at least partially is the Giddings-Strominger wormhole~\cite{Giddings:1987cg}, sometimes also referred to as a gravitational instanton. It is based on the euclidean action ($M_P=1$)
\be
S = \int \diff^4x \sqrt{g} \Big( -\frac{1}{2}R + \frac{f^2}{2}g^{\mu \nu}\partial_\mu \theta \partial_{\nu} \theta \Big).
\ee
Equivalently, one can use the dual formulation in terms of a 2-form gauge theory with field strength $H_3=dB_2$:
\be
S = \int \diff^4x \sqrt{g} \Big( -\frac{1}{2}R + \frac{1}{2f^2}\,\frac{1}{3!}
H_{\mu \nu \kappa}H^{\mu\nu\kappa} \Big)\,.
\ee
At the classical level, the duality relation is simply $H = f^2\ast \diff \theta$. However, the equivalence of the two theories extends, of course, to the full quantum systems. To see this, the dualization must be done under the path integral and care must be taken to get the signs of the kinetic terms right. The outcome is that, both in the euclidean and in the lorentzian versions, the fields have standard (non-ghostlike) kinetic terms on both sides of the duality (see~\cite{ArkaniHamed:2007js,Hebecker:2016dsw, Collinucci:2005opa, Burgess:1989da, Bergshoeff:2005zf} for details). The wormhole solution to be discussed momentarily exists only in the euclidean theory, but both in the 0-form and 2-form formulation. However, while the $B_2/H_3$ fields are real, the corresponding values of $\theta/d\theta$ are imaginary. 

Now, the relevance of an `instanton-like' euclidean solution is, of course, that it defines a saddle point of the path integral and hence a very specific, easily quantifiable contribution to the partition function. For the $B_2$ path integral, the Giddings-Strominger saddle point is then right in the standard integration domain, i.e., `on the real axis' of field space. By contrast, in the $\theta$ path integral the corresponding saddle point is `on the imaginary axis', requiring the deformation of the contour and raising the question whether such complex saddles contribute. Complex saddles are certainly known to contribute in certain cases (for a toy model relevant to the present setting see \cite{ArkaniHamed:2007js}). Thus, while we favor the (real) $B_2$ formulation for obvious reasons in what follows, there is nothing wrong in principle with the $\theta$ formulation.\footnote{
Occasionally,
the impression is raised that the $\theta$ formulation requires a wrong-sign kinetic term if one wants the wormhole solution to exist. While this perspective might technically be equivalent to what was said above, we find it conceptually misleading. In our reading, one studies a well-defined physical theory without ghost fields. It is only the desire to estimate the contribution from a certain complex saddle which leads one to work with imaginary $\theta$ temporarily.
}

After these preliminaries, let us describe the solution~\cite{Giddings:1987cg}. It can be motivated by starting from a field theory instanton and including gravitational backreaction: If an instanton couples to an axion $\theta$, the dual theory carries non-zero 3-form flux, 
\be\label{eq:charge}
\int_{S^3} H = n\,, \qquad n \in \mathbb{Z}\,,
\ee
on any sphere containing $n$ instantons (or an instanton of charge $n$). 
Placing the instanton(s) at the origin and assuming spherical symmetry, it is immediately clear that one must have
\be
H = \frac{n\,\epsilon}{2\pi^2}\,.
\ee
Here $\epsilon$ is defined as the volume form of $S^3$ in the description of 
$\mathbb{R}^4$ as $\mathbb{R}_+\times S^3$.

The above $H$ automatically satisfies the Bianchi identitiy $dH=0$ and the equation of motion $d\!*\!H=0$ (for any spherically symmetric metric). It induces a non-zero energy momentum tensor and the corresponding Einstein equation is solved by
\be
ds^2 = \Big(1 + \frac{C}{r^4} \Big)^{-1} \diff r^2 + r^2 \diff \Omega_3^2\,\,, \qquad\qquad C=-\frac{n^2}{24\pi^4f^2}\,.
\label{eq:GSwormhole}
\ee
Here $ \diff\Omega_3^2 $ denotes the round metric on the unit sphere.

\begin{figure}[ht]
	\centering
	\includegraphics[width=0.7\linewidth]{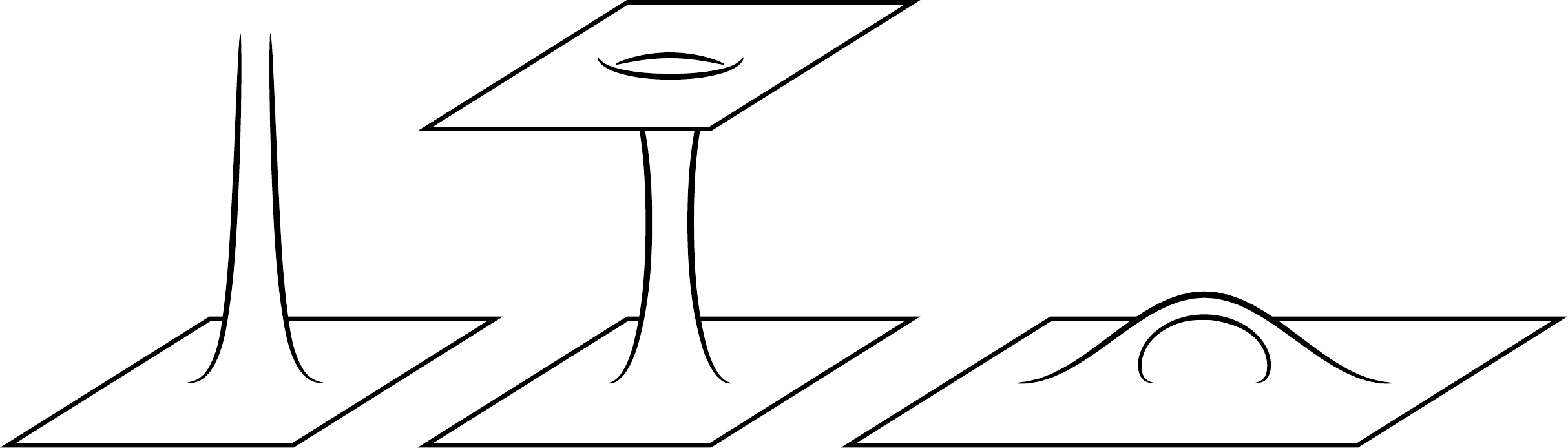}
	\caption{Wormholes: A semiwormhole (left), a wormhole connecting two distinct large asymptotically flat universes (center) and a wormhole on a single universe (right).}
	\label{fig:Wm}
\end{figure}

This geometry is asymptotically flat for $ r \to \infty $ and has a coordinate singularity at $r=r_0\equiv |C|^{1/4}$. The space given by restricting $ r \in [r_0, \infty) $ forms what is often termed a semiwormhole (see Fig.~\ref{fig:Wm}). Gluing two such solutions at the 3-spheres defined by $r=r_0$, one obtains a smooth wormhole connecting two flat universes (see Fig.~\ref{fig:Wm}). A topologically distinct, approximate solution can be obtained if the two asymptotically flat regions of Fig.~\ref{fig:Wm} are interpreted as distant parts of the same universe -- cf.~Fig.~\ref{fig:Wm}. One then has a wormhole joining two regions of the same large universe. This becomes exact in the limit that the two wormhole ends are infinitely far apart. 

The wormhole action is particularly easy to compute using the trace of the Einstein equation:
\be
S_w = \frac{1}{f^2} \int H \wedge \ast H = \frac{n^2}{2\pi^2f^2}\,\, 2\int_{r_0}^\infty \frac{dr}{r\sqrt{r^4 + C}}=\frac{\pi\sqrt{6}}{4}\cdot\frac{|n|}{f}\,.
\ee
Notice the factor 2 appearing because a wormhole consists of two solutions of the form of (\ref{eq:GSwormhole}), each restricted to $r>r_0$.

The most straightforward interpretation of this is as follows:  
Suppressed by an overall factor $\exp(-S_w)$, the partition function includes processes in which an $S^3$ baby universe supported by $H_3$-flux `bubbles off' at some space-time point $x$ and is absorbed later on at $y$ $(x,y\in \mathbb{R}^4)$. From the low-energy perspective, this is equivalent to an instanton (of charge $n$ and action $S_w/2\sim |n|/f$) at $x$ and a corresponding anti-instanton at $y$. Calculational control in semiclassical gravity requires $r_0\sim \sqrt{|n|/f}\gg 1$. This  should then give rise to a cosine potential for $\theta$ and further instanton-induced operators. It has, however, been argued that, in contrast to the instantonic situation, no such potential is induced because of the unavoidable pairing of instanons and anti-instantons \cite{Heidenreich:2015nta}. Counterarguments have been given \cite{Hebecker:2016dsw}, based essentially on the intuition that local physics is ignorant of the overall constraint on instantons vs. anti-instantons in a very large space-time. (Recall that the action stays finite as $|x-y|\to\infty$.) However, this debate is overshadowed by a much deeper issue which will permeate the rest of this review: Once one allows for wormholes, one has effectively allowed for baby-universes propagating between points $x$ and $y$. But then such baby universes must also be allowed to be part of the initial and final states of any process. More generally, there exits a `baby-universe state' in addition to our space-time and any wormhole effects (such as the naive cosine potential) depend on it.

\subsection{Dilatonic wormholes}\label{sec:dilatonic}
Before coming to the physical effects of wormholes and baby universes, we want to briefly comment on generalizations of the Giddings-Strominger solution which involve a dilaton \cite{Giddings:1987cg, Giddings:1989bq, Bergshoeff:2005zf, Heidenreich:2015nta, Hebecker:2016dsw}. This is important since such dilatons are always present in the simplest stringy models allowing for wormholes. 

Consider an action in which the axionic kinetic term depends on a further massless scalar field $\phi$, 
\be
S = \int \diff^4x \sqrt{g} \Big( -\frac{1}{2}R + \frac{1}{2}{\cal K}(\phi)g^{\mu \nu}\partial_\mu \theta \partial_{\nu} \theta + \frac{1}{2}g^{\mu \nu}\partial_\mu \phi \partial_{\nu} \phi \Big)\,,
\ee
or equivalently
\be
S = \int \diff^4x \sqrt{g} \Big( -\frac{1}{2}R + \frac{1}{2}\mathcal{F}(\phi)H_{\mu \nu \kappa}H^{\mu\nu\kappa} + \frac{1}{2}g^{\mu \nu}\partial_\mu \phi \partial_{\nu} \phi \Big)\,,
\ee
with ${\cal F}\equiv 1/(3!\,{\cal K})$. As before, spherical symmetry ensures that the equation of motion for $H$ is automatically satisfied. A new, non-trivial differential equation for the radial profile of $\phi$ arises. Remarkably, the differential equation for $g_{rr}$ (the only non-trivial part of the Einstein equation) decouples and the metric  (\ref{eq:GSwormhole}) remains a solution.\footnote{
In 
fact, the metric (\ref{eq:GSwormhole}) solves the equations of motion of
the more general action
\be 
S = \int \diff^4x \sqrt{g} \Big( -\frac{1}{2}R + \frac{1}{2}G_{IJ}(\phi)g^{\mu \nu}\partial_\mu \phi^I \partial_{\nu} \phi^J  \Big)\,,
\ee
where a set moduli $ \phi^I $ and a (non-positive-definite) metric $G_{IJ}$ on moduli space have been introduced~\cite{ArkaniHamed:2007js}. \label{fah}
}. 
We will not discuss the solution $\phi(r)$ in any detail. It is, however, interesting to note that, switching from $H_3/B_2$ to $d\theta/\theta$ for the moment, the common trajectory $\{\phi(r),\theta(r)\}$ describes a geodesic in field space. This generalizes to the case of several axionic and several non-axionic scalars (cf. Footnote~\ref{fah} and Ref.~\cite{ArkaniHamed:2007js}).

\begin{figure}[ht]
	\centering
	\includegraphics[width=0.5\linewidth]{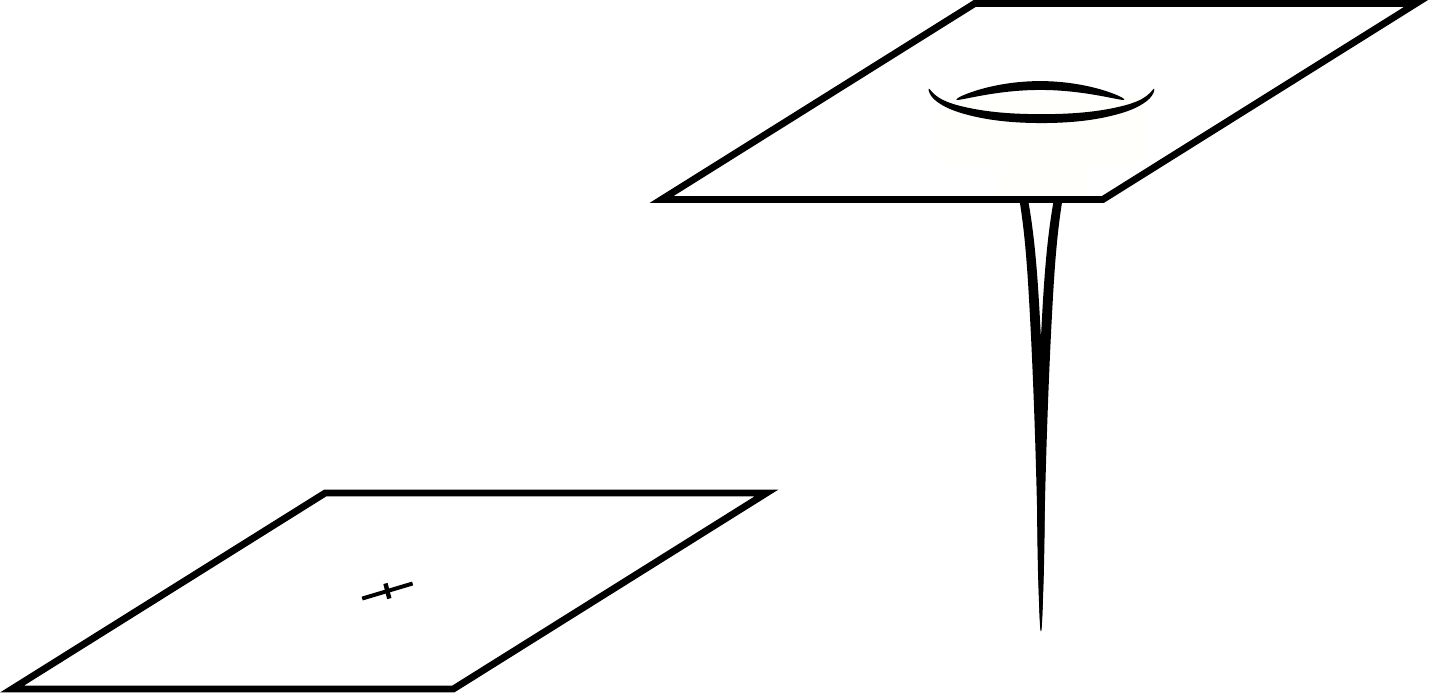}
	\caption{ Extremal (left) and Cored (right) gravitational instanton.}
	\label{fig:GrInstantons}
\end{figure}

Motivated by stringy and supergravity examples, we now restrict attention to the special case
\be\label{eq:dilatonfunction}
{\cal F}(\phi)=\frac{1}{3!\,f^2}\,\exp(-\beta\phi)\,.
\ee
Without loss of generality one can assume $\beta\ge 0$. Three different classes of solutions can be distinguished: First, as long as  $\beta<2\sqrt{2/3}$,
the Giddings-Strominger wormhole continues to exist (metric of (\ref{eq:GSwormhole}) with $C<0$). This is the case of our main interest. 
Second, there is the extremal gravitational instanton, corresponding to $C=0$. The geometry is a flat space-time with the origin removed, but $\phi$ diverges as one approaches $r=0$. Third, there are `cored gravitational instantons', corresponding to $C<0$. In this case one has a curvature singularity at $r=0$ (cf.~Fig.~\ref{fig:GrInstantons}). The last two cases have the significant drawback that they are not fully controlled within the low-energy effective theory and we will hence not discuss them further (see however~\cite{Bergshoeff:2005zf,Heidenreich:2015nta,Hebecker:2016dsw}). 

In the simplest (usually highly supersymmetric) string compactifications, axions are always accompanied by a dilatonic scalar or saxion, as above. However, the simplest models do not allow for $\beta<2\sqrt{2/3}$. Naively, one may then hope that wormholes do not arise in consistent theories of quantum gravity. But it turns out that the problem with the allowed $\beta$ range can be overcome \cite{Tamvakis:1989aq, Bergshoeff:2004pg, ArkaniHamed:2007js, Hertog:2017owm}. The underlying idea is simple: A wormhole can be charged under several axions, each with its own saxion with a certain $\beta$. The trajectory which the solution follows in the saxionic field space involves all the axions and can be characterized by a single effective $\beta$. The latter can be in the desired range even if the $\beta$-values of the ingredients were not.\footnote{
The 
necessary condition for the existence of wormholes and the way in which multiple axions help to satisfy it can also be discussed in the language of time-like geodesics in the axion/saxion field space, cf. Footnote \ref{fah}.
}
Thus, one can by now be certain that Giddings-Strominger wormholes exist in the euclidean version of supergravity theories coming from string theory. This makes all the puzzles to be discussed below even more troubling.\footnote{
A 
simpler but less rigorous argument that wormholes are `not in the swampland' can be given as follows: Surely somewhere in the string theory landscape there  exists a low-energy effective theory containing an ungauged abelian Higgs 
model. Clearly, the global $U(1)$ of this model will not be exact. The resulting effective axion will thus have a non-perturbatively generated cosine potential. This potential is in general exponentially suppressed and hence very small. The saxion, i.e. the radial direction of the complex Higgs scalar, is stabilized. Thus, wormholes based on this effective axion will exist.
}

\section{The effect of wormholes} \label{sec:WormholeEffects}
Two results of the previous section are essential for what follows. First, a dilute gas of instantons can be resummed (or `integrated out') to obtain a correction to the effective action. Second, a very similar contribution to the path integral arises in gravitational theories with an axion. The objects to be summed over are wormholes or gravitational instantons. The main novelty is that they couple to the low-energy degrees of freedom (including the background metric) at two spacetime points rather than just at one. We now want to discuss the correction to the effective action arising in this second case following \cite{Coleman:1988tj, Preskill:1988na, Coleman:1989zu}. We note that, while the specific Giddings-Strominger solution discussed above may be the simplest and best understood euclidean wormhole, the following analysis does not rely on any of its details. What matters is that the euclidean path integral includes contributions from topologies like that of Fig.~\ref{fig:Wm} (on the right). All that we will use is that they are exponentially suppressed by a sufficiently large euclidean action and that the coupling to soft field modes occurs at two uncorrelated points \cite{Hawking:1988ae,Hawking:1989vs}.

\subsection{The bilocal action}\label{bil}
We begin with a heuristic derivation of the bilocal action which captures the effect of wormholes at the semi-classical level. For this, we first recall the field theoretic partition function with instantons, Eq.~(\ref{zins}), and restrict it to the one-instanton sector for notational simplicity:
\be
Z_1=\int D\phi\,D\theta\,e^{-S[\phi,\theta]}\left(\int d^4x\,e^{-S_I+i\theta(x)}\right)\,.
\ee
Here the prefactor $M^4$ has been reabsorbed in the instanton action. (To be careful, one should then either work in Planck units or at least choose $x$ dimensionless.)

The above is unnecessarily explicit in that $\theta$ has been separated from all the other fields $\phi$. At the same time, it is oversimplified in that only the dependence of the instanton action on $\theta$ has been kept: $S_I[\theta]\equiv S_I+i\theta$. A more general version, in which $\theta$ is just one of the many fields denoted by $\phi$, reads
\be
Z_1=\int D\phi\,e^{-S[\phi]}\left(\int d^4x\,e^{-S_I[x,\phi]}\right)\,. \label{z1}
\ee
Here $S_I[x,\phi]$ is the single-instanton tunneling action for the space-time point $x$ in a background field $\phi$. It is clear that obtaining this action in a concrete model is highly non-trivial: One would have to find the analogue of the well-known instanton or wrapped-euclidean-brane solution in an, in general non-constant, background of all fields in the theory. However, we are satisfied with an approximation: the fields $\phi$ are restricted to be soft relative to the instanton scale $M$. The action can then be expanded in terms of local operators:
\be
S_I[x,\phi]=S_I+c_1\phi(x)+c_2\phi^2(x)+c_3(\partial\phi(x))^2+\cdots\,\,.
\ee
Here $S_I$ is the instanton action on the unperturbed background, say at $\phi\equiv 0$. With this, the transition to wormholes is simple. 

Indeed, the wormhole analogue of (\ref{z1}) is
\be
Z_{1\,,\,w}=\int Dg\,D\phi\,e^{-S[g,\phi]}\left(\int d^4x\,\sqrt{g(x)}\int d^4y\,\sqrt{g(y)}\,e^{-S_w[x,y,g,\phi]}\right)\,.\label{z1w}
\ee
Here $\int\! Dg$ stands for the integral over all soft (relative to the wormhole size) metrics on the topologically trivial background universe into which the wormhole is inserted. In addition, $\phi$ stands for all further fields, including the axion or the dual 2-form.\footnote{
If 
one uses specifically the Giddings-Strominger solution, the value of the axion corresponds to the one far away from the wormhole. The fast change of the axion indside the throat is not part of what we want to call the background field.
} 
As before, appealing to our restriction to soft fields and metric configurations, the wormhole action can be written as a series of local operators at $x$ and $y$:
\be
S_w[x,y,g,\phi]=S_w+c_1\phi(x)+c_1\phi(y)+c_3\phi(x)\phi(y)+\cdots+c_4(\partial\phi(x))^2(\partial\phi(y))^2+\cdots
\ee
For simplicity terms depending on a non-trivial metric background have not been displayed. It is clear that such terms, involving various curvature invariants at $x$ and at $y$ as well as products thereof, will also be present. The crucial novelty compared to the instanton case is that one is dealing with a double functional Taylor expansion and that products of local operators involving all fields will in general arise. Thus, one generically has the bilocal expression
\be
S_w[x,y,g,\phi]=S_w+\sum_{ij}\tilde{\Delta}_{ij} 
\mathcal{O}_i(x)\mathcal{O}_j(y)\,,
\ee
or, equivalently,
\be
e^{-S_w[x,y,g,\phi]}=\frac{1}{2}\sum_{ij}\Delta_{ij} \mathcal{O}_i(x)\mathcal{O}_j(y)\,.
\label{esw}
\ee
In the last expression, the exponential $\exp\left(\sum_{ij}\tilde{\Delta}_{ij}{\cal O}_i{\cal O}_j\right)$ has been expanded and  the suppression factor $\exp(-S_w)$ has been absorbed in the new coefficients $\Delta_{ij}$:
\be
\Delta_{ij}\sim e^{-S_w}\,.
\ee

Finally, one inserts (\ref{esw}) in (\ref{z1w}) and writes down analogous expressions for any number of wormholes. In doing so, the dilute gas approximation is used, i.e. that typical distances between wormhole ends are much larger than the wormhole diameter. The sum exponentiates, exactly as in the instanton case, giving
\be
Z_w=\int Dg\,D\phi\,e^{-S[g,\phi]\,+\,I}\label{zw}
\ee
with the bilocal action
\be \label{eq:BilocalAction}
I = \frac{1}{2} \int \diff^4x \sqrt{g} \int \diff^4 y \sqrt{g}\,\, \sum_{i,j} \Delta_{ij} \mathcal{O}_i(x) \mathcal{O}_j(y)\,.
\ee

\subsection{Local action involving \texorpdfstring{$\alpha$} \ \ parameters}
Following \cite{Coleman:1988tj,Preskill:1988na}, one can give the action $I$ a local form at the expense of introducing a set of auxiliary parameters $\alpha_i$. Up to some irrelevant normalization factor, one has
\be 
e^I = \prod_i\left(\int d\alpha_i\right) \exp{ \big( - \frac{1}{2}\sum_{i,j}\alpha_i \Delta^{-1}_{ij} \alpha_j \big)} \exp{ \big( \sum_i \alpha_i \int \diff^4x \sqrt{g}\,\mathcal{O}_i(x) \big)}\,.
\label{lwa}
\ee
It is natural to write the original action $S$ of our physical system using the basis of local operators as in the wormhole action $I$ above:
\be
S[g;\lambda] = \sum_i \lambda_i \int \diff^4x \sqrt{g}\,\mathcal{O}_i(x)\,.
\label{oac}
\ee
Here $ \lambda_i $ are the coupling constants. For example, $\lambda_1$, $\lambda_2$ and $\lambda_3$ could be the cosmological constant, the coefficient of the Einstein-Hilbert term, and of the $R^2$-term respectively. To minimize the notational complexity, we suppress the dependence on the non-metric fields $\phi$ here and below. Of course, all of the above holds with as many further fields as one needs.

Comparing (\ref{lwa}) and (\ref{oac}), one sees that the effect of wormholes amounts to shifting the coupling constants of the original action: $ \lambda_i \to \lambda_i - \alpha_i $. Put differently, one can use the `shifted' action $S[g;\lambda - \alpha] $, remembering of course to integrate over the 
$\alpha$ parameters. The partition function with wormhole effects included (see \eqref{zw}
and recall that we suppress $\phi$) now reads
\be
Z_w=\! \int\!\! Dg\, e^{- S[g;\lambda]+I[g]}=\! \int\!\! Dg\, D\alpha\, G(\alpha)\, e^{ -S[g;\lambda - \alpha] }=\! \int\!\! D\alpha\, G(\alpha)\left[ \int\!\! Dg\, e^{ -S[g;\lambda - \alpha]} \right],\label{zw2}
\ee
with $ G(\alpha) =  \exp{ \big( - \frac{1}{2}\sum_{i,j}\alpha_i \Delta^{-1}_{ij} \alpha_j \big)} $ the gaussian weighting factor. In the above, we also use the somewhat sloppy notation $D\alpha$ for the integration over all $\alpha_i$, in spite of the fact that the index $i$ is discrete. 

In the last expression in (\ref{zw2}), one recognizes the familiar partition function without wormholes inside the square brackets. The wormhole effect is reduced to shifting the coupling constants of that theory by $\alpha_i$. Since these $\alpha$ parameters are constants in space and time, one can take the point of view that they simply have to be measured and no relevance should be ascribed to the gaussian weight factor governing their distribution. By contrast, one may argue that statistical predictions for their values are possible, which of course involves this weight factor. This is a multiverse-type situation, discovered (and discussed by many authors) long before the string theory multiverse entered the stage. 

\subsection{Baby universes}\label{sec:babyuniverses}
\begin{figure}[ht]
	\centering
	\includegraphics[width=0.7\linewidth]{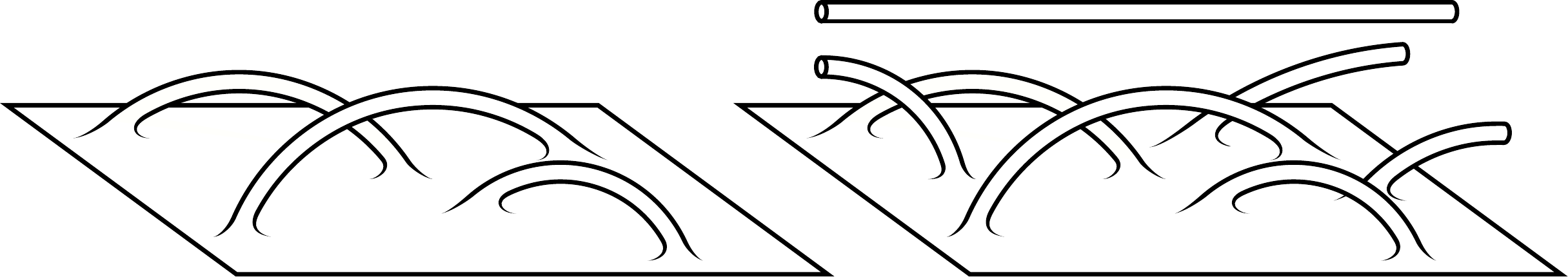}
	\caption{The effective action considered as an amplitudes (left) and an amplitude including semiwormholes (right).}
	\label{fig:MultipleWm}
\end{figure}

The physics behind $\alpha$ parameters becomes more lucid if one thinks of the wormholes in terms of $S^3$ baby universes which are emitted and absorbed by our macroscopic space-time (left hand side of Fig.~\ref{fig:MultipleWm}). To derive the corresponding formulae, one considers the situation with a single operator and hence a single $\alpha$ parameter for notational simplicity. Equation (\ref{lwa}) then reads
\be \label{eq:WormholeEffect}
e^I = \int \frac{\diff\alpha}{\sqrt{2\pi}}\, \exp{ \left( -\frac{1}{2}\alpha^2 + \alpha \sqrt{\Delta} \int \diff^4x \sqrt{g}\,\mathcal{O}(x) \right)},
\ee
obtained after rescaling $ \alpha \to \alpha \sqrt{\Delta} $ and introducing the normalization factor $1/\sqrt{2\pi}$ for later convenience.

Equation (\ref{eq:WormholeEffect}) can be viewed as a power series in ${\cal O}(x)$ encoding the sum of process in which baby universes are created and annihilated
at locations corresponding to the various values taken by $x$. All of this has of course to be inserted under the $Dg$ integral over soft background metrics. To make this manifest, one defines baby universe creation and annihilation operators $ a^\dagger $, $ a $ satisfying the usual commutation relation $ [a, a^\dagger] = 1 $.  The state with no baby universes $ |0\rangle $ is referred to as the baby universe vacuum. The normalized state with $ n $ baby universes is then given by
\be
|n\rangle = \frac{(a^\dagger)^n}{\sqrt{n!}}|0\rangle\,.
\ee
The analogues of the conventional position operator of the harmonic oscillator and its eigenstates are defined as
\be
\hat{\alpha}=a + a^\dagger\quad,\qquad\qquad\hat{\alpha}|\alpha\rangle =\alpha|\alpha\rangle\,.
\ee
Since the ground state obeys $|0\rangle \sim \int d\alpha\,\exp(-\alpha^2/4)|\alpha\rangle$, one immediately sees that
\be\label{eq:BabyOperatorsAlphas}
\langle 0 | (a + a^\dagger)^n |0\rangle = \int \frac{\diff \alpha}{\sqrt{2\pi}}\, \exp{ \left( -\frac{1}{2}\alpha^2 \right)} \alpha^n\,.
\ee
This allows one to rewrite (\ref{eq:WormholeEffect}) according to
\be \label{eq:Aplitute}
e^I = \int \frac{\diff \alpha}{\sqrt{2\pi}}\, \exp \left( -\frac{1}{2}\alpha^2\right)\,\exp \Big( \alpha\, \tilde{\cal O} \Big) = \langle 0 |\, e^{(a + a^\dagger)\,\tilde{\cal O}}\, |0\rangle,
\ee
where $\tilde{\cal O}$ is an abbreviation for $\tilde{\cal O} = \sqrt{\Delta} \int \diff^4x \sqrt{g}\mathcal{O}(x) $.

Equation (\ref{eq:Aplitute}) can be considered a convenient formal expression for a power series in $\tilde{\cal O}$. But it is much more than that: 
It formalizes the interpretation of the partition function and of the process depicted on the left hand side of Fig.~\ref{fig:MultipleWm} in terms of a baby universe Hilbert space. Equation (\ref{eq:Aplitute}) calculates the amplitude 
relating two spatial slices of the parent universe, allowing for any number of wormholes to be insterted between initial and final time. 

The most important point here is that, in this approach, it is both easy and obviously necessary to allow for more general initial and final states: There is simply no reason to treat those as baby universe vacua. For example, one can also consider the transition amplitude between states with $ n_1 $ and $ n_2 $ baby universes:
\be
\langle n_2 | e^{(a + a^\dagger)\,\tilde{\cal O}} | n_1 \rangle.
\ee
In fact, arbitrary states $ \psi_1 $ and $ \psi_2 $ can be considered, another relevant case being that of so-called $\alpha$-vacua:
\be \label{eq:WormholeEffectAlphaVacuum}
\langle \alpha | e^{(a + a^\dagger)\,\tilde{\cal O}} | \alpha \rangle = e^{\alpha\,\tilde{\cal O}}\,.
\ee
Here we ignore the divergent prefactor related to the $\delta$-function normalization of `momentum eigenstates'. 

It is easy to see that, for an arbitrary number of operators and arbitrary initial and final states, the above amplitude generalizes to 
\be \label{eq:EffectiveHamiltonian}
\langle \psi_2 | \exp{\big( \sum_i \sqrt{\Delta_{ii}} \int \diff^4x \sqrt{g}\,\mathcal{O}_i(x)(a_i + a_i^\dagger) \big)} | \psi_1 \rangle\,.
\ee
Here, a basis of local operators has been chosen such that the matrix $ \Delta_{ij} $ is diagonal. The $ a_i^\dagger $ and $ a_i $ carry the same index as the local operators and create or annihilate baby universes of type $i$. If everything is based on the Giddings-Strominger solution of lowest charge, one may think of these baby universes as of transverse spheres $S^3$ in a perturbed wormhole geometry (or some appropriate quantum superposition thereof).

The Hamiltonian $(a + a^\dagger)\,\tilde{\cal O}$ was first derived by Coleman in \cite{Coleman:1988cy} by summing explicitly over all possible wormhole and semiwormhole configurations. For completeness, we now briefly explain this computation, for the case of a single type of wormhole for simplicity. Consider a 4-manifold $M$ of the type shown in the right hand side of Fig.~\ref{fig:MultipleWm}. The initial boundary consists of a large 3-manifold parent universe and $n_1$ incoming baby universes. Of those, $n_1-r$ later on merge with $M$. The final boundary consists again of a large 3-manifold and $n_2$ outgoing baby universes, $n_2-r$ of which emerged from $M$. Thus, $r$ baby universes simply travel from the initial to the final boundary without interacting with the parent universe. Furthermore $m$ baby universes form complete wormholes on $M$. The path integral sums over all such configuration:
\be
\sum_{r, m} e^{-S}\big|_{n_1, n_2}\,. 
\ee

As before, one assumes that each semiwormhole attached to the parent universe contributes a factor $ \tilde{\mathcal{O}} = \sqrt{\Delta} \int \diff^4x \sqrt{g}\,\mathcal{O}(x) $. Taking into account the combinatorics and carrying out the summation over $m$ yields
\be
\sum_{r, m} e^{-S}\big|_{n_1, n_2}= \sqrt{n_1!} \sqrt{n_2!} \,\,e^{\tilde{\mathcal{O}}^2/2} \sum_{r=0}^{\min{(n_1,n_2)}} \frac{\tilde{\mathcal{O}}^{n_1 + n_2 - r}}{(n_1-r)!\,(n_2-r)!\,r!} = \langle n_2 | e^{ (a^\dagger + a)\tilde{\mathcal{O}} } | n_1 \rangle.
\ee
Here the second equality follows by applying Baker-Campbell-Hausdorff in the form $\exp[(a+a^\dagger)\tilde{O}]=\exp(a^\dagger\tilde{O})\exp(a\tilde{O}) \exp(\tilde{O}^2/2)$ and inserting the identity operator written as a sum over $|r\rangle \langle r|$. Thus, the language of $a$ and $a^\dagger$ introduced earlier is nothing but a convenient way of counting wormhole topologies. 

\subsection{The perspective of \texorpdfstring{$\alpha$} \ -vacua and the wormhole density}
It is clear that the appearance of $\alpha$ parameters in the path integral has the potential to change physics dramatically: Since these parameters are space-time independent, the whole universe (including its time evolution) can be thought of as a superposition of independent universes, each with a specific set of fixed $\alpha$ parameters.

This has become even more apparent in the last subsection, when the baby universe state characterized by the $\alpha$ parameters was introduced. Since all effective operator coefficients or couplings are shifted according to $\lambda_i\,\to\, \lambda_i\!-\!\alpha_i\,$, the baby universe state determines the 4d low-energy effective field theory. A whole landscape of such theories, equivalent to the space of $\alpha$-vacua, exists. At this level, every hope of predicting coupling constants from some fundamental theoretical principle appears to be lost.

The situation might not be, however, quite as bad: for transitions among baby universe vacua an integral over the $\alpha$ parameters with a very specific measure arises. This makes sense in a compact euclidean universe, for example for a large 4-sphere (or a set of large 4-spheres), where no initial or final baby universe state is required. Specifically a 4-sphere geometry is reminiscent of the Hartle-Hawking definition \cite{Hartle:1983ai} of the Wheeler-DeWitt wave function of the universe \cite{DeWitt:1967yk, Dewitt:1969ke-my}. Thus, one may think of the integral over $\alpha$ parameters (with the concrete measure derived earlier) as of a preferred wave function of the baby universe state. This point of view allows for at least a statistical prediction of effective coupling constants.

It is essential that the $\alpha$-parameters are eigenvalues of the Hamiltonian governing the interaction of our large-scale 4d world with the baby-universe state. This was derived above and it can also be seen intuitively: one can not distinguish in principle whether a wormhole attached at a given position $x$ corresponds to a baby universe being absorbed or being created. Hence one always encounters the combination $(a+a^\dagger)\,{\cal O}(x) \equiv \hat{\alpha} \,{\cal O}(x)$ in the effective Hamiltonian. When an operator coefficient is measured, one is projected to a subsector of the theory belonging to a certain eigenvalue $\alpha$. Further dynamical evolution can not change this value. 

As a consistency check one can estimate the density of wormhole ends following Preskill \cite{Preskill:1988na}. This is crucial to understand the validity of the dilute gas approximation. Returning to the perspective of the bilocal action, Sect.~\ref{bil}, one can focus on a single operator, the cosmological constant. According to (\ref{eq:BilocalAction}), the effect of wormhole insertions is then encoded in 
\be
e^I \sim e^{\frac{1}{2}V^2_4\Delta} = \sum_{n=0}^{\infty} \frac{1}{n!}\Big(\frac{1}{2}V^2_4\Delta \Big)^n,
\ee
where $ V_4 = \int \diff^4x \sqrt{g(x)} $ and $\Delta \sim e^{-S_w}$. Here the $n$-th order term corresponds to $n$ wormholes. The dominant contribution to the sum comes from terms with $ n \sim N_w \equiv \frac{1}{2}V^2_4\Delta $, such that the wormhole density in typical configurations is
\be
\frac{N_w}{V_4} = \frac{1}{2}\,V_4\,\Delta.
\ee
One arrives at the disturbing conclusion that this density grows with the volume $V_4\,$.

Fortunately, the result changes if one considers physics at fixed $\alpha$. According to (\ref{eq:WormholeEffectAlphaVacuum}) the wormhole sum is now encoded in
\be
e^{\alpha\, V_4\,\sqrt{\Delta}} = \sum_{n=0}^{\infty} \frac{1}{n!} \big(\alpha \,V_4\,\sqrt{\Delta}\big)^n.
\ee
This sum is dominated by terms with $ n \sim N_{w,\,\alpha} \equiv \alpha\,V_4 \,\sqrt{\Delta}$. A non-divergent density of wormhole ends in spacetime follows:
\be
\frac{N_{w,\,\alpha}}{V_4} = \alpha\, \sqrt{\Delta}\,.
\ee
Thanks to the supression factor $ \sqrt{\Delta} \sim e^{-S_w/2}$, this density is expected to be very small for large wormholes with a correspondingly large euclidean action. The problem encountered above in the vacuum-to-vacuum amplitude, $ |0\rangle \to |0\rangle $, appears to have been resolved. Technically, the reason is that the sum has been re-organized by combining events where a wormhole is absorbed and created at the same point: $a,\,\,a^\dagger\,\to (a+a^\dagger)$\,. However, together with the suppression factor $e^{-S_w/2}$ comes, of course, the unknown parameter $\alpha$. In the integration over $\alpha$, the problem of an overdensity of wormhole ends can in principle reappear. This is the subject of Sect.~\ref{fks}

\subsection{Multiple large universes}
\begin{figure}[ht]
	\centering
	\includegraphics[width=0.3\linewidth]{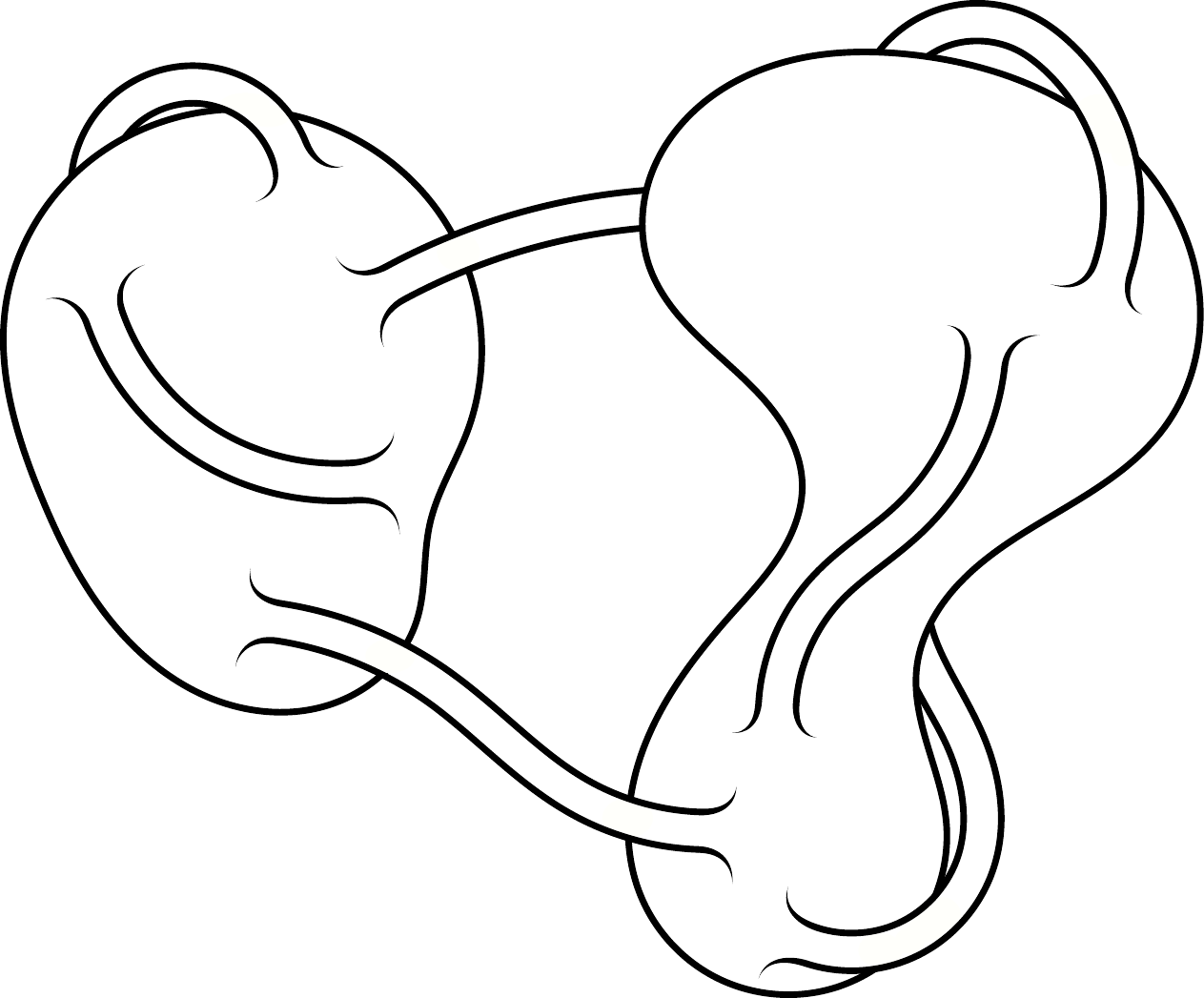}
	\caption{Large universes connected by wormholes -- figure adapted
        from Ref.~\cite{Fischler:1988ia}.}
	\label{fig:SphericalUniverses}
\end{figure}

Only the case of one large parent universe with many small-scale wormholes attached has been considered so far. It is, however, completely logical to allow for multiple large universes. Wormholes can connect one large universe to itself or to another large universe, cf.~Fig.~\ref{fig:SphericalUniverses}. When all wormholes are integrated out, the large universes become disconnected. 

Following \cite{Fischler:1988ia,Preskill:1988na}, one can single out one particular large universe and consider the expectation value of an observable $\mathcal{A}(x)$ in that universe. Keeping the values of the $\alpha$ parameters (which are common to all large universes) fixed for the moment, one has
\be
\langle \mathcal{A}(x) \rangle_\alpha \,=\,\int Dg_d\, e^{-S[g_d;\lambda - \alpha]}\mathcal{A}(x)\,.
\ee
Here $Dg_d$ (with `$d$' for disconnected) stands for the integration over all large-scale metrics, including a summation over manifolds with many components. Making this summation over the number of disconnected components explicit, 
\bea
\langle \mathcal{A}(x) \rangle_\alpha &=& \sum_N^{\infty} \frac{1}{N!}\, \prod_{n=0}^N \left( \int Dg_n\, e^{-S[g_n;\lambda - \alpha]}\right) \int Dg\, e^{-S[g;\lambda - \alpha]} \mathcal{A}(x) \nonumber \\
&=& \exp{\Big( \int Dg' e^{-S[g';\lambda - \alpha]} \Big) } \int Dg \,e^{-S[g;\lambda - \alpha]} \mathcal{A}(x)\,.
\eea
Here, in the first line, $g$ is the metric on the distinguished large universe and $g_n$ are the metrics on the other disconnected components. The second line used the fact that the sum over disconnected geometries exponentiates, introducing the variable $g'$ for the metric on a generic such component. 

Reinstating the $\alpha$-integration gives
\be
\langle \mathcal{A}(x) \rangle \,=\, \int Dg\, D\alpha\, G(\alpha)\, P(\alpha) \, e^{ -S[g;\lambda - \alpha] }\,\mathcal{A}(x)\,,
\ee
with the probability distribution
\be
P(\alpha) = \exp{\big( \int Dg\, e^{-S[g;\lambda - \alpha]} \big) }\,.
\ee
In the calculation of the partition function, i.e.  without the insertion of a local operator, no connected component is singled out. The sum over topologies then exponentiates without the need to split of one of the factors:
\be \label{eq:Partionfunction}
Z = \int D\alpha\, G(\alpha)\, P(\alpha)\,.
\ee
As discussed later, the double exponential $ P(\alpha) $ is responsible both for the initial excitement in wormhole physics (Coleman's solution to the cosmological constant problem \cite{Coleman:1988tj}) as well as for a particularly serious conceptual problem (the FKS catastrophe \cite{Fischler:1988ia,Kaplunovsky}).

\section{Phenomenological applications} \label{sec:PhenoApplications}

\subsection{Random values of couplings \texorpdfstring{\\} {}and the cosmological constant problem}\label{subsec:CosmologicalConstant}

If one accepts that euclidean wormholes contribute to the path integral, one may clearly be concerned that all of the familiar local physics will break down. The reason is that the action of the wormhole contributions does not grow with the separation of the two points where they attach to our macroscopic spacetime. The possible loss of quantum coherence has also been initially discussed in this context. However, it has quickly been established~\cite{Giddings:1988cx, Coleman:1988cy} that a local effective field theory desciption can be recovered by introducing $\alpha$ parameters in the path integral or, equivalently, $\alpha$ vacua in the canonical approach (cf. the discussion in the last chapter). 

The implications of this are nevertheless quite dramatic: All coupling constants depend on the $\alpha$ vacuum, i.e., on the a priori unknown baby universe state. This state is an unavoidable additional piece of information which has to come on top of the quantum-field-theoretic initial conditions given on a Cauchy surface of our spacetime manifold. By measuring couplings one is effectively determining some of the infinitely many $\alpha$ parameters. There seems to be no hope of predicting these couplings on the basis of a unique theory of everything, even if the latter was known to us. From a modern point of view, this is of course very similar to the situation which has anyway been widely accepted after the advent of the string theory landscape~\cite{Bousso:2000xa, Giddings:2001yu, Kachru:2003aw, Susskind:2003kw, Denef:2004ze}. In fact, both ways of randomizing coupling constants may be at work simultaneously. The familiar deep issue of the measure problem of enternal inflation (the leading candidate mechanism for populating the landscape) has a cousin in the form of the measure on or the dynamics of the baby universe state.

The above situation may be viewed as the generic phenomenological implication of euclidean wormholes or gravitational instantons. For the initial popularity of this paradigm, it was crucial that an apparently very successful attempt was made early on to derive a statistical prediction for one of the couplings - the cosmological constant~\cite{Coleman:1988tj} (for early applications of wormholes to other phenomenologically relevant couplings see~\cite{Grinstein:1988wr, Nielsen:1988kf, Choi:1989ck, Preskill:1989zu, Gilbert:1989nq}). In fact, a distribution infinitely peaked at zero was found, making the prediction exact. Subsequently many caveats were discovered such that the `cosmological constant prediction'  is not viewed as a central motivation for wormhole physics at present. Nevertheless, because of its intrinsic interest and its immense historical importance we review the argument in the remainder of this subsection (for reviews discussing this as well as other early approaches to the cosmological constant problem see~\cite{Weinberg:1988cp,Carroll:1991mt}).

The argument is due to Coleman \cite{Coleman:1988tj} and can be given using just the leading terms of the bare gravitational action:
\be
S[g] = \int \diff^4x \sqrt{g} \left( \Lambda - \frac{M_{\rm P}^2}{2}\,R + \cdots \right)
= \int \diff^4x \sqrt{g}\, \sum_i \lambda_i {\cal O}_i\,.
\ee
Here $\lambda_1=\Lambda$ and $\lambda_2=-M_{\rm P}^2/2$ characterize the cosmological constant and the Planck scale. As discussed before, including the effects of wormholes and allowing for multiple large parent universes (as in Fig.~\ref{fig:SphericalUniverses}) leads to the partition function (cf.~(\ref{eq:Partionfunction}))
\be
Z = \int D\alpha \,\exp \left( - \frac{1}{2} \sum_{i,j}\alpha_i \Delta^{-1}_{ij} \alpha_j \right)\, \exp \Bigg( \int Dg\, \exp \left( -\int \diff^4x \sqrt{g}\,\sum_i (\lambda_i-\alpha_i) {\cal O}_i \right) \Bigg)\,.
\ee
Since wormholes have been integrated out, the relevant metric in the above expression refers to a single parent universe. As argued in \cite{Coleman:1988tj}, this expression is dominated by values of $\alpha$ which correspond to $\Lambda_{\rm eff}=\lambda_1-\alpha_1>0$. Furthermore, the sum over topologies is dominated by spheres. The path integral over metrics can then be estimated in the saddle point approximation:
\be
Z = \int D\alpha \exp{ \left( - \frac{1}{2}\sum_{i,j}\alpha_i \Delta^{-1}_{ij} \alpha_j \right)}\exp\Bigg(\exp\left(-\int d^4x\,\sqrt{g}\,\left(\Lambda_{\rm eff}-\frac{M_{\rm P,\, eff}^2}{2}\,R\right)\right)\Bigg)\,,
\ee
where $M_{\rm P,\, eff}^2=M_{\rm P}^2+2\alpha_2$ and the sum is restricted to $i,j=1,2$. Thus, all one needs is the action of a 4-sphere solution with the above effective Planck scale and cosmological constant. Given that a 4-sphere of radius $r$ has volume $V_4=(8/3)\pi^2 r^4$ and scalar curvature $R=12/r^2$, this action is
\be
S_{\rm sphere}=-\frac{24\pi^2 M_{\rm P,\, eff}^4}{\Lambda_{\rm eff}}\,.
\ee
This gives
\be
Z= \int D\alpha\, \exp{ \Big( - \frac{1}{2}\sum_{i,j}\alpha_i \Delta^{-1}_{ij} \alpha_j \Big)} \exp{ \bigg[ \exp \bigg( 96\pi^2\,\frac{\big(M_{\rm P}^2/2 + \alpha_2\big)^2}{\big(\Lambda + \alpha_1\big)} \bigg) \bigg] }\,,
\label{zco}
\ee
where $\alpha_1$ was redefined $ \alpha_1 \to -\alpha_1 $.

The key point of this result is the double exponential enhancement of the measure governing the $\alpha$-parameter integration at the point where the effective cosmological constant vanishes.

As already emphasized, serious caveats exist and the above logic is nowadays generally not considered a valid solution to the cosmological constant problem.  First, the measured value of the cosmological constant is not any more consistent with zero. Second, firm evidence exists for cosmological inflation, and it is not clear how such an early quasi-de-Sitter period fits in the  argument for vanishing $\Lambda$. Finally, as we will discuss in detail in Sects.~\ref{fks} and~\ref{eqg}, the above argument may run into problems with an overdensity of wormholes (the FKS catastrophe) and the sign or negative-mode problem of euclidean quantum gravity. 

Nevertheless, reinterpretations of Coleman's mechanism have recently been explored in the context of the cosmological constant and other fine tuning problems of the Standard Model~\cite{Kawai:2011rj, Kawai:2011qb, Hamada:2014ofa, Hamada:2014xra}. The authors take a Lorentzian approach to the dynamics of multiple large universes connected by wormholes. Such a real-time formulation avoids the problems of the euclidean path integral of gravity, but at the same time modifies the conclusions obtained by Coleman. The analysis is based on the Wheeler-DeWitt wave function for a system of multiple large universes emerged in an evolving baby universe gas. By tracing out the unobserved large and baby universes, a density matrix $\rho$ describing our large universe is derived. The dependence of $\rho$ on the universe volume $z$ and on the couplings of the effective action, i.e. on the wormhole-induced $\alpha$ parameters, is studied. A problematic feature is the divergence of integrals over universe volumes $z$ arising in the density matrix calculation. 
It is treated by an IR cutoff $z_{IR}$ corresponding to a maximum universe size.  Under these assumptions, it is argued that the density matrix  $\rho$ peaks at vanishing cosmological constant, as in Coleman's mechanism, albeit with a much milder power-law dependence. This is interpreted as a prediction for a vanishing effective cosmological constant at asymptotically large times.

\subsection{Axion potentials from wormholes}

The main current phenomenological interest in wormholes lies in their interplay with axions. Axions have been an important ingredient in models of particle physics and cosmology since they were first proposed as solutions to the strong CP problem, and have found much wider applications ever since, e.g. as components of dark matter or as inflaton candidates. From a top-down perspective, axions are among the most generic outcomes of string compactifications, and are hence extremely well motivated (see, e.g.~\cite{Arvanitaki:2009fg}).  

Axions enjoy a global shift symmetry $\phi\to \phi +\epsilon$ that prevents the appearance of a potential at the perturbative level. It is only non-perturbative effects such as charged instantons and wormholes that can break these symmetries and give axions a mass. In fact, the explicit example of the Giddings-Strominger wormhole arises in the presence of axions and carries a corresponding charge given by~\eqref{eq:charge}. This is precisely the type of object required to generate an axion potential, as we review next following \cite{Rey:1989mg} (see also \cite{Alonso:2017avz}). 

Recall the discussion of Section~\ref{sec:babyuniverses} on the wormhole correction $I[g,\phi]$ to the low energy effective action of a large parent universe propagating in a plasma of baby universes. It is given by the effective Hamiltonian (\ref{eq:EffectiveHamiltonian}), which can be written in the form \cite{Rey:1989mg}
\begin{align}\label{eq:axioncorrection}
e^{I}=\langle \psi_2 |\exp\left[\sum_{n\in\mathbb{Z}} e^{-S_w(n)/2}K_n \int \diff^4x\sqrt{g(x)} \mathcal{O}_n(x)(a_n + a_{-n}^\dagger)\right]|\psi_1\rangle.
\end{align}
Here, the exponential factor $ e^{-S_w(n)/2} $ has been extracted from the matrix $ \Delta_{mn} $, making the dependence on it explicit. The remainder is denoted by $ K_n $.  The states $|\psi_i\rangle$ live in the Fock space of baby universes on which the parent universe propagates. Correspondingly, $a_n$ and $a_{n}^\dagger$ represent baby universe annihilation and creation operators. 

Baby universes associated to Giddings-Strominger wormholes carry an axionic charge given by~\eqref{eq:charge}. That is, they satisfy $[Q,a_n]=-n a_n$ and $[Q,a_n^\dagger]=n a_n^\dagger$, where $Q$ generates the $U(1)$ axionic shift symmetry. This charge is the reason why the combination $(a_n+a_{-n}^\dagger)$ appears in~\eqref{eq:axioncorrection}, generalizing equation (\ref{eq:EffectiveHamiltonian}): it is impossible for an observer in the parent universe to distinguish between the annihilation of a baby universe of charge $n$, and the creation of a baby universe of charge $-n$. These two processes hence generate the same local perturbation $\mathcal{O}_n$. Total charge conservation implies that the effective operators $\mathcal{O}_n(x)$ must be charged as well $[Q,\mathcal{O}_n]=n\mathcal{O}_n$, i.e. they transform as $ \mathcal{O}_n(x) \to e^{in\epsilon}\mathcal{O}_n(x) $ under the axionic shift $ \phi \to \phi + \epsilon f $. From this one can deduce that the local operators must be of the form $ \mathcal{O}_n(x) = e^{in\phi/f}\mathcal{O}_S(x) $, where $ \mathcal{O}_S(x) $ is a singlet. 

One can explicitly evaluate~\eqref{eq:axioncorrection} by choosing the baby universes  to be in an $\alpha$-eigenstate (introduced in Section~\ref{sec:babyuniverses}), i.e. $|\psi_1\rangle=|\psi_2\rangle=|\alpha\rangle$, with $ (a_n + a_{-n}^\dagger)|\alpha\rangle = \alpha_n |\alpha\rangle $.\footnote{Since the operator $ A_n := a_{n}+a_{-n}^\dagger $ is not Hermitian, one may worry that no basis of eigenvectors exists. To show its existence notice that the operators $ C_n: = A_n + A_n^\dagger, \bar{C}_n:= i(A_n - A_n^\dagger) $ are Hermitian. A quick calculation shows that $[C_n,\bar{C}_m] = 0 $, thus $C_n$ and $\bar{C}_n$ can be diagonalized simultaneously with an orthonormal basis. Since $ 2A_n = C_n -i\bar{C}_n $ these basis elements are also eigenvectors of $ A_n $. This also shows that the eigenvalues of $ A_n $, which are precisely the $\alpha$-parameters, will generically be complex.} The correction to the low energy action of a large parent universe propagating in such a background is hence given by 
\begin{align}\label{eq:axioncorrection2}
I&= \sum_{n\in\mathbb{Z}} e^{-S_w(n)/2}K_n \int \diff^4x\sqrt{g(x)} O_S(x)|\alpha_n|\exp{\Big(\frac{in\phi}{f} + i\delta_n\Big)} \\ \nonumber
&= \sum_{n\in\mathbb{N}_0} e^{-S_w(n)/2}K_n \int \diff^4x\sqrt{g(x)} O_S(x)|\alpha_n|\cos{\Big(\frac{n\phi}{f} + \delta_n\Big)}
\end{align}
where $ \alpha_n = |\alpha_n|e^{i\delta_n} $. Of course, it is easy to consider propagation between more general baby universe states. For example, $|\psi_1\rangle=|\psi_2\rangle=|0\rangle$ would lead to an integral of~\eqref{eq:axioncorrection2} over $\alpha_n$ with a Gaussian measure analogous to~\eqref{eq:Aplitute}. 

The operator $\mathcal{O}_S$ can be expanded in a set of singlet operators, e.g. $\mathcal{O}=1+a \mathcal{R}+\ldots$. Of these, the most interesting one is the unit operator, which leads to a potential for the axion. Taking into account only wormholes with charge $ n = \pm1 $ the induced potential is of the form
\be\label{eq:axionpotential}
V_{w}(\phi) \sim \frac{|\alpha_1|}{r_0^4}e^{-\frac{S_w}{2}}\cos{\bigg(\frac{\phi}{f}+\delta\bigg)}.
\ee
The coefficient of the potential is hard to calculate in general. In most cases (in particular for $f<M_P$) its precise value is not very relevant due to the dominant exponential suppression $ e^{-S_w/2} $. In the following, whenever an explicit estimate is needed, we will follow~\cite{Alonso:2017avz} and use the wormhole neck radius $ r_0^4 = C = (24\pi^4f^2M_P^2)^{-1} $ as in~\eqref{eq:axionpotential}. At this stage, if no other potential exists, the phase $ \delta \equiv \delta_1 $ is unphysical and could be absorbed by a shift in the axion field. Generically, in the presence of other terms in the potential, there is no reason why $ \delta_1 $ should not appear. The dimensionless parameter $|\alpha_1|$ depends on the baby universe state and is not predicted by the theory. For explicit evaluations one can assume that $ |\alpha_1| $ is an order one parameter~\cite{Alonso:2017avz}. A possible justification could be the expectation value $ \int d|\alpha| e^{-|\alpha|^2}|\alpha|^2 = \sqrt{\pi}/4 $ of order one. It is not unreasonable to use the Gaussian distribution since the latter appears when considering propagation between baby universe $|0\rangle$-vacua. In principle, however, the $\alpha$-parameters could take any value.

\subsection{Superplanckian axions}\label{sec:superplanckian}

\subsubsection{Large field excursions and inflation}

One of the most interesting applications of axions is inflation (see, e.g.~\cite{Baumann:2009ds,Baumann:2014nda,Westphal:2014ana} for reviews with emphasis on stringy contexts). The perturbative shift symmetry that axions enjoy makes them ideal inflaton candidates in models of large field inflation. In these constructions, the inflaton traverses distances in field space larger than the Planck scale, $\Delta \phi \gtrsim M_P$.  Generically, such large field displacements imply a high UV sensitivity of the model since higher-order terms in the potential, $\Delta V(\phi)\sim \phi^{n+4}/\mu^n_{UV}$, become relevant (here $\mu_{UV}\lesssim M_P$ is a UV cutoff scale).  This may clash with the slow-roll requirement of a smooth potential. Successful models of large field inflation hence demand a fine control of UV corrections, as it is indeed provided by axions. 

One of the main reasons for the current interest in large field models is their prediction of observable primordial tensor modes in the CMB. These are parametrized by the tensor-to-scalar ratio $r$. Under mild assumptions, the Lyth bound~\cite{Lyth:1996im} relates $r$ to the inflaton field displacement,
\be
\Delta \phi\gtrsim \left(\frac{r}{0.01}\right)^{1/2} M_P\,.
\ee
The current experimental bounds~\cite{Ade:2015tva,Ade:2015lrj,Array:2015xqh} are $r <0.07$ (95\% confidence level, Planck, BICEP2/Keck-Array combined), with near future experiments expected to strengthen this bound significantly. The  combination of these searches with the Lyth bound and the UV sensitivity of large field inflation provides an ideal playground for testing UV features of effective field theories and possibly quantum gravity. 

As already emphasized, the main challenge facing large field models is 
the control of UV corrections. Symmetries are required to avoid a drastic tuning of higher dimension operators. This is naturally realized by axions since, due to the shift symmetry $\phi\to \phi +\epsilon$, the axion potential vanishes automatically. This symmetry is mildly broken by non-perturbative effects, such as instantons and wormholes, which generate a periodic potential of the form
\be\label{eq:axionV}
V(\phi)= \Lambda^4 \sum_{n} e^{-S_n}\cos\left(\frac{n\phi}{f}+\delta_n\right)= \Lambda^4 e^{-S_1}\cos\left(\frac{\phi}{f}+\delta_1\right)+\ldots\,.
\ee
Here $\Lambda$ is a typical UV scale and $n$-dependent non-exponential prefactors have been suppressed. As discussed previously, gauge instantons and axionic wormholes induce such potentials (eqs.~\eqref{eq:instantonpotential} and~\eqref{eq:axionpotential}, respectively). Different harmonics correspond to instantons/wormholes of different instanton number/axionic charge $n$.

The idea of {\it natural inflation}~\cite{Freese:2014nla} is to use the $n=1$ term in~\eqref{eq:axionV}. Neglecting higher harmonics is justified in many cases due to the expectation that $e^{-S_n}\ll e^{-S_1}$ for $n>1$. Slow roll inflation then requires $f > M_P$ (notice that the maximum field displacement of the canonically normalized axion is $2\pi f$). In this simplest form, models of natural inflation are disfavoured by Planck~\cite{Ade:2015tva, Ade:2015lrj,Array:2015xqh}. However, this can be remedied by small corrections, e.g. from higher harmonics. More importantly, natural inflation continues to play the role of a `benchmark model' exemplifying in the simpest way the interplay between UV theory constraints and observations. Our considerations also have applications in models of axion monodromy~\cite{Silverstein:2008sg, McAllister:2008hb}.

One might try to use wormholes to generate the inflationary potential, but one immediately runs into difficulties. Semiclassically, the charge-$n$ wormhole action is $S_n\sim n M_P/f$. In the regime of interest, $f\gtrsim M_P$, higher harmonics are not sufficiently suppressed, $e^{-S_{n+1}}\not\ll e^{-S_n}$, at least for terms with $n\lesssim f/M_p$. Thus, there is no hierarchy between the first few terms in the series~\eqref{eq:axionV} and hence no perturbative control.

A closely related and more profound issue is the fact that the lowest charge instantons are microscopic and subject to strong corrections. The spectrum of microscopic instantons does not need to resemble the classical spectrum of macroscopic wormholes (just like the spectrum of charged elementary particles does not resemble the spectrum of Reissner-Nordstrom black holes). This suggests that the dominant axion potential will be generated by some microscopic non-perturbative effect, over which one has little control, and macroscopic wormholes will only induce higher corrections. The ideal situation for inflation would then take the form
\be\label{eq:inflation}
V(\phi)=\Lambda_{inf}^4 \cos\left(\frac{\phi}{f}\right)+\sum_{n\,>\,n_c}\Lambda_{w}^4 e^{-S_{w}(n)/2}\cos\left(\frac{n\phi}{f}+\delta_n\right)\,.
\ee
Here $\Lambda_{inf}$ is the scale of the inflationary potential, generated by a microscopic instanton. The sum is only over macroscopic wormholes, i.e. those whose radius of curvature is larger than the cutoff lenght. 

Given this setup, one may ask how important the wormhole  contribution is~\cite{Montero:2015ofa,Hebecker:2015rya}. To have a successful model of inflation, it should be subdominant:
\be\label{eq:constraint}
\frac{\Lambda_{w}^4e^{-S_{w}(n)/2}}{\Lambda_{inf}^4}\ll 1\,.
\ee
Because of the exponential dependence, this constraint is highly sensitive to the wormhole action $S_w(n)=(\pi\sqrt{6}/4)\,|n|M_P/f$. The dependence on the prefactor $\Lambda_{w}$ is much milder and one can, as in~\eqref{eq:axionpotential}, write $\Lambda_w^4\sim r_{n}^{-4}=24\pi^4 f^2 M_P^2/n^2$, where $r_n$ is the radius of the $S^3$ at the neck of the wormhole. The constraint~\eqref{eq:constraint} becomes
\be\label{eq:bound2}
\frac{\Lambda_{w}^4e^{-S_{w}(n)/2}}{\Lambda_{inf}^4}\sim \frac{e^{-(3\pi^3/2)\,r_n^2M_P^2}}{r_{n}^4\Lambda_{inf}^4}\ll1\,.
\ee

This bound takes its tightest form for the wormholes with lowest charge. One should, however, only consider those which are controlled in effective field theory, i.e. whose neck radius $r_n$ is larger than a UV cutoff scale $r_n\gtrsim\mu_{UV}^{-1}$. This condition defines $n_c$ in (\ref{eq:inflation}). 
The constraint can now be further rewritten in terms of the cutoff \cite{Hebecker:2015rya,Hebecker:2016dsw}
\be 
\frac{e^{-(3\pi^3/2)\,M_P^2/\mu_{UV}^2}}{\Lambda_{inf}^4/\mu_{UV}^4}\ll1\,.
\label{ineq}
\ee
One sees that, parametrically, inflation is in trouble in theories with a high cutoff, $\mu_{UV}\sim M_P$. The reason is that one has an ${\cal O}(1)$ number in the exponent, hence an ${\cal O}(1)$ numerator, and a parametrically small denominator. However, taking into account the surprisingly large numerical prefactor $3\pi^3/2$ and the value $\Lambda^4_{inf}\sim 10^{-8}M_P$ relevant for phenomenological large field inflation, the conclusion changes dramatically. One finds that the inequality~\eqref{ineq} is saturated at $\mu_{UV}\simeq 2.5\,M_P$ (corresponding to $r_n\simeq 0.4 M_P$). Thus, even the smallest controlled wormhole solutions appear to be harmless~\cite{Hebecker:2016dsw}.\footnote{
The 
exponentials in~\eqref{eq:bound2} and (\ref{ineq}) are highly sensitive to the precise definition of the cutoff or, equivalently, the critical radius $r_c$ above which wormholes are considered `macroscopic'. This was analysed in~\cite{Hebecker:2016dsw} using string compactifications with $g_s=1$ and self-dual compactification radius $R$. Equating the (appropriate power) of the wormhole $S^3$ volume with the volume of the compact torus, $(2\pi^2 r_c^3)^2\equiv (2\pi R)^6$, one obtains a suppression of $e^{-S/2}\sim 10^{-68}$. Imposing instead that the great circle of the $S^3$ be equal to the torus $S^1$s, $2\pi r_c\equiv 2\pi R$, the suppression becomes $e^{-S/2}\sim 10^{-13}$. In neither of these cases are macroscopic wormholes able to affect inflation. Nevertheless, minor modifications of the definition of $r_c$ could change this conclusion.} 

\subsubsection{The weak gravity conjecture (WGC)}\label{sec:WGC}

The inflationary potential~\eqref{eq:inflation} is perfectly acceptable from a (bottom-up) effective field theory perspective. As  just discussed, macroscopic wormholes do not affect this potential significantly. However, one may be concerned that the contribution from smaller wormholes was removed by hand, and this is, at least naively, the dominant one. To argue for generic constraints coming from this regime, where one loses semiclassical control, one has to resort to ideas about the quantum gravity {\it swampland}.

The concept of the swampland~\cite{Vafa:2005ui} refers to the set of apparently consistent low-energy effective field theories which are, nevertheless, inconsistent with a UV completion in quantum gravity. It arises most naturally in string theory, where it represents the complement, in the space of effective field theories, of the vast {\it landscape} of string compactifications. Effective theories in the swampland are those that cannot arise from a UV-complete fundamental theory, and in particular from string theory.

Several criteria have been conjectured to discern whether a given theory belongs to the swampland. Most of them refer to properties of the spectra of operators charged under gauge symmetries. The simplest and perhaps most solidly founded of the swampland conjectures are the statements that every symmetry must be local and that the whole lattice of corresponding gauge charges consistent with charge quantization must be populated (see, e.g.~\cite{Banks:2010zn}). That is, for every symmetry there must exist a gauge potential ($A_\mu$ in the case of a one-form), and there must exist states carrying every possible set of charges (every integer charge for a single $U(1)$ with an appropriate normalization). 

A more stringent, albeit more speculative conjecture is the WGC~\cite{ArkaniHamed:2006dz}: It states that at least some of the charged states must be super-extremal, that is, their charge-to-mass ratio must be larger than that of the corresponding extremal gravitational solution:
\be\label{eq:WGC}
z_{WGC}\equiv \left(\frac{q}{m}\right)_{WGC}\geq \left(\frac{Q}{M}\right)_{ext}\,.
\ee
This is generally described as the statement that ``gravity is always the weakest force'', since when~\eqref{eq:WGC} is satisfied, the gauge repulsion of two distant equal-charge objects dominates their gravitational attraction.  In case of a single $U(1)$, the extremal object corresponds to an extremal Reissner-Nordstrom black hole, which in appropriate units satisfies $z_{ext}=M_P^{-1}$.\footnote{
The
WGC has been motivated by the requirement that no stable bound states with arbitrary charge should exist. Super-extremal objects implement this requirement by permitting otherwise stable extremal black holes to decay through Schwinger pair production. It remains to be rigorously proven, however, that this requirement arises from fundamental consistency conditions. Unfortunately, the exciting and active field of the WGC lies outside the scope of this review and we limit ourselves to the axionic version and the consequences for natural inflation since this directly relates to our main subject.
} 
Since macroscopic gravitational solutions cannot be super-extremal (super-extremal black holes contain a naked singularity, violating cosmic censorship), one expects~\eqref{eq:WGC} to be satisfied by microscopic objects. For such states, quantum corrections can become relevant, pushing them away from extremality. 

Now, what does all of this have to do with axions, wormholes and inflation? In general, abelian gauge theories arise from $p$-form gauge fields under which $p$-dimensional objects (i.e. whose world-volumes are $p$-dimensional) are charged. The swampland conjectures, and in particular the WGC, are expected to hold for all possible $p$~\cite{ArkaniHamed:2006dz}.  The case of particles with electric charge corresponds to $p=1$, strings coupled to a two-form field to $p=2$ and, most relevant for our interests, axions can be understood as $p=0$ gauge fields, to which `zero-dimensional' instantons/wormholes couple. This interpretation can be made manifest by considering a standard one-form gauge field in 5d, reduced on a circle to 4d. The component of the gauge field along the circle, the Wilson line, becomes a periodic axion in 4d, whose periodicity reflects the higher dimensional gauge symmetry. In this way, one can relate the mass $m$ and charge $q$ of a 5d particle to the action $S_n$ and axionic charge $n$ of a 4d instanton, respectively. The WGC~\eqref{eq:WGC}, when applied to axions is hence expected to read
\be\label{eq:axionWGC}
\left(\frac{n}{S_n} \right)_{WGC}\geq \left(\frac{N}{S_N}\right)_{ext}\,.
\ee
Just like extremal black holes satisfy $M/Q\sim e M_P$ with $e$ the gauge coupling, one generally expects {\it extremal instantons} to satisfy $S_N/N\sim M_p/f$. If the WGC~\eqref{eq:axionWGC} is satisfied  by the instanton of lowest charge $n=1$, this means that $S_1f \lesssim M_P$. This is incompatible with the basic requirements of large field natural inflation ($f\gtrsim M_P$) in regimes of perturbative control, $S_{inst}\gtrsim 1$.\footnote{The loss of perturbative control has a particularly nice interpretation in string theory compactifications, where one generically finds that trans-planckian axions $f\gtrsim M_P$ arise only in regimes where either the string coupling becomes strong, or Kaluza-Klein/winding modes become light~\cite{Banks:2003sx}.} Setups in which the instanton that satisfies the WGC is not the one of lowest charge have been proposed as a loophole to this strong constraint~\cite{Rudelius:2015xta,Brown:2015iha,Brown:2015lia} and are being actively investigated~\cite{Hebecker:2015rya}.\footnote{
More 
generally, current approaches to large field axion inflation can be roughly divided into multi-axion~\cite{Kim:2004rp,Dimopoulos:2005ac} and monodromy~\cite{Silverstein:2008sg,McAllister:2008hb,Marchesano:2014mla,Blumenhagen:2014gta,Hebecker:2014eua} models (also useful in the relaxion mechanism~\cite{Graham:2015cka}). The WGC and related swampland ideas can be generalized to such setups, and lead to interesting phenomenological features and constraints. The strength of these depends on subtleties in the precise formulation of the WGC and is being intensely debated.} 

The main difficulty in making the requirement~\eqref{eq:axionWGC} more precise is to properly identify what one means by an extremal instanton/wormhole. In setups where the axion arises from a 5d gauge field, one can see that the higher dimensional extremal black holes correspond to the extremal instantons introduced in Section~\ref{sec:dilatonic}. However, these setups always involve a dilaton field (the radius of the compactification circle) for which the coupling parameter $\beta$ of eq.~\eqref{eq:dilatonfunction} is $\beta=2\sqrt{2/3}$. Recall from Sec.~\ref{sec:dilatonic} that wormhole solutions only exist for $\beta<2\sqrt{2/3}$. Since the main interest (at least for inflation) is in the case where the dilaton has been stabilized and disappears from the low energy theory, i.e.~$\beta=0$, the relation to higher dimensional black holes is lost, along with a rigorous notion of an extremal instanton/wormhole. 

Hence, with our current understanding, some amount of guesswork is required to properly interpret~\eqref{eq:axionWGC} in a pure axion-Einstein theory. Following~\cite{Hebecker:2016dsw}, we will assume that, on the right hand side of the bound, one needs to use the classical action of a macroscopic wormhole. With this interpretation, the WGC states that some microscopic `wormhole' has a charge-to-action ratio larger than its macroscopic counterpart, i.e. that $S_n\leq (\pi\sqrt{6}/4)\,|n|M_P/f$.

Finally, we return to the effective model of natural inflation~\eqref{eq:inflation}. As discussed before, the sum over macroscopic wormholes is generically suppressed strongly enough to be ignored. Ideally, one could hope that the uncontrolled microscopic wormholes somehow disappear from the low energy theory. However, the WGC implies\footnote{
More
precisely, this requires one of the stronger forms of the conjecture. For example, one may demand that the instanton satisfying the bound $S_n/n<M_P/f$ has $n=1$ or at least $n\sim{\cal O}(1)$.
} 
quite the opposite, namely that (at least some) microscopic wormholes/instantons are less suppressed than their macroscopic counterparts and inflation is strongly affected. 

A possible caveat to this conclusion is the implicit assumption that all instantons enter the potential with ${\cal O}(1)$ prefactors. This, however, is not in principle required by the WGC. The smoothness of the inflationary potential may be preserved if the coefficients of dangerous corrections vanish or are highly suppressed, i.e. if $\Delta V \ll \Lambda_{inf}^4$ (see~\cite{delaFuente:2014aca} for a model potentially realising this possibility).

To discuss this point more generally, one can split the correction to the potential according to
\begin{equation}
\Delta V=\Delta V_1+\Delta V_2\quad\mbox{with}\quad
\begin{array}{ll}
\Delta V_1(\phi)\sim \sum_n r_n^{-4}e^{-S_n} & \text{for   } ~~r_n \gg \mu_{UV}^{-1}\\
\\
\Delta V_2(\phi)\sim \sum_n r_n^{-4}\left(\mu_{UV} r_n\right)^{\alpha}e^{-S_n} &   \text{for   }  ~~r_n \lesssim \mu_{UV}^{-1}\,,
\end{array}\label{two}
\end{equation}
with $\alpha>0$. Here $\Delta V_1$ comes from macroscopic instantons or wormholes and is harmless, as explained above. By contrast, $\Delta V_2$ comes from their microscopic counterparts and is dangerous according to the WGC. The reason is that small, low-charge instantons are not exponentially suppressed,  $e^{-S_n}\sim {\cal O}(1)$ for $n\sim {\cal O}(1)$. However, the prefactor of those instantons can be smaller than the naively expected $r_n^{-4}$. This has been has been parametrized by including a factor $(\mu_{UV}r_n)^\alpha$. 

As an example, let the microscopic instantons be gauge instantons of some non-abelian 4d gauge theory. The presence of charged fermions of mass $m$ does not affect the contribution of large instantons (relative to $1/m$). By contrast, the contribution of small instantons is suppressed by $(mr_n)^\alpha$ with $\alpha$ proportional to the number of fermions (as in the lower line in (\ref{two})). An analogous suppression can arise in the case of brane-instantons due to the presence of fermionic zero-modes. These are generally lifted by the SUSY breaking required for inflation. The formula (\ref{two}) is grossly oversimplified in that just a single threshold, $\mu_{UV}$, occurs. It is only intended to illustrate how the smallness of corrections could in principle come about. Indeed, one sees that $\Delta V\ll V_{inf}\sim H^2 M_P^2$ may be satisfied (for appropriate $\alpha$) together with $H\lesssim \mu_{UV}\ll M_P$. Finding specific implementations of such a mechanism remains challenging.

To summarize: effects of macroscopic wormholes in the low energy Einstein-axion theory are in general not strong enough to constrain models of natural inflation. Nevertheless, expectations based on the WGC place potentially strong bounds on such models. In particular, trans-Planckian axion decay constants are expected to arise only in regimes where perturbative control is lost, e.g. where microscopic wormholes/instanton spoil slow-roll conditions. Several possible ways around such constraints exist and are being actively studied. Whether such loopholes can be implemented in specific (string theory) setups is the subject of ongoing research.

\subsection{Subplanckian axions and Goldstone bosons}

Sub-Planckian axions $f<M_P$ are not suitable to accommodate inflation but they are extremely interesting in other phenomenological setups. Again, it is their shift symmetry and the resulting exponential suppression of their masses that makes them stand out among the plethora of fields relevant at low energy.

Let us repeat here for convenience the wormhole induced axion potential~\eqref{eq:axionpotential}:
\be\label{eq:sub-planckpotential}
V_{w}(\phi) \approx \frac{|\alpha_1|}{r_0^4}e^{-S_w/2}\cos{\bigg(\frac{\phi}{f}+\delta\bigg)}.
\ee
The mass induced by this potential is given by
\begin{align}
m^2 = 24\pi^4M^2_P|\alpha_1|e^{-S_w/2}= 24\pi^4M_P^2|\alpha_1|e^{-\frac{\pi\sqrt{6}}{8}\frac{M_P}{f}}\,.\label{mf}
\end{align}
In contrast to trans-Planckian axions, for a decay constant smaller than the Planck scale the wormhole contribution is strongly suppressed through the exponential $ e^{-S_w/2} $. This  ensures that the axion is very light, making it suitable for many phenomenological applications. The exponential dependence implies that small changes in $ f $ drastically change the value of $ m $, allowing for a wide range of values for the mass. This observation will be a recurring theme in the following. 

Another feature specific to sub-Planckian axions is that even wormholes of unit charge are macroscopic, in the sense that the size of their throat $r_0$ is larger than the Planck scale. This is a rather peculiar property, but it is necessary for~\eqref{eq:sub-planckpotential} to be trustworthy. More in general, in an effective theory with UV cutoff $\mu_{UV}$, the validity of~\eqref{eq:sub-planckpotential} requires $r_0^{-1}\sim \sqrt{fM_P}< \mu_{UV}$. If the cutoff scale becomes too low, one may expect sizeable corrections to the action of the wormholes with lowest charge.\footnote{The scale $\mu_{UV}\sim\sqrt{fM_P}$ itself arises in the context of the (magnetic) WGC as an intrinsic UV cutoff. The unit charge wormhole lies precisely at this scale, and is hence potentially subject to relatively sizeable corrections to its action.} The results described in this section assume the validity of~\eqref{eq:sub-planckpotential}, with $S_w$ taking its classical value. The important caveat just mentioned should nevertheless be taken into account when interpreting these results.

We proceed now to review potential phenomenological applications of axions with an induced wormhole potential of the form~\eqref{eq:sub-planckpotential}. Significant parts of our discussion follows~\cite{Alonso:2017avz}.\footnote{Mild discrepancies with the results of ref.~\cite{Alonso:2017avz} arise from the inclusion of a (Gibbons-Hawking-York) boundary contribution to the action of a semi-wormhole in~\cite{Alonso:2017avz}. Our perspective is that of a summation over full wormholes, where no such contribution arises. The semi-wormhole factor $e^{-S_w/2}$ appears only effectively through a rewriting.}

\subsubsection{Black hole superradiance and bosenovas}

As just explained, axions play a special role in testing quantum gravity, especially wormhole or baby universe effects. The reason is their extremely suppressed potential. Furthermore, assuming that the relevant $\alpha$ parameters take their natural ${\cal O}(1)$ value, the potential and hence the mass are predicted in terms of the decay constant. 

However, a generic (in particular non-QCD) axion is hard to observe. One classical possibility is black hole superradiance~\cite{Damour:1976kh, Detweiler:1980uk, Zouros:1979iw}. This term characterizes the energy deposition by a spinning black hole into a light scalar field, not-necessarily an axion, of suitable mass. The relevance for the discovery of axions has been emphasized in the context of the `string axiverse'~\cite{Arvanitaki:2009fg} and continues to receive much attention (see e.g.~\cite{Arvanitaki:2014wva, Arvanitaki:2016qwi, Brito:2017wnc, Brito:2017zvb, Cardoso:2018tly}). A recent discussion in the wormhole context appears in~\cite{Alonso:2017avz}.

The dependence of superradiance on the most important physics parameters are easily explained. Consider a spinning black hole with mass $M$, angular momentum $J$ and typical radius $R\equiv M/8\pi M_P^2$. One generally uses the spin parameter $a=J/M$ to characterize its rotation, with $a=R$ correspnding to extremality.\footnote{
Intuitively, 
$a$ is the radius which a shell with mass $M$, rotating at the speed of light, would need to have to generate $J$. It can not exceed the Schwarschild radius corresponding to $M$.
}
Superradiance is a classical instability which draws energy from the black hole and deposits it in the field oscillations of a light scalar, localized in a spherical region outside the horizon. Very roughly, one may think in terms of (classical analogues of) electron shells of an atom being populated by this scalar. The effect relies on the black hole being near extremality and on the Compton wavelength of the axion being comparable to the black hole radius, $R\sim 1/m$. 

It is instructive to consider what happens if this latter condition is not met~\cite{Arvanitaki:2009fg}: For an extremal black hole and $R \gg 1/m$, the instability time scale is given by \cite{Zouros:1979iw} 
\be
\tau \simeq (10^7\,R)\,\exp{(1.84\,mR)}\,.
\ee
In this regime, the Compton wave length is small and only modes with a large angular excitation superradiate. But such modes experience a high and thick centrifugal barrier, leading to an exponential suppression of the rate $1/\tau$. For subcritical $a$ the exponential suppression is even stronger. In the opposite regime $R\ll 1/m$, one has \cite{Detweiler:1980uk}
\be
\tau = 24\,R\,(R/a)\,(mR)^{-9}\,.
\ee
In this limit, low modes are available for superradiance. However, the potential well is now very wide and the modes spread out. One may say that the scalar's Compton wavelength is too large such that the small overlap with the black holes induces a suppression. 

Efficient superradiance hence requires a relation between the black hole mass and the axion Compton wavelength. Stellar black holes ($2-100M_\odot $) correspond to axion masses of $10^{-13}-10^{-10}$eV and supermassive black holes ($10^6-10^8M_{\odot}$) to $10^{-19}-10^{-16}$eV. The crucial signal for an axion in one of these regions would be gaps in the spectrum of rotating black holes. At present, spin and mass observations of stellar black holes exclude the range \cite{Arvanitaki:2014wva}
\be
6 \times 10^{-13}\text{eV} \lesssim m \lesssim 2 \times 10^{-11}\text{eV}.
\ee
Note that a detection of axion-induced superradiance is also possible through gravitational waves. The gravitational wave signals from, e.g., axion annihilation or axion transitions may be detected by future experiments at LIGO, VIRGO and at LISA \cite{Arvanitaki:2014wva, Arvanitaki:2016qwi, Cardoso:2018tly, Brito:2017zvb, Brito:2017wnc}.

In our context, i.e. for a pure-quantum-gravity potential, the relation (\ref{mf}) between mass and decay constant may in principle provide information beyond the generic axion case. Of course, the mass is subject to the uncertainties from the $\alpha$ parameter and fluctuation determinant. However, as can be seen by solving (\ref{mf}) for $f$,
\be
f=\frac{M_P\,(\pi\sqrt{6}/8)}{\ln(24\pi^4 M_P^2|\alpha_1|/m^2)}\,,
\ee
the sensitivity to these uncertainties is extremely week. Indeed, for $|\alpha_1|=1$ the above excluded mass window corresponds to the surprisingly narrow range $1.23\times 10^{16}$GeV$\lesssim f \lesssim 1.28\times 10^{16}$GeV. Thus, under the above assumptions, an axion discovery at the edge of the present mass window would imply a very precise determination of $f$. Similarly, the mass window $10^{-19}$eV$\lesssim m \lesssim 10^{-16}$eV accessible via supermassive black holes translates to $1.06\times 10^{16}$GeV$\lesssim f \lesssim 1.13\times 10^{16}$GeV. However, the key question is then whether an independent measurement of $f$ for such a `quantum gravity axion' is conceivable. 

It turns out that the answer to this question is positive: To measure the mass, it suffices to study superradiance at linear order. However, to get independent information about $f$, higher-order terms in the $\cos(\phi/f)$ potential have to be probed. This is possible, for example, in the context of the so-called bosenova. The term derives from analogous condensed matter phenomena~\cite{RIS_0}. In a bosenova, the self-interactions of the growing axion cloud around the black hole lead to a dynamical collapse: Part of the extracted energy is ejected and the rest returned to the black hole. This may happen multiple times until enough spin has been extracted from the black hole and superradiance (at least for the given level) is lost~\cite{Yoshino:2012kn}. 

Among the observable effects are a continuous gravitational wave signal as well as bursts of gravitational waves. For the continuous case, an analysis based on a possible axion cloud of the stellar black hole Cygnus X-1 was reported in~\cite{Yoshino:2014wwa}. Assuming that the LIGO upper limit is similar to that for gravitational waves from rotating distorted neutron stars, an expected exclusion range was derived. For $1.1\times 10^{-12}\text{eV} < m < 2.5\times 10^{-12}\text{eV}$, it restricts $f$ to lie below $10^{15}$ - $10^{16}$GeV. This can be understood intuitively since, as $f$ grows, the bosenova cuts off the superradiance instability at higher axion densities, leading to larger signals. 
Notice that the bound on $f$ is in the range relevant for wormhole induced potentials as discussed above. Realistic detection prospects exist also for gravitational wave bursts \cite{Yoshino:2015nsa}. Present limitations of the theoretical analysis are related to the need for including backreaction and extending certain parts of the numerics from the Schwarzschild to the Kerr solution (for details see e.g.~\cite{Yoshino:2015nsa}).

\subsubsection{QCD axion}

For the QCD axion an interesting observation can be made~\cite{Alonso:2017avz}. The total potential, including the contribution from the usual QCD instantons, is given by
\be\label{eq:QCDpotential}
V(\phi) = -\Lambda_{QCD}^4\cos{ \bigg( \frac{\phi}{f} \bigg) } - \frac{1}{r_0^4}e^{-\frac{S_w}{2}}\cos{\bigg(\frac{\phi}{f}+\delta\bigg)},
\ee
where the axion $ \phi $ is defined such that the QCD instanton induced potential is minimized at $\phi=0$. The phase $ \delta $ is redefined accordingly and is generically non-zero since there is no obvious reason for the two terms in the potential to have the same minimum. Furthermore, the $|\alpha_1|$ parameter has been set to one.

The dependence of the axion mass on the decay constant is interesting. With increasing $f$, the QCD contribution decreases while the wormhole one grows. Hence, the axion mass features a minimum as a function of $f$. 
It is not unreasonable to expect, on theoretical grounds, that gravitational effects are subdominant with respect to gauge contributions. This requires that $f\lesssim 1.4\times10^{16} \,\text{GeV}$. This bound can also be derived from phenomenological considerations.
The the phase of the wormhole contribution implies a shift of the minimum of the potential and hence a non-zero QCD $\theta$-parameter $\theta_{\text{eff}}$. The experimental bound $ \theta_{\text{eff}} \lesssim 10^{-10} $ coming from the neutron electric dipole moment constrains the wormhole contribution. Specifically, assuming $\sin(\delta)\sim \mathcal{O}(1)$, one finds a bound on the decay constant $f\lesssim 1.2\times 10^{16}\, \text{GeV}$. One might have suspected that the tight requirement~$ \theta_{\text{eff}} \lesssim 10^{-10} $ would lead to a stronger bound on $f$. This is not the case, however, due to the strong exponential dependence of the wormhole contribution. 

In the regime of small wormhole corrections, one can expand the potential~\eqref{eq:QCDpotential} and obtain the  axion mass and effective $\theta_{\text{eff}}$ parameter as functions of $f$, $\Lambda_{QCD}$ and $M_P$:
\begin{align}
m^2 &\approx \frac{\Lambda^4_{QCD}}{f^2} + 24\pi^4M^2_P\cos(\delta)\exp\Big(-\frac{\pi\sqrt{6}}{8}\frac{M_P}{f}\Big) \label{eq:QCDmass} \\ 
\theta_{\text{eff}} &\approx
24\pi^4\sin(\delta)\frac{f^2M_P^2}{\Lambda_{QCD}^4} \exp\Big(-\frac{\pi\sqrt{6}}{8}\frac{M_P}{f}\Big)\,.
\end{align}
The minimal mass obtained from~\eqref{eq:QCDmass} is $m\gtrsim 4 \times 10^{-9} \,\text{eV}$. Notice that the bound coming from superradiance described above is irrelevant in this case.

\subsubsection{Axions as dark matter}

Despite its success on scales larger than $ 10 \text{kpc} $, the scale of stellar distributions in typical galaxies, it is not clear yet if the cosmological $ \Lambda$CDM model is consistent with observations at smaller distances \cite{Weinberg:2013aya}. The tension arises from the cuspy halo cores and an abundance of satellite galaxies predicted by numerical simulations but incompatible with observations. Using an extremely light scalar field with mass $ 10^{-22} -10^{-21} \text{eV}$, it is possible to construct a model of dark matter with the same large scale predictions as CDM, in which, however, these problems are absent. The key idea here is that the large Compton wavelength of a light particle can suppress the formation of structures on sufficiently small scales. This dark matter model goes by the name of Fuzzy Dark Matter (FDM) \cite{Hu:2000ke}.

Because of its extreme lightness, an axion with the induced potential~\eqref{eq:sub-planckpotential} is an ideal candidate for FDM. Information about the possible values of $ f $ can  be obtained by reproducing the observed relic abundance via the misalignment mechanism~\cite{Preskill:1982cy, Abbott:1982af, Dine:1982ah}. Assuming an initial misalignment angle of order one $\theta_i=\phi_i/f\approx 1$, the axion contribution to today's energy density (normalized by the critical energy density) is given by \cite{Arvanitaki:2009fg, Kim:2015yna} (see also \cite{Hui:2016ltb})
\be\label{eq:energydensity}
\Omega_a h^2 \approx 0.1 \bigg( \frac{f}{10^{17}\text{GeV}} \bigg)^2 \Big( \frac{m}{10^{-22}\text{eV}} \Big)^{\frac{1}{2}}
\ee
where $ h = 0.678 $ is the dimensionless Hubble parameter. Requiring that the axion accounts for (a large fraction of) the measured dark matter energy density, i.e. that $\Omega_a h^2\approx 0.1$, implies a relation between the axion mass and its decay constant. For the FDM range of masses $ 10^{-22} \lesssim m \lesssim 10^{-21}\, \text{eV}$ the axion decay constant must lie in the range $ 5.6 \times 10^{16} \lesssim f \lesssim 10^{17} \,\text{GeV} $. 

It is interesting to compare these relations to those predicted by a wormhole induced mass (again using $|\alpha_1|\sim\mathcal{O}(1)$ as a benchmark)~\cite{Alonso:2017avz}. Plugging~\eqref{mf} into~\eqref{eq:energydensity}, one obtains that the correct relic density is obtained when $f\approx 10^{16}\,\text{GeV}$, which corresponds to an axion mass $m\approx 7\times10^{-19}\,\text{eV}$. While still valid as a candidate for dark matter, this mass is above the appropriate regime for the FDM scenario $m\lesssim 10^{-21}\, \text{eV}$. For the FDM setup, the exponential suppression $e^{-S_w/2}$ is too strong to obtain the full dark matter relic abundance. 

This conclusion is rather general and relates to the WGC described in section~\ref{sec:WGC}: Consider a generic non-perturbative axion mass $m^2= M_P^2\, e^{-S_i}$. A mass in the FDM range $m\lesssim 10^{-21}\, \text{eV}$ requires $S_i\gtrsim 220$. At the same time, obtaining the right relic abundance through eq.~\eqref{eq:energydensity} requires a rather large decay constant $f\gtrsim 5.6 \times 10^{16}\,\text{GeV}$. These two estimations combined lead to the interesting but potentially troublesome relation $f S_i \gtrsim 5 M_P$. The situation is similar to that of natural inflation described in section~\ref{sec:WGC}: demanding the production of enough FDM pushes instanton effects into the sub-extremal range $f S_i\gtrsim M_P$. This regime conflicts with the WGC which requires the presence of super-extremal (and hence naively dominant) instantons.

Of course, this conclusion is subject to several caveats. First, the exponential dependence on the instanton action makes the constraint highly sensitive to the precise extremality bound that enters the WGC. As previously discussed, the WGC for  wormhole generated potential suggests $f S_i \leq \sqrt{6}\pi M_P/8\approx 0.96 M_P$. Other setups (e.g. axio-dilaton instantons) provide slightly different numerical bounds, but none of them seem to prevent the conflict. Second, the dark matter abundance expressed by eq.~\eqref{eq:energydensity} assumes an initial angle of axion misalignment $\theta_i=\phi_i/f\approx 1$. Larger initial displacements can lead to an enhanced axion density. Specifically, for generic $-\pi<\theta_i<\pi$, equation~\eqref{eq:energydensity} should read
\be
\Omega_a h^2 \approx 0.1 \bigg( \frac{f}{10^{17}\text{GeV}} \bigg)^2 \Big( \frac{m}{10^{-22}\text{eV}} \Big)^{\frac{1}{2}} \,\theta_i^2 f(\theta_i)\,.
\ee
The function $f(\theta_i)$ accounts for anharmonicities of the potential when the initial value of the axion is close to maximum $\theta_i\to \pi$, where it diverges.  Using the approximate analytic expression for $f(\theta_i)$ given in~\cite{Visinelli:2009zm}, one can estimate that an initial tuning $\theta_i\approx 0.91 \pi$ leads to the correct relic abundance for $m\approx10^{-21}\,\text{eV}$ and $f S_i\approx M_p$. 

A third caveat is the fact that, as discussed around equation~\eqref{two}, the energy scale at which instantons generate a mass may be lowered if a UV cutoff $\mu_{UV}$ exists below the Planck scale, e.g. due to the pressence of fermionic modes.  
Consider an axion mass of the form $m^2=\mu_{UV}^2 e^{-S_i}$, and assume that the instantons saturate the bound $f S_i\approx M_P$. It is easy to see that the linear dependence of $m$ on $\mu_{UV}$ (as opposed to its exponential dependence on $S_i$) requires an extremely low instanton scale. In fact, for generic initial misalignment $\theta_i=1$, the cutoff scale should be $\mu_{UV}\approx 10^{-12}\, \text{eV}$. When the potential is generated by wormholes, a similar suppression could be in principle achieved by tuning the $|\alpha_1|$-parameter. 

Finally, as in its applications to large field inflation,  mild formulations of the WGC allow for loopholes in which sub-extremal instantons dominate the potential, and hence avoid the above constraints. In particular, systems with multiple axion are being actively investigated in this respect~\cite{Bachlechner:2017zpb}.

The above mechanisms can quite possibly reconcile axions as candidates of FDM with the WGC. It is nevertheless very interesting that such models, motivated mainly by their phenomenological applications, are probing  quantum gravity constraints.

Whether fuzzy or not, axions and their induced gravitational potentials provide well-motivated dark matter candidates. Further phenomenological features, such as the formation and stability of substructures (e.g. axion stars or oscillons) also depend on the ratio of $f$ and $m$. These will hopefully be experimentally probed in the near future. Moreover, direct detection experiments such as CASPEr~\cite{JacksonKimball:2017elr} and HeXeniA~\cite{HeXeniA} can also be expected to test the regime of extremely small (QCD-) axion masses in the foreseeable future (see also~\cite{Alvarez:2017kar}). 

In summary, there exists by now a whole set of promising phenomenological directions probing very light scalars, especially axions, which relate in a non-trivial way to quantum gravity and gravitational instantons.

\section{Conceptual issues} \label{sec:IssuesQuestions}
The discussion of wormholes presented so far has glossed over some fundamental questions which may change our perspectives on, or even invalidate, several of the results described in previous sections. Some of these issues were recognized and thoroughly discussed immediately after the first wormhole solutions were constructed, while others have been raised more recently, when wormholes have been considered in string and holographic setups. It is fair to say, nevertheless, that none of them has been fully understood yet. It is possible that the correct interpretation of wormholes and topology change will remain obscure until a controllable non-perturbative description of quantum gravity is found. It may well be, on the other hand, that the puzzles posed by wormholes can guide us in the pursuit of such a theory.

\subsection{FKS catastrophe}\label{fks}
Following Coleman's intriguing proposal for a wormhole-based solution to the cosmological constant problem \cite{Coleman:1988tj}, Fischler, Kaplunovsky and Susskind have argued that an inconsistency may be hidden in the underlying calculation \cite{Fischler:1988ia,Kaplunovsky}. Concretely, they extended Coleman's argument by including an $R^2$ term in the action and by allowing for Wilsonian renormalization group (RG) running. As a result, they found an overdensity of wormholes, even of those with large radius. 

The analysis of \cite{Fischler:1988ia} follows that of Coleman (cf.~Sect.~\ref{subsec:CosmologicalConstant}) very closely: The starting point is the action
\be
S[g] = \int \diff^4x \sqrt{g} \left( \Lambda - \frac{M_{\rm P}^2}{2}\,R +\gamma R^2+ \cdots \right) = \int \diff^4x \sqrt{g}\, \sum_i \lambda_i {\cal O}_i\,
\label{act3}
\ee
with $\lambda_1=\Lambda$, $\,\,\lambda_2=-M_{\rm P}^2/2$ and $\lambda_3=\gamma$. The partition function, including wormholes and multiple large universes, reads
\be
Z = \int D\alpha \,\exp \left( - \frac{1}{2} \sum_{i,j}\alpha_i \Delta^{-1}_{ij} \alpha_j \right)\, \exp \Bigg( \int Dg\, \exp \left( -\int \diff^4x \sqrt{g}\,\sum_i (\lambda_i-\alpha_i) {\cal O}_i \right) \Bigg)\,.
\ee
As before, the integral over metrics is performed in the saddle point approximation. This amounts to evaluating the action of (\ref{act3}) on a sphere of radius $r$ and extremizing in $r$. But, on dimensional grounds, the $R^2$ part of the action, evaluated on a sphere, gives an $r$-independent contribution. Hence the euclidean de-Sitter solution of Sect.~\ref{subsec:CosmologicalConstant} remains entirely unchanged. The resulting partition function is an exact copy of (\ref{zco}), except that the $R^2$ part has to be added to the saddle-point action in the double exponent:
\be
Z= \int D\alpha\, \exp{ \Big( - \frac{1}{2}\sum_{i,j}\alpha_i \Delta^{-1}_{ij} \alpha_j \Big)} \exp{ \bigg[ \exp \bigg( 96\pi^2\,\frac{\big(M_{\rm P}^2/2 + \alpha_2\big)^2}{\big(\Lambda + \alpha_1\big)}+(\gamma+\alpha_3) \bigg) \bigg] }\,.\label{zpro}
\ee
Again, $\alpha_1$ has been redefined $ \alpha_1 \to -\alpha_1 $, and $\gamma$ and $\alpha_3$ have been rescaled to avoid  numerical prefactors.

Now, by the same mechanism that drives $\alpha_1$ to $-\Lambda$ (Coleman's solution of the cosmological constant problem), the parameter $\alpha_3$ is driven to infinity. This will turn out to be problematic. To explain the issue, the Wilsonian RG perspective is useful.

Start with the effective action at some UV length scale $\rho_{\rm UV}$. The wormholes to be integrated out come in all sizes $\rho>\rho_{\rm UV}$. Indeed, even in the simple Giddings-Strominger case with a single axion, the different wormhole charges give rise to a discrete set of wormholes of different radii. Thus, one can think of going down in energy in a renormalization-group-like way: One first integrates out wormholes of sizes $\rho\in [\rho_{\rm UV},\rho_1]$, then those with $\rho\in [\rho_1,\rho_2]$, and so on (with $\rho_{\rm UV}<\rho_1<\rho_2<\cdots$). Very schematically, the previous formulae can be adjusted to this perspective by
\be
\int D\alpha\qquad\to\qquad \prod_{\rho}\int D\alpha(\rho)
\ee
and
\be
\lambda_i-\alpha_i\qquad\to \qquad \lambda_i-\sum_\rho\alpha_i(\rho)\,.
\label{liri}
\ee
The above wormhole-induced change of the couplings follows from iterating the basic step
\be
\lambda_i(\rho+\Delta\rho)=\lambda_i(\rho)-\alpha_i(\rho)\,.
\ee

In addition to (and intertwined with) this stepwise renormalization by wormholes, the usual RG running takes place. According to \cite{Fischler:1988ia}, the combined effect may be described by a set of modified RG equations,
\be
\frac{\diff \tilde{\lambda}_i(\rho)}{\diff \ln (\rho)} = -\beta \big( \tilde{\lambda}_i(\rho) \big) - \tilde{\alpha}_i(\rho)\,,\label{rg}
\ee
where $\tilde{\lambda}_i=\lambda_i\rho^{\text{dim}(\lambda_i)}$ and $\tilde{\alpha}_i= \alpha_i(\rho)\rho^{\text{dim}(\alpha_i)}$ are the dimensionless coupling constants and $\alpha$ parameters respectively. The first part of (\ref{rg}) is just the standard general form of an RG equation, the additional $\tilde{\alpha}_i$ terms encode the wormhole effect. A redefinition of the $\alpha_i$ is necessary when deriving this from the above stepwise procedure (i.e. when taking the continuum limit $\Delta\rho\to 0$). This is left implicit here.

It will be useful for what follows to spell out the leading terms in the $\beta$ function for the simple three-operator model considered:
\begin{gather}
\frac{\diff \tilde{\Lambda}}{\diff \ln (\rho)} = 4\tilde{\Lambda}+c_1 +\tilde{\Lambda}/\tilde{M}_P^2+\gamma/\tilde{M}_P^2 + \dots + \tilde{\alpha}_1(\rho) \label{laga} \\
\frac{\diff \tilde{M}_P^2}{\diff \ln (\rho)} = 2\tilde{M}_P^2 +c_2+ \tilde{\Lambda}/\tilde{M}_P^2 + \gamma/\tilde{M}_P^2 + \dots + \tilde{\alpha}_2(\rho) \\
\frac{\diff \tilde{\gamma}}{\diff \ln (\rho)} = c_3 + \tilde{\Lambda}/\tilde{M}_P^2 + \gamma/\tilde{M}_P^2 + \dots + \tilde{\alpha}_3(\rho)\,.\label{gaa3}
\end{gather}
Here the two leading terms $4\tilde{\Lambda}$ and $2\tilde{M}_P^2$ correspond to the naive scaling dimension of the operators. The terms $c_i$ arise through the quartic, quadratic and logarithmic divergence of the three operator coefficients in question. The numerical prefactors of all other terms have been suppressed for brevity.\footnote{
A 
very naive way to derive, for example, the first equation is to write the loop corrected cosmological constant in the schematic form $\Lambda=\Lambda_0+c_1\mu^4+\gamma\mu^6/M_P^2+\Lambda\mu^2/M_P^2$. Here the correction terms correspond to the usual one-loop quartic divergence and the leading one-loop tadpole diagrams involving $\gamma$ and $\Lambda$ itself. The expression $(\partial/\partial\ln\mu)(\Lambda/\mu^4)$ gives our desired perturbative $\beta$-function if one identifies the regulator scale $\mu$ with $1/\rho$. Explicit formulae for such $\beta$-functions have more recently appeared in the context of `asymptotic safety', see e.g. \cite{Reuter:1996cp, Litim:2003vp, Falls:2014tra}.
}

As the above discussion shows, the Wilsonian procedure of integrating out high-scale perturbative fluctuations and wormholes induces a dependence of each effective coupling constant $\lambda_i$ on {\it all} the $\alpha$ parameters. The relevant distribution function, e.g. in~(\ref{zpro}), hence becomes
\be \label{eq:P(a)}
P(\alpha) = \exp \bigg( 24\pi^2 \frac{M_P(\alpha)^4}{\Lambda(\alpha)} + \gamma(\alpha) \bigg) \,,
\ee
where in the FKS truncation $\alpha\equiv\{\alpha_1,\alpha_2,\alpha_3\}$ and the analysis is restricted to the single-universe-case for simplicity (hence no double-exponent). We suppress the further complication that one needs a different $\alpha_i$ for each $\rho$, for the whole range of $\rho$. It is sufficient to consider higher scales as having been integrated out, such that $P(\alpha)$ is interpreted as governing the physics at some low effective scale $1/\rho$, with a single set of $\alpha$ parameters, $\alpha_i=\alpha_i(\rho)$, all belonging to that scale. 

The next key point is to understand how the $\alpha$ parameters are related to the wormhole density. To see this, return to the simple toy model with only one wormhole type and thus one $ \alpha$-parameter. Consider the Taylor-expansion of the $\alpha$ distribution:
\be
P(\alpha) = \sum_{n=0}^\infty c_{n}\, \alpha^{n}.
\ee
Under the integral over the $ \alpha$-parameters, this can be rewritten using baby universe operators, cf.~(\ref{eq:BabyOperatorsAlphas}). Thus, the $n$th term in the expansion corresponds to an amplitude with the insertion of $n$ baby universe operators. It represents a configuration with $n$ wormhole ends.
The average number of such wormhole ends is then given by
\be
\overline{N} = \frac{1}{P(\alpha)} \sum_{n=0}^\infty n\, c_{n}\, \alpha^{n} = \frac{1}{P(\alpha)}\,\alpha \frac{\partial P(\alpha)}{\partial \alpha},
\ee
where $P(\alpha)$ appears in the denominator for normalization. Utilizing (\ref{eq:P(a)}) now gives
\be
\overline{N} \sim - \frac{M_P(\Lambda)^4}{\Lambda(\alpha)^2}\,\alpha\frac{\partial \Lambda(\alpha)}{\partial \alpha} + \frac{M_P(\Lambda)^2}{\Lambda(\alpha)}\,\alpha\frac{\partial M_P(\Lambda)^2}{\partial \alpha} + \alpha\frac{\partial \gamma(\alpha)}{\partial \alpha}\,,
\ee
where several unwieldy numerical prefactors were suppressed. 

The curvature-squared of the classical 4-sphere solution is $\sim \Lambda/M_P^2$. Dividing by the corresponding volume, $V_4\sim M_P^4/\Lambda^2$, gives the wormhole density
\bea
\nu = \frac{\overline N}{V_4} \sim  - \alpha\frac{\partial \Lambda(\alpha)}{\partial \alpha},
\eea
where only the leading term in the limit of small $ \Lambda $ have been kept. It is easy to see that the above logic goes through also in the case of multiple $\alpha$ parameters, giving
\be
\nu  \sim - \sum_i \alpha_i\frac{\partial \Lambda(\alpha)}{\partial \alpha_i}.
\ee
Since $\alpha_3$ is driven to infinity, the third term will dominate this expression. The relevant $\alpha_3$ dependence of $\Lambda$ follows from the $\gamma$ term on the r.h. side of (\ref{laga}). This is clear since $\gamma$ involves an additive $\alpha_3$ contribution according to (\ref{gaa3}). Thus,
\be
\nu \sim \alpha_3\, \frac{\partial}{\partial \alpha_3}\,\Lambda \sim
\alpha_3\,\frac{\partial}{\partial \gamma} \left(\frac{\gamma}{M_P^2\rho^2} \cdot\frac{1}{\rho^4}\right)\sim \frac{\alpha_3}{M_P^2\rho^6}\,.\label{dens}
\ee
However, the maximum attainable density of wormholes of size $\rho$ is given by $\nu \sim \rho^{-4}$. Since $\alpha_3$ is driven to large values, (\ref{dens}) will saturate this bound, corresponding to the maximal $\alpha_3$ value 
\be
\alpha_3 \sim M_P^2\rho^2\,.
\ee
It follows that the path integral is dominated by close packing configurations. Moreover, this effect persists as $\rho$ increases, contrary to the  expectation that large wormholes should be suppressed.

Arguments against this so-called FKS or giant wormhole catastrophe were raised in \cite{Preskill:1988na,Coleman:1989ky}, but both proposed resolutions were critizised by Polchinski in \cite{Polchinski:1989ae}. According to \cite{Preskill:1988na}, small wormholes can, when they are packed sufficiently densely, `crowd out' larger wormholes. This `excluded volume' resolution has been criticised in \cite{Polchinski:1989ae} on the grounds that it violates the Wilsonian RG perspective: The effect of small wormholes should not be more drastic than to change the parameters of the effective action at lower energies. Moreover, an explicit toy model calculation was presented to demonstrate that the proposed excluded volume mechanism fails to suppress large wormholes. 

The argument of \cite{Coleman:1989ky} is related but different at the technical level. Here, it is suggested that small wormholes induce charge violating interactions which are sufficiently strong to destabilize larger wormholes. From a microscopic perspective, small wormholes `bleed off' the charge stabilizing the large ones. While this mechanism can be consistent with a Wilsonian RG perspective, it is clearly peculiar to the Giddings-Strominger and related wormhole solutions for which charge (or 3-form flux) is essential. Polchinski \cite{Polchinski:1989ae} argues against this resolution on the grounds that our focus on classical saddle points is merely due to our technical inability to treat more general topology-changing transitions (e.g. euclidean wormholes which do not solve the classical equations of motion). If included, such more general wormholes will not fall victim to the destabilization effect of \cite{Coleman:1989ky}, reinstating the FKS catastrophe. 

Finally, as emphasized in \cite{Hawking:1989vs}, the divergence of the measure $P(\alpha)$ in certain regions of the $\alpha$ parameter space calls for regularization. Depending on the cutoff procedure, different preferred values for the $\alpha$ parameters and hence the effective couplings may be obtained. This can affect both the original argument for vanishing $\Lambda$ as well as the `infinite force' driving $\alpha_3$ to infinity and leading to the giant wormhole problem.

\subsection{Euclidean quantum gravity and negative modes}\label{eqg}
The most immediate suspicion that wormholes should give rise to is that they are based on a very poorly understood sum over four-geometries and topologies, described by the euclidean path integral of quantum gravity. As is well known, this formulation suffers from serious technical and interpretational pitfalls.

Of course, the non-renormalizability of quantum gravity implies that the effective description in terms of the Einstein-Hilbert action will break down at some UV scale (e.g. the string scale) at which new degrees of freedom (excited string modes) will become important. This should not, however, pose an obstacle as long as considerations are restricted to wormholes whose size is much larger than the UV scale, i.e. $\rho\gg \ell_{\text UV}$.

Much more worrisome is the fact that the euclidean version of the Einstein-Hilbert action is unbounded from below. Consider a conformal transformation $g_{\mu\nu}\to \Omega^2 g_{\mu\nu}$, under which 
\be\label{eq:conformal}
S =  -\frac{1}{2} \int \diff^4x \sqrt{g}\, R ~~  \longrightarrow ~~ S=  -\frac{1}{2} \int \diff^4x \sqrt{g} \, \Omega^2R - 3 \int \diff^4x \sqrt{g} g^{\mu\nu}\nabla_\mu \Omega \nabla_\nu\Omega
\ee
By choosing a rapidly varying conformal factor $\Omega$, one can make the action arbitrarily large and negative, even when the original metric $g_{\mu\nu}$ satisfies the equations of motion ($R_{\mu\nu}=0$ in the absence of a cosmological constant). As a consequence, saddle points of the action, including the Giddings Strominger wormhole,  necessarily possess negative modes.

This infamous conformal factor problem has been the subject of much debate, and several prescriptions have been given in order to avoid it. The most common approach, that of Gibbons, Hawking and Perry (GHP)~\cite{Gibbons:1978ac}, amounts to a rotation in the path integral contour such that the conformal factor of the metric takes imaginary values (see~\cite{Schleich:1987fm,Hartle:1988xv,Mazur:1989by} for further discussions). This prescription provides us with a satisfactory action which is bounded from below, but has dramatic consequences for Coleman's argument for a vanishing cosmological constant described in section~\ref{subsec:CosmologicalConstant} (and perhaps more generally for Baum and Hawking's mechanism~\cite{Baum:1984mc,Hawking:1984hk}, of which Coleman's is a particular implementation). The vanishing of the cosmological constant arises from divergent probability amplitudes of the form $P(\alpha)\sim \exp\left[\exp\left(\frac{1}{4\Lambda(\alpha)\kappa^4(\alpha)}\right)\right]$, whose ultimate origin is the conformal factor problem. A complex contour of integration leads to a better defined euclidean quantum gravity, but it results in a crucial change of sign $P(\alpha)\sim \exp\left[\exp\left(-\frac{1}{4\Lambda(\alpha)\kappa^4(\alpha)}\right)\right]$ or $P(\alpha)\sim \exp\left[-\exp\left(\frac{1}{4\Lambda(\alpha)\kappa^4(\alpha)}\right)\right]$, depending on how  the contour rotation is precisely implemented. Either way, these amplitudes give no explanation of the smallness of the cosmological constant~\cite{Polchinski:1988ua,Fischler:1989ka}.

The conformal factor problem also obscures the correct interpretation of wormholes in a different respect. It is well known that, in non-gravitational theories, minima of the euclidean action give a real contribution to the ground state energy, breaking the degeneracy of classically equivalent vacua, e.g. by inducing non-perturbative potentials for axions. In the presence of a negative mode the corresponding contribution to the energy becomes pure imaginary (from the one-loop determinant contribution), signalling an instability of a classical vacuum against tunnelling~\cite{Coleman:1978ae,Coleman:1987rm}. These statements, however, do not generalize straightforwardly to gravitational theories where there is no direct correlation between the euclidean path integral and the WKB prescription, and so the correct interpretation of negative modes remains unclear in this case~\cite{Lavrelashvili:1985vn,Tanaka:1992zw,Lavrelashvili:1998dt,Tanaka:1999pj,Khvedelidze:2000cp,Lavrelashvili:1999sr,Gratton:2000fj,Hackworth:2004xb,Lavrelashvili:2006cv,Dunne:2006bt,Brown:2007sd,Yang:2012cu,Battarra:2013rba,Lee:2014uza,Lee:2014yud,Koehn:2015hga}.

The conformal factor problem would naively suggest that there is an infinite number of negative modes around wormhole solutions. The gravitational action, however, is largely redundant due to its invariance under diffeomorphisms. In order to properly count the number of the negative modes, one should carefully fix the gauge and take constraints into account to identify the physical degrees of freedom of the theory. The negative modes in the conformal sector of the metric are expected to be removed in the process, possibly by the GHP or similar prescriptions. There is an important caveat, however, when one tries to apply this procedure to wormholes. The gauge constraints can only be properly taken into account in the real-time theory, around solutions of the lorentzian equations of motion. Topologically non-trivial manifolds such as wormholes do not admit non-degenerate metrics, and hence cannot represent such real solutions.\footnote{In two dimensions this is known as the Anderson-DeWitt problem~\cite{Anderson:1986ww} (see also~\cite{Strominger:1994tn}), but it is generic to higher dimensions as well~\cite{Horowitz:1990qb}.}  

These subtleties in the interplay between gauge redundancies and constraints, and the transition to euclidean space, have led to contradictory statements regarding the role of negative modes around wormhole solutions. Rubakov and Shvedov have argued in~\cite{Rubakov:1996cn} that, after implementing the GHP rotation, one physical negative-action deformation of the Giddings Strominger wormholes exists. This was interpreted as an instability of large parent universes against decay by emission of baby universes. It has been argued~\cite{Barvinsky:1997fk}, however, that such a negative deformation, being in the conformal sector of the metric, should correspond to a gauge degree of freedom and hence disappear from the spectrum. In fact, an alternative computation in which physical modes were identified in the Lorentzian theory (where the wormhole solution is complex), has more recently found no negative modes~\cite{Alonso:2017avz}. The issue becomes even more obscure in the presence of extra scalar fields (such as the dilatons of section~\ref{sec:dilatonic}), where scalar and metric deformations are intertwined \cite{Kim:1997dm,Khvedelidze:2000cp}, or in the presence of a cosmological constant. The appropriate interpretation of negative modes around wormhole solutions is hence still an open question. 

The above considerations make it clear that the path integral approach to quantum gravity and the role played by gravitational instantons are still obscure. Our degree of understanding of different issues is quite disparate. While still mysterious in many aspects, the euclidean path integral has illuminated important setups of quantum gravity, several of which involve non trivial topologies (including the description of thermodynamic properties of black holes~\cite{Gibbons:1976ue}, the instability of hot flat space against black hole nucleation~\cite{Gross:1982cv}, or the instability of the Kaluza-Klein vacuum~\cite{Witten:1981gj}). It seems hence quite likely that topology change through euclidean wormholes is unavoidable and, following the arguments of section~\ref{sec:WormholeEffects}, will induce corrections in the low energy effective action. The interpretation of the resulting path integral is however still much open to debate. 

Alternative formulations of effective quantum gravity will ultimately be necessary to illuminate these issues. Recently, an approach to the Lorentzian path integral based on Picard-Lefshetz theory has been used in~\cite{Feldbrugge:2017kzv,Feldbrugge:2017fcc,DiazDorronsoro:2017hti,Feldbrugge:2017mbc,DiazDorronsoro:2018wro} to explore certain aspects of quantum gravity. In this approach, wormholes would correspond to complex extrema of the Einstein-Hilbert-axion action. Picard-Lefschetz theory would then determine how the contour in the path integral is to be deformed into the complexified field space, and which saddles contribute to the path integral. It would be interesting to understand in this framework what the role played by gravitational instantons and wormholes is.

In order to shed some light on the conceptual problems raised by wormholes, we describe in the following sections toy models in setups where topology change is better understood, namely, theories of gravity in lower dimensions. 
Although some of the simplifications that arise in such theories surely hide crucial aspects of quantum gravity in four and higher dimensions, they allow us to understand some fundamental aspects of wormholes in relatively controlled settings. 

\subsection{One-dimensional universes: Feynman diagrams}\label{sec:1d}
In the next four subsections (Sects.~\ref{sec:1d}~-~\ref{sec:non-crit}), we discuss the dynamics of the baby universe state and its interplay with the dynamics of our large universe. More precisely, almost all of this discussion will be in the context of toy models, the most developed and complex of which rely on 2d quantum gravity~\cite{Polchinski:1989fn,Banks:1989qe,Banks:1990mk,Hawking:1990ue,Lyons:1991im,Cooper:1991vg}. Such baby-universe and quantum-cosmology toy model calculations have been performed in the context of non-critical string theory~\cite{Polchinski:1989fn,Banks:1990mk,Cooper:1991vg} and will be the subject of Sect.~\ref{sec:non-crit}. However, to prepare the stage, we will start with 1d quantum gravity in the present section~\cite{Strominger:1988ys,Hawking:1990ue}, and its Wheeler-DeWitt formulation with baby universes \cite{Banks:1988je,Strominger:1988ys,Giddings:1988wv,Fischler:1989ka,Cooper:1991vg} in Sect.~\ref{sec:WdW}. Two-dimensional quantum gravity corresponding to critical string theory~\cite{Lyons:1991im,Hawking:1990ue} will be described in Sect.~\ref{sec:2d}.

As promised, we now start with the simplest case following~\cite{Strominger:1988ys,Hawking:1990ue}. Consider the one-dimensional diffeomorphism invariant theory with action
\be\label{eq:worldline}
S[X,e]=\int d\tau  \left(e^{-1}g_{\mu\nu}\dot{X}^\mu \dot{X}^\nu - e m^2 \right)\,.
\ee
This obviously describes a free particle moving in a target space of $D$ dimensions with metric $g_{\mu\nu}$. Upon quantization of the fields $X^\mu$, one is dealing the quantum mechanics of that particle.

The interest here, however, is in interpreting this as a theory of gravity in one dimension. In this sense, one can refer to the particle as the universe, with euclidean worldline element given by $ds^2=e^2 d\tau^2$ and $D$  matter fields $X^\mu$. The parameter $m^2$ hence corresponds to a one dimensional cosmological constant. Of course, such a toy model lacks many interesting features that arise in higher dimensions (to begin with, the Ricci scalar vanishes identically in one dimension, and there is no corresponding Einstein-Hilbert term in~\eqref{eq:worldline}). However, studying topology change in the one dimensional model can illuminate some points that are obscure in higher dimensions. 

The theory described by eq.~\eqref{eq:worldline} is gauge invariant under local time reparametrizations. One can conveniently fix the gauge such that $e=N$, where $N$ is constant.\footnote{In one dimension the vielbein $e(t)$ coincides with the lapse function $N(t)$. The gauge is fixed such that this becomes a constant $N(t)\equiv N$.}  It measures the proper length of the worldline and hence, in a path integral approach, it must be integrated over together with the matter fields:
\be\label{eq:1dpath}
\langle X_f|X_i\rangle_0=\int_0^\infty dN \int_{X_i}^{X_f} {\cal D}X \exp\left[-\int_0^1 d\tau\left(N^{-1}g_{\mu\nu}\dot{X}^\mu \dot{X}^\nu+N m^2\right)\right]\,.
\ee
Here (euclidean) time was chosen to run from $\tau_i=0$ to $\tau_f=1$. The subscript zero indicates that this corresponds to a path integral of a single-component universe, i.e. a single line in the absence of wormholes or baby universes. 

Just as in higher dimensions, the action in~\eqref{eq:1dpath} is unbounded from below if the target space metric $g_{\mu\nu}$ has Minkowskian signature. The negative mode arises in this case from the matter field $X^0$ associated to the target-space time direction. The solution is clear here: one needs to Wick rotate the target spacetime metric (i.e. $X^0\to i X^0$). That is, one considers the propagation of the euclidean one-dimensional universe (particle) through a euclidean D-dimensional target spacetime. From now on, hence, $g_{\mu\nu}$ is considered to to have euclidean signature.

The path integral in~\eqref{eq:1dpath} can be carried out explicitly, yielding~\cite{Strominger:1988ys,Hawking:1990ue}
\be
\langle X_f|X_i\rangle_0=\int d^DP\,  \frac{e^{i P(X_f-X_i)}}{P^2+m^2}
\ee
where scalar products are taken with the target space metric $g_{\mu\nu}$. This is of course nothing but the euclidean propagator of a free (as indicated by the subscript) scalar of mass $m$ in $D$ dimension.

\begin{figure}[ht]
	\centering
	\begin{subfigure}{0.4\linewidth}	
		\centering
	\includegraphics[width=\linewidth]{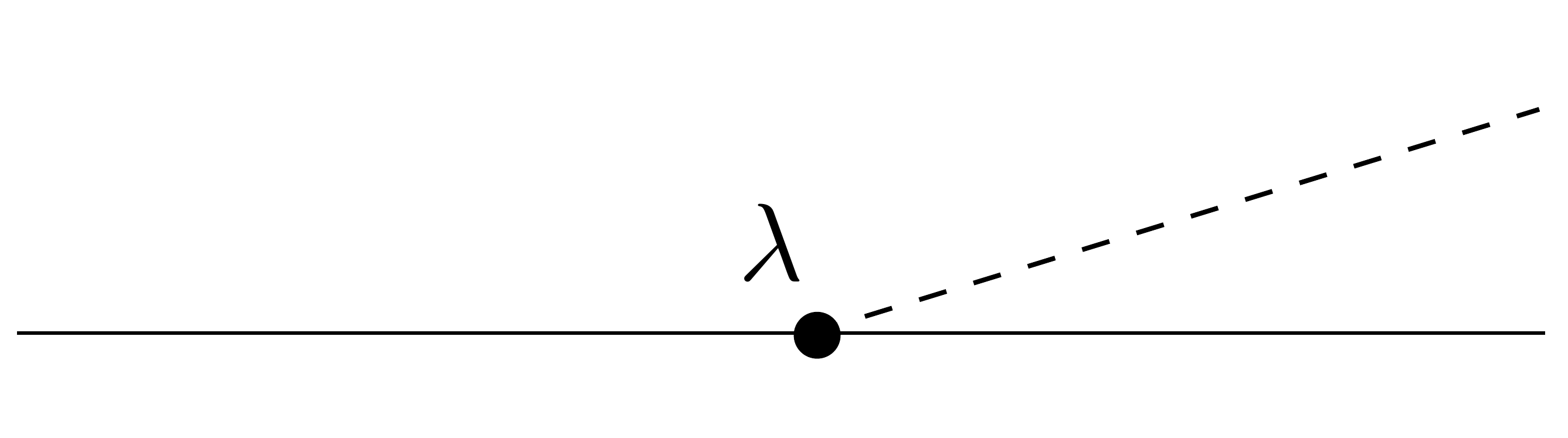}
	\end{subfigure}
	\qquad
	\begin{subfigure}{0.4\linewidth}	
		\centering
	\includegraphics[width=\linewidth]{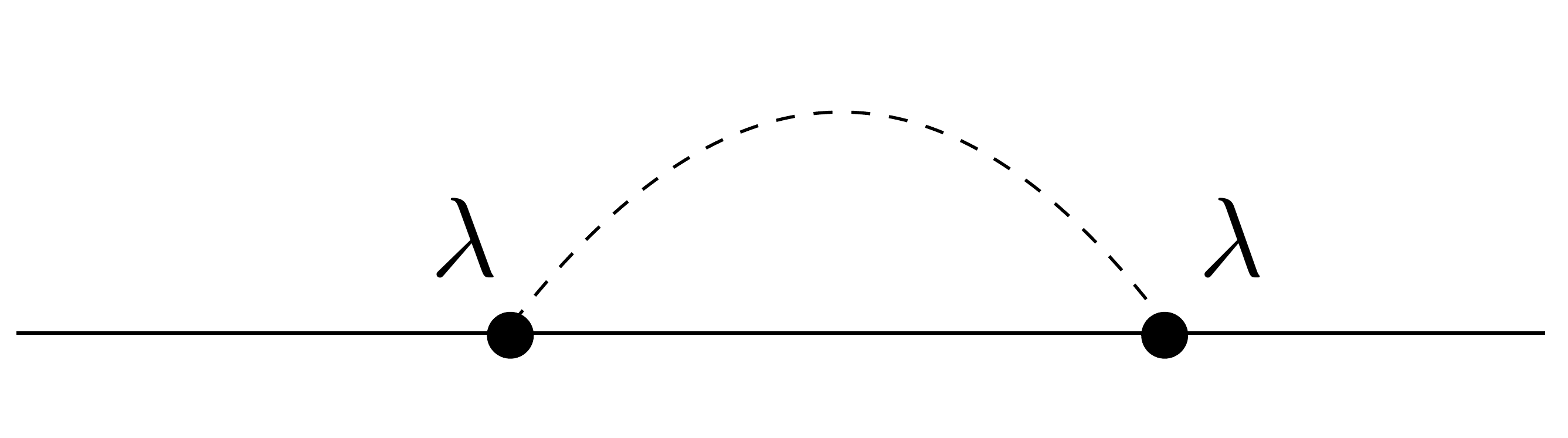}
	\end{subfigure}		
	\caption{Topologically non-trivial processes: Left - a baby universe (dotted line) is emitted from a parent universe (solid line). Right - a wormhole is represented by the emission and absorption of a baby universe line by a parent universe. }
	\label{fig:onedim}
\end{figure}

In order to discuss topology change and the emission of baby universes, one can introduce in the path integral (the sum over one-geometries) processes such as those shown in the figure~\ref{fig:onedim}. To reflect as closely as possible the higher dimensional case, one would like to implement topology change as a process in which small baby universes are nucleated from large ones. Unfortunately, one dimensional universes are pointlike and there is no notion of big or small. One can, however, introduce baby universes as a different species of particles (universes) with much smaller mass than the parent universe. For concreteness, introduce a single type of baby universe with zero mass: $m_b=0$.

The effect of a single wormhole on a parent universe propagator (right diagram of figure~\ref{fig:onedim}) is given by
\begin{eqnarray}
\langle X_f|X_i\rangle_{1}&=&\int_0^\infty dN \int_{X_i}^{X_f} {\cal D}X e^{-S[X,N]}\left(-\lambda^2 N^2\int_0^1d\tau_1\int_0^1 d\tau_2 \langle X(\tau_2)|X(\tau_1)\rangle_{0,b}\right)\\
&=&\int_0^\infty dN \int_{X_i}^{X_f} {\cal D}X e^{-S[X,N]}\left(-\lambda^2 N^2\int_0^1d\tau_1\int_0^1 d\tau_2 \int d^DP\,  \frac{e^{i P[X(\tau_2)-X(\tau_1)]}}{P^2}   \right)\nonumber
\end{eqnarray}
where $\lambda$ controls the coupling between parent and baby universes. Upon summing over arbitrary numbers of wormholes, their contribution exponentiates in a standard fashion to yield
\be\label{eq:fullpropagator}
\langle X_f|X_i\rangle=\sum_{n=0}^\infty \frac{1}{n!}\langle X_f|X_i\rangle_{n}=\int_0^\infty dN \int_{X_i}^{X_f} {\cal D}X e^{-S[X,N]-I[X,N]}
\ee
where $S$ is given by~\eqref{eq:worldline}, and $I$ is the bilocal wormhole contribution:
\be
I[X,N]=\lambda^2 \int \frac{d^DP}{P^2} \left(\int d\tau_2\, N \, e^{iP X(\tau_2)}\right)\left(\int d\tau_1\, N\, e^{-iP X(\tau_1)}\right)\,.
\ee
As in previous discussions of bilocal operators, one can introduce (complex) $\alpha$ parameters to make the action local, at the expense of having variable coupling constants:
\be
e^{-I[X,N]}=\int {\cal D}\alpha(P) e^{-\int d^DP\, P^2 |\alpha(P)|^2}\exp{\left[\lambda  N \int_0^1 d\tau  \int d^D P \left(\alpha(P)  e^{i PX}+\text{c.c.}\right)\right]}\,.
\ee

All this discussion resembles closely the description of wormholes in four dimensions. The one-dimensional theory has split into super selection sectors, labelled by $\alpha(P)$, which determine an infinite set of new couplings on the worldline. The target space momentum $P$ labels the different species of wormholes, in analogy to the index $i$ of the generic wormhole parameters $\alpha_i$ in previous sections. Following the analogy with higher dimensional wormholes, one would affirm that the couplings $\alpha(P)$ have a probability distribution $e^{-|\alpha|^2}$. 

One can also use the advantageous perspective of a parent universe as a particle propagating in $D$ dimensions. As mentioned before, the sum over parent universe one-geometries, in the absence of baby universes, is nothing but the propagator of a free scalar field $\Phi(X)$ in $D$ dimensions. The sum over non-trivial one-geometries is represented naturally by the sum over (connected) Feynman diagrams, where the field $\Phi(X)$ has a cubic coupling to a light baby-universe scalar field $\phi(X)$. That is, all the results described previously can be derived from a quantum field theory in $D$-dimensional target space, with action:
\be\label{eq:superspaceaction}
{\cal S}[\Phi,\phi]=\frac{1}{2}\int d^DX\sqrt{g}\left(g^{\mu\nu}\partial_\mu \Phi\partial_\nu\Phi+g^{\mu\nu}\partial_\mu \phi\partial_\nu\phi+m^2 \Phi^2+\lambda\Phi^2\phi\right)\,.
\ee
It is in fact easy to check that equation~\eqref{eq:fullpropagator} is reproduced by
\be
\langle X_f|X_i\rangle=\int {\cal D}\Phi \,{\cal D}\phi\, \,\Phi(X_i)\Phi(X_f) e^{-{\cal S}[\Phi,\phi]}\,.
\ee
One can furthermore see that the $\alpha(P)$ parameters induced in the worldline effective action by wormholes are nothing but the Fourier modes of the baby universe field
\be
\phi(X)=\int d^DP\,\alpha(P) \,e^{-iPX}\,.
\ee
The $D$-dimensional space on which this $(\Phi,\phi)$-theory lives is nothing but the superspace (in Wheeler's sense) of the one-dimensional universe. 

\subsection{The Wheeler-DeWitt perspective}\label{sec:WdW}

It is also instructive to take a canonical rather than path integral approach to wormholes. The basic ingredient in the canonical treatment of quantum gravity is the Wheeler-DeWitt (WDW) equation, which imposes time reparametrization invariance as a constraint on the wave function of the universe. We follow the discussion of~\cite{Strominger:1988ys}. 

In a one dimensional theory with action given by~\eqref{eq:worldline}, a single universe is described by a wave function on superspace $\Phi(X)$. After fixing the gauge $e(\tau)=N$, the action is invariant under time translations $\tau\to \tau+const$, which are generated by the Hamiltonian
\be
H=g^{\mu\nu}P_\mu P_\nu/4+m^2\,.
\ee
Here, $P_\mu$ are the canonical momenta for the matter fields $P_\mu=\frac{2}{N}g_{\mu\nu} \dot{X}^\nu$. Invariance of the quantum theory under these transformations is imposed by the WDW equation
\be \label{eq:1dWdW}
H\,\Phi(X)=0
\ee
where $P_\mu=-i\,\nabla_\mu$. Equation~\eqref{eq:1dWdW} should describe the dynamics of a one-dimensional (pointlike) universe, i.e. its propagation in target spacetime or superspace. It is not, however, a Schr\"odinger-type equation, but rather a Klein-Gordon equation in a (possibly curved) $D$-dimensional background. This, together with our goal of describing a system of an arbitrary number of universes, naturally suggests that $\Phi(X)$ should be treated like a quantum field in superspace rather than as the wave function of a single universe. 

With this interpretation, the linear WDW equation describes the dynamics of a free quantum field, which acts on the Fock space of an arbitrary number of universes propagating in superspace.\footnote{
Despite 
being a free theory, interesting dynamics, such as universe production, can arise if the target spacetime metric $g_{\mu\nu}$ is curved~\cite{Fischler:1989ka}.
} 
One expects that~\eqref{eq:1dWdW} only represents the leading approximation to a theory of interacting universes. In fact,   such a theory was already introduced in the previous section. The superspace action~${\cal S}[\Phi,\phi]$ of equation~\eqref{eq:superspaceaction} describes the dynamics of a parent universe field $\Phi(X)$, interacting with a baby universe field $\phi(X)$ through a $\lambda \phi \Phi^2$ coupling. The resulting equation of motion for $\Phi$,
\be
\left(\nabla^2-m^2\right)\Phi=\lambda \Phi \phi\,,
\ee
indeed generalizes the WDW equation to the case of interacting universes. 

A meta-observer capable of measuring different multi-universe states would straightforwardly interpret this theory as a  quantum field theory of point-like particles propagating in $D$-dimensional spacetime. However, the interpretation is much more subtle for an observer living on the worldline of a single parent universe propagating in a background of baby universes. Such an interpretation was described in section~\ref{sec:1d}: The sum over one-geometries (Feynman diagrams) derived from the superspace action~\eqref{eq:superspaceaction} is reproduced by the worldline action modified by an infinite set of $\alpha$-parameters, representing the baby universe field $\phi(X)$. In order for the single parent universe approximation to be valid, one has to make sure that the background metric in superspace is adiabatic, and that interactions among universes are small.

In the classical limit of the superspace theory, one can consider the baby universe to be in an eigenstate that satisfies the baby universe equations of motion. That is, one can replace $\phi(X)$ by a solution $\alpha(X)$ of the equation
\be\label{eq:babyeom}
\nabla^2 \alpha= 0
\ee
where the backreaction of parent universes has been neglected.\footnote{Of course, this equation would in general be modified by self interaction terms coming from a baby universe potential $V(\phi)$.} The gravitational worldline theory of the parent universe in such a classical baby universe background is given by the action:
\be
S=\int d\tau \left(\frac{1}{N}\dot X^2-N m- N\lambda \alpha(X)\right)
\ee
The worldline observer would measure a potential given by $\alpha(X)$, which in turn is determined by the superspace equations of motion~\eqref{eq:babyeom}. 

Of course, when quantum fluctuations of the baby universe field are taken into account, the effective coupling constants induced on the parent universe worldline theory are no longer deterministic and are subject to the superspace quantum uncertainty principle. One would conclude, for example, that there is an intrinsic indeterminacy in the worldline potential $\alpha(X)$ once its first derivative has been measured to a finite accuracy. The interpretation of these quantum uncertainties in the worldline couplings is still somewhat obscure. 

The above logic should generalize to higher dimensional theories. In the two-dimensional case, which will be considered in more detail in the following sections, the setup is just string theory. The WDW operator implementing time-reparametrization invariance corresponds to the worldsheet Hamiltonian $H$.\footnote{More generally, the BRST operator which implements full reparametrization invariance.\label{foot:BRST}}  The WDW equation is then $H\Phi(X)=0$, where $\Phi(X)$ is the wave function of a single universe, a function on superspace. In order to discuss multiple universes and topology change, one promotes $\Phi$ to a quantum field, and interprets the WDW equation as the linearized equation of motion of the corresponding superspace theory (a string field theory). This step is sometimes referred to as `third quantization'. Topology change arises when one introduces interactions between string fields, leading to a non-linear generalization of the WDW equation. A two dimensional observer would interpret this theory as a gravitational theory on a genus zero worldsheet, with couplings determined by the background configuration of baby-strings. These would represent, in turn,  target space fields such as the metric, two-form and the dilaton. In the classical limit of the superspace theory, this background could be in a classical state satisfying the equations of motion, which would in turn lead to a determination of the worldsheet couplings. However, just as in one dimension, quantum fluctuation in superspace would lead to an instrinsic uncertainty in such couplings.

One can try to understand four dimensional wormholes in analogy to the previous discussions. The main idea is to promote the wave function of the universe to a field in superspace, and to interpret the WDW equation as the linearized equation of motion of the corresponding quantum field theory. Non-linearities arise when interactions are introduced. These would represent the effects of wormholes in the multi-universe theory. The qualitative picture should be similar to the lower dimensional cases. Four dimensional observers measure coupling constants that are determined by a background of baby universes that propagate in superspace. In the classical limit of superspace, these couplings are determined by the corresponding equations of motion, but in the quantum theory they are subject to the uncertainty principle. 

Unfortunately, the infinite dimensional superspace of four dimensional universes is too complicated for this approach to be tractable in practice. One can drastically simplify the problem by reducing superspace to a finite number of dimensions, e.g. in mini-superspace approximations. An analysis of such setups, with emphasis of phenomenological implications, has been performed in~\cite{Fischler:1989ka} (see also~\cite{Giddings:1988wv}). Baby universes are modelled as small spheres, interacting with large universes through non-linear terms in the WDW equation. The main phenomenological focus is on the cosmological constant problem, for which the outcome appears to be negative: While a variant of the Baum-Hawking-Coleman enhancement at $\Lambda_{eff}\to 0$ is recovered, it occurs for empty and cold universes rather than for inflationary or big-bang cosmologies. A way beyond this negative result would require a  non-standard re-interpretation of boundary conditions in the WDW equations in the multi-universe setting.

\subsection{The two-dimensional case: critical strings}\label{sec:2d}
The one dimensional theory described above is useful in many respects to understand wormhole properties in higher dimensions. It still lacks, however, important ingredients, some of which appear in the much richer context of two dimensional quantum gravity. 

The way non trivial one-topologies in the path integral were introduced was rather {\it ad hoc}. In two dimensions, on the contrary, the sum over non-trivial topologies arises naturally. It is the basis of (perturbative) string theory. Furthermore, the superspace of one dimensional theories of gravity is finite dimensional, in contrast to the infinite dimensional superspace of worldsheet and higher dimensional theories.

Hence, one would like to discuss string theories as two dimensional models of quantum gravity. As is well known, in critical string theory the two-dimensional metric can be (locally) gauged away, and the resulting theory contains only the matter fields $X^{\mu}$, with $\mu=1,\ldots,D$, as physical degrees of freedom.\footnote{We are only considering bosonic degrees of freedom, e.g. by restricting attention to bosonic string theory with $D=26$.}  In conformal gauge, the two-dimensional action is given by
\be
S_P[X]=\frac{1}{2\pi\alpha'}\int d^2z \,\left( \partial X^\mu \bar{\partial} X_\mu+R \Phi_0\right)\,.
\ee
For simplicity,  the $D$-dimensional background on which the string propagates has a flat metric and constant dilaton, and all other fields (tachyon, two-form, etc) vanish.

A single spherical universe with $g$ handles (=wormholes) attached corresponds to worldsheets of genus $g>0$. Just like in the general treatment of previous sections, one can take a dilute wormhole approximation and replace these wormholes with local operators at each endpoints. One can show that these operators are nothing else than the standard vertex operators of string theory~\cite{Lyons:1991im}. 

These vertex operators are in one to one correspondence with target spacetime fields. In the dilute gas approximation, only the lowest string modes contribute significantly. These are, other than possible tachyons, the target space metric, two-form and dilaton fields. They correspond to the traceless symmetric, anti-symmetric, and trace parts of the local vertex operators
\be
V^{\mu\nu}(K;z) = \partial X^{\mu}\bar\partial X^{\nu} e^{iKX}
\ee
where $X(z)$ are the embedding functions of the worldsheeet into target space, and $K$ is a target spacetime momentum. 

As usual, upon summing over wormhole contributions with such vertex operators attached to each end, one gets a bilocal contribution to the two-dimensional effective action,
\be
\Delta S= \int d^DK \left[\left(\int d^2z_1\, V^{\mu\nu}(K;z_1)\right) \Delta_{\mu\nu\rho\sigma}(K) \left(\int d^2z_2 \, V^{\rho\sigma}(K;z_2)\right) \right]\,,
\ee
where $\Delta_{\mu\nu\rho\sigma}(K)$ is the wormhole action, which is nothing but the $D$-dimensional target space propagator of massless gravitons, two-forms and dilatons. 

Once again, one can introduce a set of $\alpha_{\mu\nu}(K)$ parameters to turn this into a local contribution to the worldsheet action. The resulting path integral is:
\be\label{eq:2dZ}
Z=\int {\cal D}X e^{-S_P[X]-I[X]}
\ee
where $S_P$ is the original Polyakov action, and the wormhole contribution is given by a path integral
\be
e^{-I[X]} = \int {\cal D}\alpha_{\mu\nu}(K) \exp\left[\int d^DK \, \alpha \Delta^{-1}\alpha^*\right] \exp\left[\int d^DK \alpha_{\mu\nu}\partial X^{\mu}\bar\partial X^{\nu} e^{iKX}+\mathrm{c.c}\right]\,.
\ee
It is important to notice that, since wormholes have been integrated out, the path integral~\eqref{eq:2dZ} is only over a sphere, which is to be interpreted as the parent universe. The effects of worldsheets with higher genus are encoded in the wormhole contribution $I[X]$ via $\alpha$-parameters. 

From the parent worldsheet point of view, wormholes have introduced a randomness in the coupling constants. Of course, this has a natural interpretation in target space: The $\alpha$-parameters, which can be conveniently denoted $\{\alpha_{\mu\nu}(K)\}=\{G_{\mu\nu}(K), B_{\mu\nu}(K), D(K)\}$, simply describe the background of metric, two-form and dilaton fields on which the string propagates.

So far only the dominant wormhole contributions, coming from  massless string modes, have been considered. Massive modes will of course contribute to terms of higher dimension in the effective action, introducing an infinite set of $\alpha(K)$-parameters. Their quantization, i.e. the path integral over this infinite set of target space fields, should lead to string field theory (this interesting relation goes beyond the scope of this review).

\subsection{Two dimensional quantum cosmology}\label{sec:non-crit}
In this section we would like to consider two dimensional quantum cosmology in baby universe backgrounds as a toy model of the four-dimensional case. String theory in critical dimensions is not ideal for this purpose since the worldsheet metric can be gauged away. One can nevertheless follow~\cite{Polchinski:1989fn, Banks:1989qe, Banks:1990mk, Cooper:1991vg, DaCunha:2003fm} and can consider a generally covariant theory with scalar matter fields $X^i$, with $i=1,\ldots,D$ and general target space dimension $D$:
\be
S=\frac{1}{8\pi}\int d^2\sigma\sqrt{\gamma}\left[\gamma^{ab}\partial_a X\cdot \partial_b X+\omega R + \lambda\right]\,.
\ee
Here $\gamma_{ab}$ is the worldsheet metric, $\lambda$ is the cosmological constant, and the topological $\omega R$-term counts the genus of the worldsheet. The signature of the $D$-dimensional $X$-space is taken to be euclidean. It is useful to fix the gauge such that the metric becomes $\gamma_{ab}=e^{\phi}\hat{\gamma}_{ab}$, where $\hat{\gamma}_{ab}$ is an arbitrary fiducial metric. The path integral over worldsheet metrics reduces to that over the Liouville field $\phi$, with an action determined by the conformal anomaly~\cite{Polyakov:1981rd}:
\be\label{eq:S2d}
S=\frac{1}{8\pi}\int d^2\sigma \sqrt{\hat{\gamma}}\left\{\hat\gamma^{ab}\partial_a X\cdot \partial_b X+\lambda e^{\phi}+\frac{26-D}{12}\left[\hat{\gamma}^{ab}\partial_a\phi\partial_b\phi+2\hat R\phi\right]\right\}\,.
\ee
Here $\omega$ has been reabsorbed by a shift of $\phi$ and a rescaling of $\lambda$. The equations of motion for $\phi$ are solved by metrics of constant curvature $R(\gamma)=R(e^{\phi}\hat{\gamma})\sim \lambda$, supporting the interpretation of $\lambda$ as a two-dimensional cosmological constant.

Notice that in~\eqref{eq:S2d} the action for the metric degree of freedom $\phi$ takes the same form as that for the matter fields $X^i$. One can naturally interpret $\{\phi, X^i\}$ as parametrising a $D+1$-dimensional target space on which the string propagates. Interestingly, the target spacetime has euclidean signature for $D<26$, and lorentzian for $D\geq 26$.\footnote{For the Weyl invariant case of the critical string $D=26$ the Liouville mode $\phi$ is a gauge degree of freedom and disappears from the spectrum. Of course this is the best studied case. Lower central charges $D\leq 1$ have also received much attention in the context of matrix models, see e.g.\cite{Klebanov:1991qa,Ginsparg:1993is,Klebanov:1994kv,Martinec:2004td}.} In the latter case, the Liouville mode $\phi$ plays the role of a time-like coordinate in target space. It is this situation that most closely resembles gravitational theories in four dimensions~\cite{Polchinski:1989fn}.

With this interpretation, equation~\eqref{eq:S2d} corresponds to a subset of a more general class of 2d gravitational theories, where all $D+1$ scalars enter on equal footing,
\be
S=\frac{1}{8\pi}\int d^2\sigma \sqrt{\hat\gamma}\left[T(X)+\left(\hat{\gamma}^{ab}G_{\mu\nu}(X)+i \epsilon^{ab}B_{\mu\nu}(X)\right)\partial_aX^\mu\partial_b X^\nu+2\hat{R}\Phi(X)+\ldots\right] \,
\ee
with $X^0$ corresponding to the Liouville mode $\phi$. The function $T(X)$ plays the role of a cosmological constant. Preserving two-dimensional diffeomorphism invariance at the quantum level is equivalent to conformal invariance and imposes strong constraints on the couplings $\{T,\,G_{\mu\nu},\Phi, B_{\mu\nu},\ldots\}$, namely the vanishing of their $\beta$-functions. These constraints correspond, in $(D+1)$-dimensional target space, to the equations of motion of a tachyon, the metric, and the dilaton fields (setting $B_{\mu\nu}=0$ for simplicity):
\begin{eqnarray}
\nabla^2T-\nabla\Phi\cdot \nabla T&=&V'(T)\,,\\
\nabla^2\Phi-2\left(\nabla\Phi\right)^2&=&\frac{1}{6}(D-25)+V(T)\,,\\
R_{\mu\nu}-\frac{1}{2}G_{\mu\nu}R&=&-2\nabla_\mu\nabla_\nu\Phi+G_{\mu\nu}\nabla^2\Phi+\nabla_\mu T\nabla_\nu T-\frac{1}{2}G_{\mu\nu}(\nabla T)^2
\end{eqnarray}
where $V(T)=-T^2+\ldots$ is the target space tachyon potential. 

These equations describe the dynamics of the background on which the string propagates. In our context, this background is the ``baby universe state'' surrounding our spacetime. It is a condensate of baby universes in the same sense that the string target space is a condensate of string states. Since the background equations of motion arise from the requirement of diffeomorphism invariance of the worldsheet, they should contain the 2d WDW equation. Non-linearities in these equations go beyond the standard WDW framework and reflect baby universe interactions. In other words, they come from topology change.

A solution is given by the linear dilaton background
\be\label{eq:2dsolution}
T(X)=0\,,\qquad G_{\mu\nu}=\eta_{\mu\nu}\,,\qquad \Phi(X)=-\sqrt{\frac{D-25}{12}}X^0\,.
\ee
Notice that the dilaton controls the string coupling $g_s\sim e^{\Phi}$. The semiclassical regime is realized in the limit $D\to\infty$ for positive $X^0$. At early time $X^0$, the theory is strongly coupled. In the solution~\eqref{eq:2dsolution}, the tachyon is balanced on top of its potential. This vanishing of the two-dimensional cosmological constant is obviously unstable against condensation of tachyons. In the linearized approximation $V(T)=-T^2$, this happens with a homogeneous profile (in the limit $D\to \infty$)
\be
T(X^0)=\lambda e^{\sqrt{\frac{12}{D}}X^0}\,.
\ee
This solution is valid for small values of $T(X)$. Higher order terms in the tachyon potential $V(T)$ soon become relevant as the tachyon rolls down, but are hard to compute. It is conceivable that these terms produce a minimum away from zero, leading to a stable solution with constant $T$. It has been argued~\cite{Cooper:1991vg} that this stability, i.e. the absence of growing modes in the WDW equation, will be interpreted by the worldsheet observer as the vanishing of the cosmological constant.

\subsection{Wormholes in AdS/CFT}
In the last few sections we have discussed the interpretation and effects of wormholes in low dimensional theories, where they are relatively well understood. However, given the simplicity of these models, in particular of their gravitational sectors, one should be very cautious when trying to extrapolate conclusions to four dimensional setups. In order to properly tackle the puzzles of wormholes, one needs to study them directly in theories of quantum gravity in higher dimensions. For this, one of the main tools presently at our disposal is the AdS/CFT correspondence. 

Superstring theories in asymptotically AdS spacetimes are dual to conformal field theories living on the boundary~\cite{Maldacena:1997re,Witten:1998qj,Gubser:1998bc,Aharony:1999ti}. The partition function of the CFT should be encoded in a sum over all geometries with the correct asymptotics, possibly including topologically non-trivial ones. If wormhole configurations can be embedded in the low energy supergravity theories that arise in string theory AdS compactifications, one should arguably be able to interpret their effects, and in particular the $\alpha$-parameters they induce, on the field theory side. 

This, however, poses severe problems~\cite{Bergshoeff:2005zf, ArkaniHamed:2007js, Hertog:2017owm, Ruggeri:2017grz}. It has been argued in~\cite{ArkaniHamed:2007js} that AdS wormholes clash with locality of the boundary field theory. The cluster decomposition principle implies that for boundary operators ${\cal O}_1$ and ${\cal O}_2$ separated by a large (Euclidean) time $T$, the CFT correlator can be decomposed as
\be\label{eq:CFT}
\langle {\cal O}_1 {\cal O}_2\rangle=\langle {\cal O}_1 \rangle \langle{\cal O}_2\rangle +{\cal O}(e^{-ET})
\ee
where $E$ is non-zero if the vacuum of the theory is unique. (The argument can also be extended to cases with a finite set of vacua.) Using the AdS/CFT dictionary, the correlators in~\eqref{eq:CFT} should be reproduced on the gravity side by a path integral over geometries. If these include wormholes, $\alpha$-parameters correct the effective couplings. Hence, the two point function on the left hand side of~\eqref{eq:CFT} should be given by
\be\label{eq:AdS}
\langle {\cal O}_1 {\cal O}_2\rangle =\int D\alpha \,e^{-\alpha \Delta^{-1} \alpha} \langle {\cal O}_1 {\cal O}_2\rangle_\alpha=\int D\alpha \,e^{-\alpha \Delta^{-1} \alpha} \langle {\cal O}_1 \rangle_\alpha \langle{\cal O}_2\rangle_\alpha+{\cal O}(e^{-E_\alpha T})\,,
\ee
where the correlators in the integrand are to be computed in the AdS gravitational theory with $\alpha$-shifted couplings. The second equality assumes the factorization (at large $T$ and for fixed $\alpha$) on the AdS side of the duality. We expect this not to be problematic, at least in the classical limit.\footnote{Notice that, for some values of $\alpha$, massless modes could arise, for which $E_\alpha=0$, and the corrections in~\eqref{eq:AdS} would not be exponentially suppressed. This caveat may affect the above argument, although it is not likely that it could reconcile the different structures of~\eqref{eq:AdS} and~\eqref{eq:factors}.} One can similarly compute the expectation values on the right hand side of~\eqref{eq:CFT}:
\be\label{eq:factors}
\langle {\cal O}_1\rangle \langle {\cal O}_2\rangle=\int D\alpha_1 \,e^{-\alpha_1 \Delta^{-1} \alpha_1} \langle {\cal O}_1 \rangle_{\alpha_1} \int D\alpha_2 \,e^{-\alpha_2 \Delta^{-1} \alpha_2} \langle{\cal O}_2\rangle_{\alpha_2}\,.
\ee
Equations~\eqref{eq:AdS} and~\eqref{eq:factors} are inequivalent in general, in contradiction with the locality requirement stated in~\eqref{eq:CFT}. To see this explicitly, assume that ${\cal O}_1$ and ${\cal O}_2$ are actually the same operator, just inserted at different times $t_1$ and $t_2$. Then~\eqref{eq:AdS} gives the expectation value of $\langle {\cal O}\rangle_{\alpha}^2$, interpreted as a function of $\alpha$ and using a  Gaussian probability distribution $P(\alpha)=e^{-\alpha \Delta^{-1}\alpha}$. By contrast,~\eqref{eq:factors} corresponds to the square of the expectation value of $\langle {\cal O}\rangle_{\alpha}$, with the same $\alpha$-distribution. These are equal only if $\langle {\cal O}\rangle_{\alpha}$ is independent of $\alpha$, i.e. if the expectation values computed in AdS are independent of the couplings affected by wormholes.

Another problem is that the presence of wormholes in AdS can result in a violation of the BPS bound on the boundary super Yang-Mills theory~\cite{Bergshoeff:2005zf}. Bulk axions source the $F\wedge F$ operator on the boundary, while the accompanying dilaton (always present in supersymmetric string compactifications) sources the gauge kinetic operator $F\wedge \tilde{F}$. It can be shown~\cite{Bergshoeff:2005zf} that wormholes correspond, on the CFT side, to configurations that violate the BPS bound, namely, for which $\langle |F-\tilde{F}|^2\rangle <0$. These are obviously inconsistent, and pose a problem to the correct interpretation of wormholes in the holographic framework.

One might hope that string theory prevents the presence of wormholes in holographic setups where these paradoxes arise. In fact axions are always accompanied by dilatons in superstring compactifications and, as discussed in Section~\ref{sec:dilatonic}, the existence of regular wormholes solutions depends crucially on their coupling. While the first wormhole constructions in AdS string compactifications indeed were singular~\cite{Rey:1998yx,Maldacena:2004rf}, regular solutions have been obtained more recently~\cite{ArkaniHamed:2007js,Hertog:2017owm,Ruggeri:2017grz}. These analysis suggest that wormholes do exist in controlled holographic setups and hence represent a sharp paradox in AdS/CFT.

The correct resolution of this paradox is still not understood. One possibility is that some mechanism in string theory prevents topology change in holographic setups. One would need to understand in this case how such a mechanism is implemented and if it applies more generally to every string compactification. It could also be that wormholes exist but their effect on the effective action is not given in terms of $\alpha$-parameters (e.g. because of issues with negative modes discussed in Section~\ref{eqg}). Finally, another possibility is that the holographic dictionary, or the correct understanding of the strongly coupled CFT, would encode the $\alpha$-parameters in a so far unknown manner. It is conceivable that the CFT could develop a vacuum degeneracy in its strong coupling regime which is not directly sees and which is only accessible through the $\alpha$ parameters of the gravity dual. Alternatively, one would recover the correct factorization of two-point functions~\eqref{eq:CFT} if one considered the AdS theory to be in an $\alpha$ eigenstate. It is however unknown how the CFT would encode the appropriate value of $\alpha$, or if there is a preferred  $\alpha$ in string compactifications. 

Let us now turn to a related apparent puzzle that arises when a wormhole connects two different AdS spacetimes rather than two distant regions of one AdS space. Such a geometry contains two boundaries and is hence dual to a pair of CFTs. Since the boundaries are disconnected, one naively expects CFT correlation functions of the type $\langle {\cal O}_1(x_1) {\cal O}_2(x_2)\rangle_{\text{CFT}}$, where $x_1$ and $x_2$ belong to different boundaries, to factorize as $\langle {\cal O}_1(x_1)\rangle_{\text{CFT}_1}\, \langle {\cal O}_2(x_2)\rangle_{\text{CFT}_2}$. But this contradicts the gravity side computation: Here, the presence of the wormhole, which connects the two dual AdS spaces, leads to non-trivial correlators between operators on the different boundaries. This problem is similar to the one described above, around~\eqref{eq:CFT}-\eqref{eq:factors}.

\begin{figure}[ht]
	\centering
	\includegraphics[width=0.27\linewidth]{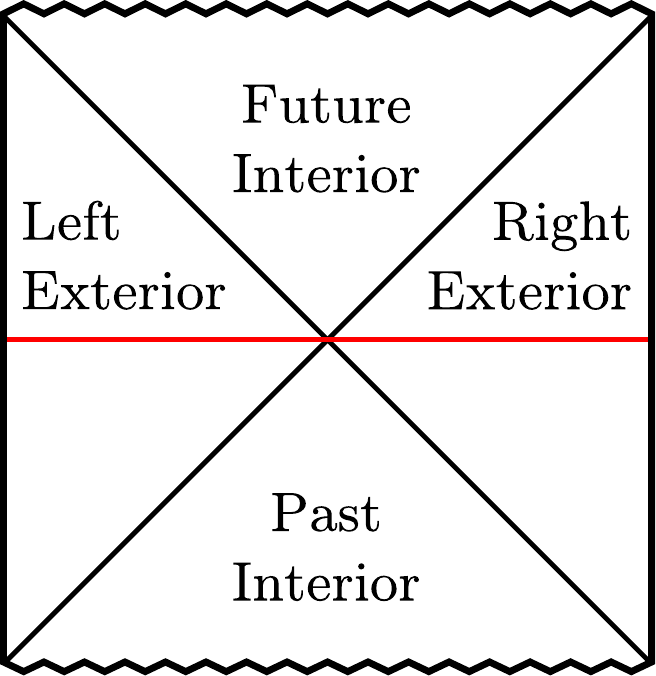}
	\caption{Extended AdS-black hole with two boundaries connected by a wormhole or Einstein-Rosen bridge (horizontal line). 
}
	\label{fig:blackhole}
\end{figure}

In lorentzian signature the resolution of this puzzle is well known~\cite{Maldacena:2001kr}: AdS geometries with multiple boundaries always contain horizons that separate the different boundaries~\cite{Galloway:1999br}.\footnote{See, however,~\cite{Fujita:2011fn, Arias:2010xg}.} The prototypical example is an extended AdS-black hole which has two asymptotic AdS regions connected by a non-traversable wormhole or Einstein-Rosen bridge (see Fig.~\ref{fig:blackhole}). This geometry is dual to a pair of CFTs in an {\it entangled} state, the correlators of which hence do not factorize. Furthermore, the entanglement entropy of each boundary CFT is related to the entropy of the horizon that separates the boundaries. This can be explicitly checked with the Ryu-Takayanagi (RT) or the covariant Hubeny-Rangamani-Takayanagi prescription~\cite{Ryu:2006bv,Ryu:2006ef,Hubeny:2007xt}: The entanglement entropy of a spacelike region $A$ in the CFT is computed in the bulk by the area of a co-dimension-two minimal surface with boundary anchored on $\partial A$. As an example one can take $A$ to be one of the boundaries of an AdS-black hole geometry. Since each CFT lives on a sphere one has $\partial A=0$. The surface measuring the entanglement entropy of $A$ then detaches from the boundary and moves into the bulk, becoming precisely the black hole horizon and hence measuring its area. This relation between Einstein-Rosen bridges and entanglement entropy has led to the remarkable conjecture, known as ER=EPR~\cite{Maldacena:2013xja}, which says that entangled states (even microscopic ones) are generally described by wormholes.

While lorentzian wormholes, including their description in AdS/CFT, are a fascinating subject (see e.g.~\cite{ Visser:1995cc, Maldacena:2018lmt}), the focus of the present review is different: We are interested in euclidean wormholes, their interpretation as tunnelling events, and the resulting  contribution to the effective actions of gravitational theories. Unfortunately, it is not immediately clear how to carry over the above discussion, especially the elegant resolution of the paradox, to this setting. 

\begin{figure}[ht]
	\centering
	\includegraphics[width=0.45\linewidth]{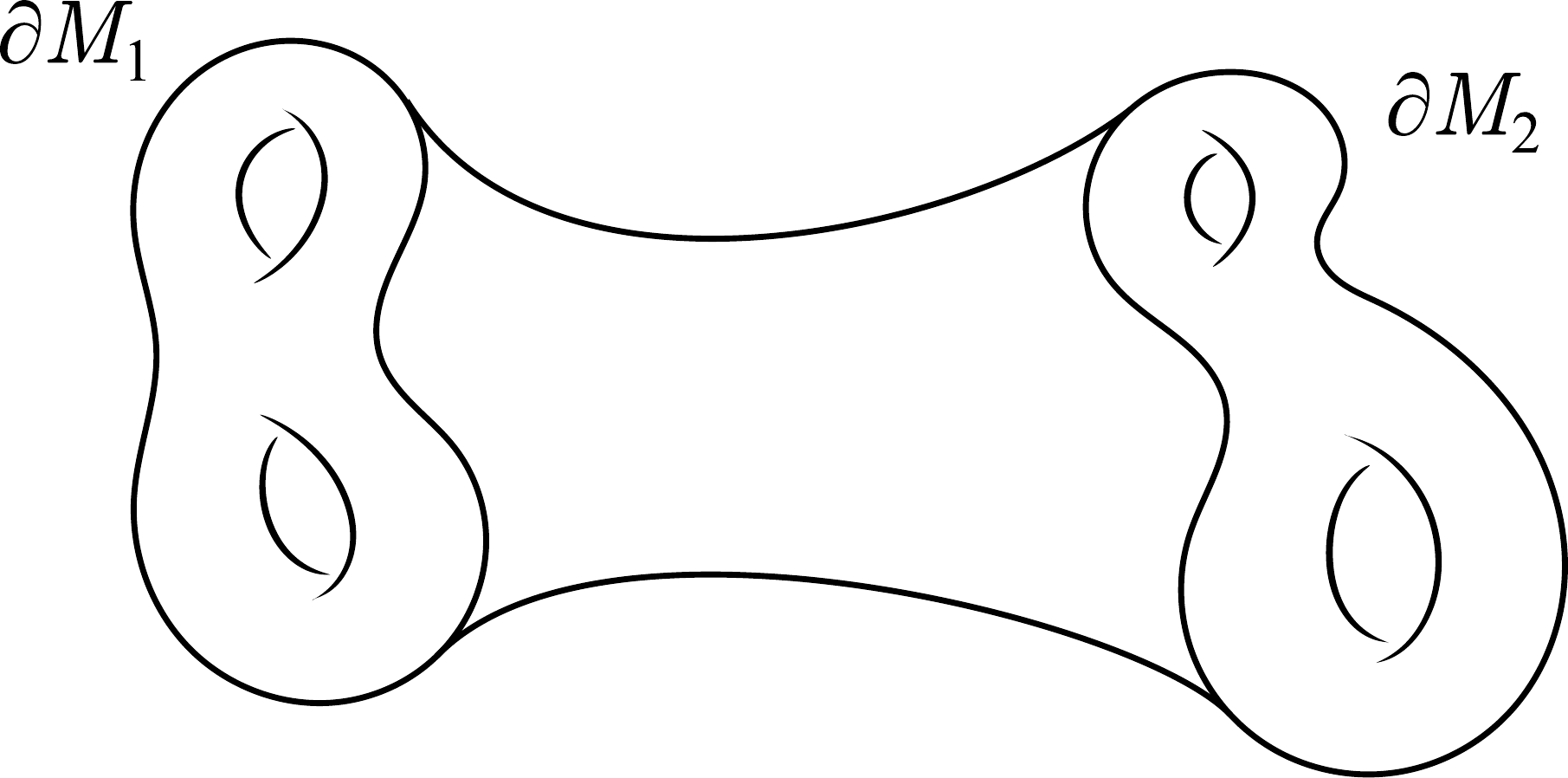}
	\caption{Illustration of a 3d euclidean AdS wormhole geometry with the two boundary components $\partial M_1$ and $\partial M_2$ corresponding to Riemann surfaces \cite{Maldacena:2004rf, Hubeny:2007xt}.} \label{adsw}
\end{figure}

A promising way forward may be to consider euclidean rather than lorentzian wormholes which connect AdS spaces with two disconnected boundary components. The latter correspond to two euclidean CFTs \cite{Maldacena:2004rf}. Euclidean wormholes are different from their lorentzian counterparts in that they do not posses a horizon separating the two boundaries. In fact, the simplest examples are obtained by starting with global AdS and modding out a discrete symmetry. In this case there are no matter fields supporting the wormhole throat and one can not think of the wormhole ends as being localized at arbitrary points inside AdS spaces. Rather, the whole AdS space is the wormhole (cf.~Fig.~\ref{adsw}).

Due in particular to the absence of horizons, the relation of these wormholes to entanglement entropy is not immediately clear~\cite{Maldacena:2004rf}. 
However, Hubeny, Rangamany and Takayanagi~\cite{Hubeny:2007xt} have made a very intriguing suggestion (conceptually related to their time-dependent generalization of RT) for interpreting such euclidean wormholes in terms of entangled CFT states. The idea is to focus on a CFT state corresponding, for example, to a 1-cylce $A$ in $\partial M_1$ in Fig.~\ref{adsw}. In this case the entangling surface is the minimal-length 1-cycle in the bulk to which $A$ can be deformed. This relates to the minimal width of the wormhole at its waist. Interesting extensions include those to multiboundary wormholes~\cite{Balasubramanian:2014hda}, to situations with inflating wormhole interiors~\cite{Fischetti:2014uxa}, and many others (see, e.g.~\cite{Mandal:2014wfa, Maxfield:2014kra}). In our context, the crucial question is whether such an entanglement interpretation of euclidean wormholes holds the key to resolving the problems described above. In particular, it is tantalizing to think that some generalization of ER=EPR can be applied to Euclidean wormholes, perhaps giving them a description in terms of `entangled instantons'. Can one hope to obtain a satisfactory interpretation of Coleman's $\alpha$-parameters in holographic setups in this way? At a more modest level, the mere existence of well-established entangelement entropy interpretations of euclidean AdS wormholes strengthens the case of this review by making it less likely that such exotic objects can be dismissed altogether. Summarizing, it seems clear that AdS/CFT correspondence and holographic entanglement entropy suggest promising avenues to resolving the long-standing puzzles posed by wormholes.

\section{Conclusions}\label{conc}
We have reviewed a number of issues, both theoretical and phenomenological, arising in the context of gravitational instantons, euclidean wormholes and baby universes. The more recent interest in this old subject is related to the weak gravity conjecture, which is further connected to the interplay between charged microscopic objects and charged black branes. In the special case of the axion or $0$-form gauge field, this is the interplay between microscopic instantons and gravitational instantons or wormholes.

The latter case is, however, very special. Indeed, if one insists that the macroscopic charged objects are not UV-sensitive, then cored (i.e. singular) gravitational instantons are excluded and the objects to be considered are the Giddings-Strominger wormholes. Those can be interpreted as tunneling processes in which an $S^3$ baby universe is emitted in some region of 4d non-compact space-time and re-absorbed at an arbitrarily distant location. Allowing such processes unavoidably introduces a baby-universe state, characterized by so-called $\alpha$ parameters, into our description of reality. This is a form of indeterminacy reminiscent of that induced by the string theory landscape. It is, however, of very different conceptual origin and potentially more severe in that parameters are scanned in a continuous way. While the axionic euclidean wormhole solution of Giddings and Strominger played a prominent role in the inception of this picture, it is really not that central: All one needs is some form of topology change.

Crucially, not only the instanton-induced axionic cosine potential is affected, but all coupling constants of the 4d effective theory. Historically, Coleman's suggested solution to the cosmological constant problem played a crucial role in this discussion. This was based on the attempt to integrate over the $\alpha$ parameters together with the 4d geometry, producing a probabilistic distribution of $\Lambda$-values infinitely peaked at zero. However, this has become less believable due to severe technical problems and the fact that arguably a cold and empty universe is predicted. 

The more modest recent discussions of phenomenology have mostly been limited to the axionic cosine potential, under the assumption that the relevant $\alpha$ parameters take their arguably natural ${\cal O}(1)$ value. For (effective) axions with $f>M_P$, this is relevant in the context of large-field inflation, where wormholes could in principle have a sizeable impact in the inflaton potential~\cite{Montero:2015ofa,Hebecker:2016dsw}. However, it turns out that in this regime only wormholes with large 3-form flux are semiclassically controlled. As the UV cutoff is lowered, the required charge grows together with the wormhole action, and the induced potential falls exponentially. Thus, bounds independent of microscopic instantons and the weak gravity conjecture are hard to obtain. The potentially strong constraints on large-field inflation arising from the weak gravity conjecture are being intensely studied, and are one of the main reasons for the current interest in wormhole physics~\cite{Montero:2015ofa,Heidenreich:2015nta,Hebecker:2016dsw}.

By contrast, for small-$f$ axions, even minimally charged wormholes have radii above $M_P^{-1}$ and are semiclassically controlled. This leads to interesting limits on, or even predictions of, axion masses for axions without (or with highly suppressed) microscopic instantons~\cite{Alonso:2017avz}. Such bounds have immediate phenomenological relevance for black hole superradiance and light or ultralight dark matter. In the specific case of the QCD axion, the wormhole-induced potential starts to compete with the QCD-instanton effects at $f\sim 10^{16}$ GeV, potentially spoiling the solution of the strong CP-problem at such relatively large decay constants.

While the above phenomenological considerations are intriguing and deserve further developement, it is important to emphasize that deep conceptual issues remain unresolved. First, the Giddings-Strominger wormhole is a solution of euclidean quantum gravity and the status of the latter is unclear. This is in particular due to the negative modes associated with the conformal factor. Also, the question of whether the Giddings-Strominger solution has negative modes beyond those generically present in euclidean gravity, and how they should be interpreted, is being controversially discussed. However, we consider it unlikely that arguments along those lines can be strong enough to entirely forbid womhole-type tunneling events. Indeed, in quantum mechanics, as a rule of thumb `anything that can happen will happen', even without a stable euclidean saddle point. In this case one needs to understand which role, if any, is played by topology change, and what are the resulting effects on the low-energy effective field theory (e.g. whether $\alpha$ parameters arise). 

More drastically, one could argue that topology change may be strictly forbidden. Indeed, in lorentzian signature no smooth and everywhere defined metric can exists on a space-time `with a handle'. Thus, if one wants to think of the corresponding tunneling trajectory directly in the lorentzian theory, one is forced to deal with (mildly) singular points. We can not rule out that this will probe unknown UV-features of the theory which will rule out the desired transitions. However, it must also be said string theory as our best candidate for a theory of quantum gravity is built on topology change in 2d and includes many examples of well-understood and controlled topology change in 10d. Thus, we find a generic censorship of topology change unlikely. 

Accepting wormholes as a feature of the theory, the problems do unfortunately not end: One has to deal with the $\alpha$ parameters and simply integrating over them together with the metric may lead to problems. One extreme instance of this is known as the Fischler-Kaplunovsky-Susskind catastrophe, which states that under reasonable assumptions the density even of large wormhole ends in 4d space becomes large and the dilute-gas approximation breaks down. 

Stepping back and considering the role of $\alpha$ parameters from a more fundamental perspective, one discovers that simply integrating over them is too simplistic. Indeed, the proper approach is the Wheeler-DeWitt equation describing the full dynamics of a superposition of many large universes interacting with a wormhole baby universe `gas' surrounding them. A helpful 1-dimensional analogy which we described is that of a heavy particle (electron) emitting and absorbing light particles (photons), the cloud of which represents a background field. The latter corresponds to the $\alpha$ parameters, which hence have their own quantum dynamics. The Wheeler-DeWitt equation in this case encodes a standard quantum field theory. A more elaborate toy model takes the point of view of an observer living on the worldsheet of a string that propagates through target space. To this 2d observer, the sum over worldsheet topologies of string theory represents a sum over wormholes, and his $\alpha$ parameters correspond to target space fields (metric, dilaton, etc.). Thus, understanding the values of $\alpha$ parameters amounts to studying string field theory. Very interesting investigations of this setting have been undertaken in the context of `2d quantum cosmology'~\cite{Cooper:1991vg}. In particular, in the context of non-critical strings, insights into issues such as the emergence of time or the evolution of cosmological parameters (in particular the cosmological constant) and their interplay with wormholes appear to be within reach. 

Unfortunately, even in these toy models, firm conclusions are hard to come by. Furthermore, the deep differences between one- or two-dimensional theories of gravity and higher-dimensional ones make the extrapolation of results highly speculative~\cite{Giddings:1988wv,Fischler:1989ka}. It is conceivable that some mechanism forbids wormholes in four dimensions while allowing them in two dimensions. However, we are not aware of such a constraint. It is hence crucial to obtain insight directly in higher dimensions. A powerful tool we have at hand is the AdS/CFT correspondence. In this context, wormholes pose a new type of puzzles. It has been argued that, while wormholes can be embedded in AdS string compactifications, their interpretation in terms of $\alpha$ parameters lead to problems in the boundary field theory, such as non-localities or violations of the BPS bound~\cite{Bergshoeff:2005zf,ArkaniHamed:2007js,Hertog:2017owm,Ruggeri:2017grz}. The resolution of this conflict remains to be understood.

To summarize: the existence and effects of wormholes in theories of gravity remains, after almost forty years, an important but enigmatic subject with both deep fundamental issues and potential phenomenological applications to be explored. Despite new insights into quantum gravity and string theory, progress in our understanding of wormholes has been slow. Our picture remains rather incomplete. Whether topology change (at low energy) is required or forbidden in four and higher dimensions remains to be conclusively settled. Either possibility opens new questions to be addressed. If wormholes exists, their effects lead, as we have discussed, to several puzzles to be resolved. If wormholes are absent altogether, the censorship mechanism at work needs to be understood. Furthermore, in this case one should also ask what are the model- and UV-independent objects (gravitational instantons) that break global axionic shift symmetry. We believe that these questions deserve further investigation. 

\section*{Acknowledgments}
We would like to thank W.~Cottrell, P.~Henkenjohann, J.~Jaeckel, E.~Kiritsis, P.~Mangat, M.~Montero, F.~Rompineve, G.~Shiu, M.~Sloth, S.~Theisen, T.~Van Riet and L.~Witkowski for useful discussions.  We would also like to thank one of the referees for drawing our attention to the interesting issue of the entanglement interpretation of euclidean wormholes in AdS. This work was supported by the DFG Transregional Collaborative Research Centre TRR~33 ``The Dark Universe''. We acknowledge financial support by Deutsche Forschungsgemeinschaft within the funding programme Open Access Publishing, by the Baden-W\"urttemberg Ministry of Science, Research and the Arts and by Ruprecht-Karls-Universit\"at Heidelberg.

\bibliography{Bibliography}\bibliographystyle{utphys}

\end{document}